\documentclass[emulateapj]{aastex62}

\usepackage[]{graphicx}
\usepackage{amsmath,amssymb,color}
\definecolor{red}{rgb}{1.0,0.0,0.0}

\newcommand{\mjup}{\mbox{$M_{\rm Jup}$} }

\begin{document}

\title{The Gemini Planet Imager Exoplanet Survey: 
\\Giant Planet and Brown Dwarf Demographics From 10--100 AU}

\author[0000-0001-6975-9056]{Eric L. Nielsen}
\affiliation{Kavli Institute for Particle Astrophysics and Cosmology, Stanford University, Stanford, CA 94305, USA}

\author[0000-0002-4918-0247]{Robert J. De Rosa}
\affiliation{Kavli Institute for Particle Astrophysics and Cosmology, Stanford University, Stanford, CA 94305, USA}

\author[0000-0003-1212-7538]{Bruce Macintosh}
\affiliation{Kavli Institute for Particle Astrophysics and Cosmology, Stanford University, Stanford, CA 94305, USA}

\author[0000-0003-0774-6502]{Jason J. Wang}
\altaffiliation{51 Pegasi b Fellow}
\affiliation{Department of Astronomy, California Institute of Technology, Pasadena, CA 91125, USA}
\affiliation{Department of Astronomy, University of California, Berkeley, CA 94720, USA}

\author[0000-0003-2233-4821]{Jean-Baptiste Ruffio}
\affiliation{Kavli Institute for Particle Astrophysics and Cosmology, Stanford University, Stanford, CA 94305, USA}

\author[0000-0002-6246-2310]{Eugene Chiang}
\affiliation{Department of Astronomy, University of California, Berkeley, CA 94720, USA}

\author[0000-0002-5251-2943]{Mark S. Marley}
\affiliation{NASA Ames Research Center, Mountain View, CA 94035, USA}

\author{Didier Saumon}
\affiliation{Los Alamos National Laboratory, P.O. Box 1663, Los Alamos, NM 87545, USA}

\author[0000-0002-8711-7206]{Dmitry Savransky}
\affiliation{Sibley School of Mechanical and Aerospace Engineering, Cornell University, Ithaca, NY 14853, USA}

\author[0000-0001-5172-7902]{S. Mark Ammons}
\affiliation{Lawrence Livermore National Laboratory, Livermore, CA 94551, USA}

\author[0000-0002-5407-2806]{Vanessa P. Bailey}
\affiliation{Jet Propulsion Laboratory, California Institute of Technology, Pasadena, CA 91109, USA}

\author[0000-0002-7129-3002]{Travis Barman}
\affiliation{Lunar and Planetary Laboratory, University of Arizona, Tucson AZ 85721, USA}

\author{C{\'e}lia Blain}
\affiliation{National Research Council of Canada Herzberg, 5071 West Saanich Rd, Victoria, BC, V9E 2E7, Canada}

\author{Joanna Bulger}
\affiliation{Subaru Telescope, NAOJ, 650 North A{'o}hoku Place, Hilo, HI 96720, USA}

\author[0000-0001-6305-7272]{Jeffrey Chilcote}
\affiliation{Kavli Institute for Particle Astrophysics and Cosmology, Stanford University, Stanford, CA 94305, USA}
\affiliation{Department of Physics, University of Notre Dame, 225 Nieuwland Science Hall, Notre Dame, IN, 46556, USA}

\author[0000-0003-0156-3019]{Tara Cotten}
\affiliation{Department of Physics and Astronomy, University of Georgia, Athens, GA 30602, USA}

\author[0000-0002-1483-8811]{Ian Czekala}
\altaffiliation{NASA Hubble Fellowship Program Sagan Fellow}
\affiliation{Department of Astronomy, University of California, Berkeley, CA 94720, USA}
\affiliation{Kavli Institute for Particle Astrophysics and Cosmology, Stanford University, Stanford, CA 94305, USA}

\author{Rene Doyon}
\affiliation{Institut de Recherche sur les Exoplan{\`e}tes, D{\'e}partement de Physique, Universit{\'e} de Montr{\'e}al, Montr{\'e}al QC, H3C 3J7, Canada}

\author[0000-0002-5092-6464]{Gaspard Duch\^ene}
\affiliation{Department of Astronomy, University of California, Berkeley, CA 94720, USA}
\affiliation{Univ. Grenoble Alpes/CNRS, IPAG, F-38000 Grenoble, France}

\author[0000-0002-0792-3719]{Thomas M. Esposito}
\affiliation{Department of Astronomy, University of California, Berkeley, CA 94720, USA}

\author{Daniel Fabrycky}
\affiliation{Department of Astronomy \& Astrophysics
University of Chicago
5640 S. Ellis Ave., Chicago IL, 60615}

\author[0000-0002-0176-8973]{Michael P. Fitzgerald}
\affiliation{Department of Physics \& Astronomy, University of California, Los Angeles, CA 90095, USA}

\author[0000-0002-7821-0695]{Katherine B. Follette}
\affiliation{Physics and Astronomy Department, Amherst College, 21 Merrill Science Drive, Amherst, MA 01002, USA}

\author[0000-0002-9843-4354]{Jonathan J. Fortney}
\affiliation{Department of Astronomy, UC Santa Cruz, 1156 High St., Santa Cruz, CA 95064, USA}

\author[0000-0003-3978-9195]{Benjamin L. Gerard}
\affiliation{University of Victoria, 3800 Finnerty Rd, Victoria, BC, V8P 5C2, Canada}
\affiliation{National Research Council of Canada Herzberg, 5071 West Saanich Rd, Victoria, BC, V9E 2E7, Canada}

\author[0000-0002-4144-5116]{Stephen J. Goodsell}
\affiliation{Gemini Observatory, 670 N. A'ohoku Place, Hilo, HI 96720, USA}

\author{James R. Graham}
\affiliation{Department of Astronomy, University of California, Berkeley, CA 94720, USA}

\author[0000-0002-7162-8036]{Alexandra Z. Greenbaum}
\affiliation{Department of Astronomy, University of Michigan, Ann Arbor, MI 48109, USA}

\author[0000-0003-3726-5494]{Pascale Hibon}
\affiliation{European Southern Observatory, Alonso de Cordova 3107, Vitacura, Santiago, Chile}

\author[0000-0001-8074-2562]{Sasha Hinkley}
\affiliation{University of Exeter, Astrophysics Group, Physics Building, Stocker Road, Exeter, EX4 4QL, UK}

\author[0000-0001-8058-7443]{Lea A. Hirsch}
\affiliation{Kavli Institute for Particle Astrophysics and Cosmology, Stanford University, Stanford, CA 94305, USA}

\author[0000-0001-9994-2142]{Justin Hom}
\affiliation{School of Earth and Space Exploration, Arizona State University, PO Box 871404, Tempe, AZ 85287, USA}

\author[0000-0003-1498-6088]{Li-Wei Hung}
\affiliation{Natural Sounds and Night Skies Division, National Park Service, Fort Collins, CO 80525, USA}

\author{Rebekah Ilene Dawson}
\affiliation{Department of Astronomy \& Astrophysics, Center for Exoplanets and Habitable Worlds, The Pennsylvania State University, University Park, PA 16802}

\author{Patrick Ingraham}
\affiliation{Large Synoptic Survey Telescope, 950N Cherry Ave., Tucson, AZ 85719, USA}

\author[0000-0002-6221-5360]{Paul Kalas}
\affiliation{Department of Astronomy, University of California, Berkeley, CA 94720, USA}
\affiliation{SETI Institute, Carl Sagan Center, 189 Bernardo Ave.,  Mountain View CA 94043, USA}

\author[0000-0002-9936-6285]{Quinn Konopacky}
\affiliation{Center for Astrophysics and Space Science, University of California San Diego, La Jolla, CA 92093, USA}

\author{James E. Larkin}
\affiliation{Department of Physics \& Astronomy, University of California, Los Angeles, CA 90095, USA}

\author[0000-0002-1228-9820]{Eve J. Lee}
\affiliation{TAPIR, Walter Burke Institute for Theoretical Physics, Mailcode 350-17, Caltech, Pasadena, CA 91125, USA}

\author{Jonathan W. Lin}
\affiliation{Department of Astronomy, University of California, Berkeley, CA 94720, USA}

\author{J\'er\^ome Maire}
\affiliation{Center for Astrophysics and Space Science, University of California San Diego, La Jolla, CA 92093, USA}

\author[0000-0001-7016-7277]{Franck Marchis}
\affiliation{SETI Institute, Carl Sagan Center, 189 Bernardo Ave.,  Mountain View CA 94043, USA}

\author[0000-0002-4164-4182]{Christian Marois}
\affiliation{National Research Council of Canada Herzberg, 5071 West Saanich Rd, Victoria, BC, V9E 2E7, Canada}
\affiliation{University of Victoria, 3800 Finnerty Rd, Victoria, BC, V8P 5C2, Canada}

\author[0000-0003-3050-8203]{Stanimir Metchev}
\affiliation{Department of Physics and Astronomy, Centre for Planetary Science and Exploration, The University of Western Ontario, London, ON N6A 3K7, Canada}
\affiliation{Department of Physics and Astronomy, Stony Brook University, Stony Brook, NY 11794-3800, USA}

\author[0000-0001-6205-9233]{Maxwell A. Millar-Blanchaer}
\affiliation{Jet Propulsion Laboratory, California Institute of Technology, Pasadena, CA 91109, USA}
\affiliation{NASA Hubble Fellow}

\author[0000-0002-1384-0063]{Katie M. Morzinski}
\affiliation{Steward Observatory, 933 N. Cherry Ave., University of Arizona, Tucson, AZ 85721, USA}

\author[0000-0001-7130-7681]{Rebecca Oppenheimer}
\affiliation{Department of Astrophysics, American Museum of Natural History, New York, NY 10024, USA}

\author{David Palmer}
\affiliation{Lawrence Livermore National Laboratory, Livermore, CA 94551, USA}

\author{Jennifer Patience}
\affiliation{School of Earth and Space Exploration, Arizona State University, PO Box 871404, Tempe, AZ 85287, USA}

\author[0000-0002-3191-8151]{Marshall Perrin}
\affiliation{Space Telescope Science Institute, Baltimore, MD 21218, USA}

\author{Lisa Poyneer}
\affiliation{Lawrence Livermore National Laboratory, Livermore, CA 94551, USA}

\author{Laurent Pueyo}
\affiliation{Space Telescope Science Institute, Baltimore, MD 21218, USA}

\author{Roman R. Rafikov}
\affiliation{Centre for Mathematical Sciences, Department of Applied Mathematics and Theoretical Physics, University of Cambridge, England}

\author[0000-0002-9246-5467]{Abhijith Rajan}
\affiliation{Space Telescope Science Institute, Baltimore, MD 21218, USA}

\author[0000-0003-0029-0258]{Julien Rameau}
\affiliation{Institut de Recherche sur les Exoplan{\`e}tes, D{\'e}partement de Physique, Universit{\'e} de Montr{\'e}al, Montr{\'e}al QC, H3C 3J7, Canada}

\author[0000-0002-9667-2244]{Fredrik T. Rantakyr\"o}
\affiliation{Gemini Observatory, Casilla 603, La Serena, Chile}

\author[0000-0003-1698-9696]{Bin Ren}
\affiliation{Department of Physics and Astronomy, Johns Hopkins University, Baltimore, MD 21218, USA}

\author{Adam C. Schneider}
\affiliation{School of Earth and Space Exploration, Arizona State University, PO Box 871404, Tempe, AZ 85287, USA}

\author[0000-0003-1251-4124]{Anand Sivaramakrishnan}
\affiliation{Space Telescope Science Institute, Baltimore, MD 21218, USA}

\author[0000-0002-5815-7372]{Inseok Song}
\affiliation{Department of Physics and Astronomy, University of Georgia, Athens, GA 30602, USA}

\author[0000-0003-2753-2819]{Remi Soummer}
\affiliation{Space Telescope Science Institute, Baltimore, MD 21218, USA}

\author{Melisa Tallis}
\affiliation{Kavli Institute for Particle Astrophysics and Cosmology, Stanford University, Stanford, CA 94305, USA}

\author{Sandrine Thomas}
\affiliation{Large Synoptic Survey Telescope, 950N Cherry Ave., Tucson, AZ 85719, USA}

\author[0000-0002-4479-8291]{Kimberly Ward-Duong}
\affiliation{School of Earth and Space Exploration, Arizona State University, PO Box 871404, Tempe, AZ 85287, USA}

\author[0000-0002-9977-8255]{Schuyler Wolff}
\affiliation{Leiden Observatory, Leiden University, P.O. Box 9513, 2300 RA Leiden, The Netherlands}

\begin{abstract}

We present a statistical analysis of the first 300 stars observed by the Gemini Planet Imager Exoplanet Survey (GPIES). 
This subsample includes six detected planets and three brown dwarfs; from these detections and our contrast curves
we infer the underlying distributions of substellar companions 
with respect to their mass, semi-major axis, and host stellar mass.  We uncover a strong correlation between planet occurrence rate and
host star mass, with stars $M_*>1.5M_{\odot}$ more likely to host
planets with masses between 
2--13\mjup and semi-major axes of 3--100au at 99.92\% confidence.  
We fit a double power-law model in planet mass ($m$) and semi-major axis ($a$) for planet populations around high-mass stars ($M_*>1.5M_\odot$) of the form $d^2N/(dm\,da){\propto}m^\alpha \, a^\beta$, finding $\alpha=-2.4\pm0.8$ and $\beta=-2.0\pm0.5$, and an integrated
occurrence rate of 9$^{+5}_{-4}$\% between 5--13\mjup and 10--100au.  A significantly lower occurrence rate is obtained for brown dwarfs around all stars, with 0.8$^{+0.8}_{-0.5}$\% of stars hosting a brown dwarf companion between 13--80\mjup and 10--100au.  Brown dwarfs also appear to be distributed differently in mass and semi-major axis 
compared to giant planets; whereas giant planets follow a bottom-heavy mass distribution and favor smaller semi-major axes, brown dwarfs exhibit just the opposite behaviors. Comparing to studies of short-period giant planets from the RV method, our results are consistent with a peak in occurrence of giant planets between $\sim$1--10au. We discuss how these trends, including the preference of giant planets for  high-mass host stars, point to formation of giant planets by core/pebble accretion, and formation of brown dwarfs by gravitational
instability.

\end{abstract}


\section{Introduction}

Since giant extrasolar planets ($\gtrsim$1 M$_{\rm Jup}$) are typically the easiest to detect for most search techniques, there is a long record of studies that investigate the characteristics of the underlying population.  A principal scientific goal is to distinguish between different planet formation and evolution theories by measuring planet occurrence rates with respect to the properties of the planets (e.g., mass, semi-major axis, eccentricity) and host stars (e.g., mass and metallicity).  A correlation between giant planet frequency and stellar metallicity for FGK stars was among the first empirical findings using radial velocity (RV) data  \citep{gonzalez:1997,fischer:2005}.  \citet{tabachnik:2002} and \citet{cumming2008} analyzed datasets of RV-detected planets and completeness to derive occurrence rates for giant planets, fitting power law distributions to the population in mass and semi-major axis.  \citet{johnson:2007} and \citet{johnson:2010} observed a correlation between RV giant planet occurrence rate and stellar host mass, with more massive stars more likely to host giant planets with $a<2.5$ au, though there is uncertainty in deriving a main sequence mass for subgiants \citep{lloyd:2011,johnson:2013,lloyd:2013}.  Results from asteroseismology suggest spectroscopically-derived stellar masses are too high \citep{johnson:2014,campante:2017}, but not significantly so \citep{north:2017,stello:2017}. Thus the underlying correlation appears robust: for RV-detected giant planets the occurrence rate is proportional to $M_*^\gamma$, with $\gamma = 1.05^{+0.28}_{-0.24}$ \citep{ghezzi:2018}.

Direct imaging in the infrared targets young, nearby stars to detect self-luminous giant planets at high contrast at separations of $\lesssim 1''$ from their parent stars.  This technique is sensitive to giant planets at wider separations ($\gtrsim5$ au), and has thus far been detecting planets primarily around higher-mass stars ($\sim$1.5-1.8 M$_\odot$, e.g., \citealt{Marois:2010b}, \citealt{Lagrange:2010}), despite the lower intrinsic luminosity of lower-mass stars making it easier to image planetary-mass companions.  While higher-mass stars hosting most directly imaged planets is suggestive that the RV trend of rising giant planet fraction with increasing stellar mass holds at wider separations, it has yet to be robustly demonstrated  \citep{wahhaj13,galicher:2016,bowler:2016}.  \citet{lannier16} found that while imaged intermediate mass-ratio companions (e.g., brown dwarfs) appeared equally common around higher-mass stars (spectral types A and F) and lower-mass stars (M dwarfs), for low mass ratio companions (such as giant planets) there was a 74.5\% probability that the distributions around the higher-mass and lower-mass stars were different.  With new observations and a larger statistical sample we reexamine this question.

Brown dwarfs, intermediate in mass between stars and giant planets, have typically been detected in greater numbers by imaging surveys compared to giant planets, despite relatively low brown dwarf occurrence rates \citep{metchev09}.  Recent advances in high contrast imaging instrumentation \citep{Macintosh:2014js,Beuzit:2008}, observing techniques \citep{Marois:2006df}, and data reduction \citep{Lafreniere:2007,Soummer:2012ig,Pueyo2016,Wang:2015th} have led to detections of planets orbiting HR 8799 \citep{Marois:2008ei, Marois:2010b}, $\beta$ Pictoris \citep{Lagrange:2009hq}, HD 95086 \citep{Rameau:2013dr, Rameau:2013ds}, 51 Eridani \citep{Macintosh:2015ew}, and HIP 65426 \citep{Chauvin:2018vm}, among others.  \citet{brandt14} proposed that such wide separation (10--100 au) giant planets and brown dwarfs are part of the same underlying population.  This work tests this hypothesis with new observations.

\section{Survey Description}

The Gemini Planet Imager (GPI) is an instrument optimized to directly detect planets at high contrast within $\sim1''$ radius from their parent star.  Coupling a woofer/tweeter extreme adaptive optics system \citep{poyneer:2014,bailey:2016} with an apodized Lyot coronagraph \citep{soummer:2007,savransky:2014} and an integral field spectrograph \citep{chilcote:2012,larkin:2014}, GPI routinely reaches contrasts between $\sim 10^{-5}$ and $10^{-6}$ at $0.4''$ from bright stars (m$_I\lesssim$ 8.0 mag) after post-processing \citep{Macintosh:2014js,Ruffio2017}.  The coronagraphic inner working angle (IWA) is 2.8$\lambda/D$, or 0.12$''$ in $H$-band, though in practice the contrast near the IWA after postprocessing depends on additional factors such as the total sky rotation achieved during the observations of each star.

The Gemini Planet Imager Exoplanet Survey (GPIES) utilizes GPI for a 600-star survey, begun in late 2014 and set to conclude in 2019 \citep{macintosh:2018}.  Based at Gemini South at Cerro Pachon, Chile, GPIES is targeting the youngest, closest stars in the sky in a systematic survey for giant planets between 10--100 au.  In particular, the goal of the planet-finding campaign is to directly measure the planet occurrence rate, determine how planet occurrence correlates with stellar properties, probe the atmospheres of giant planets and brown dwarfs, and determine the orbital properties of substellar companions. The debris disk portion of GPIES is described in depth in Esposito et al. (2019, in prep.).

Here we present the results from the first 300 stars observed by GPIES.  As the first half of the planned survey, 300 stars represents the natural midcourse point to consider the occurrence rate of giant planets partway through the survey.  The 300th star was observed on UT 2016-09-21, and in this work we consider only GPIES campaign observations on or before this date.  Candidate companions from direct imaging observations can be reduction artifacts, background stars, or bound companions.  Second epoch observations confirm that a source is astrophysical and not spurious, and common proper motion indicates whether the star is a bound companion or a background star (e.g., \citealt{DeRosa:2015jl}, \citealt{nielsen:2017}).  For these 300 stars we have completed second epoch observations of all but seven of the 55 candidate companions, and most of these remaining seven have spectra consistent with background stars;  follow-up observations are ongoing for the stars observed after this date.  Thus the 300-star sample represents an essentially complete sample where candidate companions have been classified into PSF reduction artifacts, background stars, or bound companions.

\subsection{Target Selection and Properties}
\begin{figure}[ht]
\includegraphics[width=\columnwidth,trim=0 10.0cm 0 8cm]{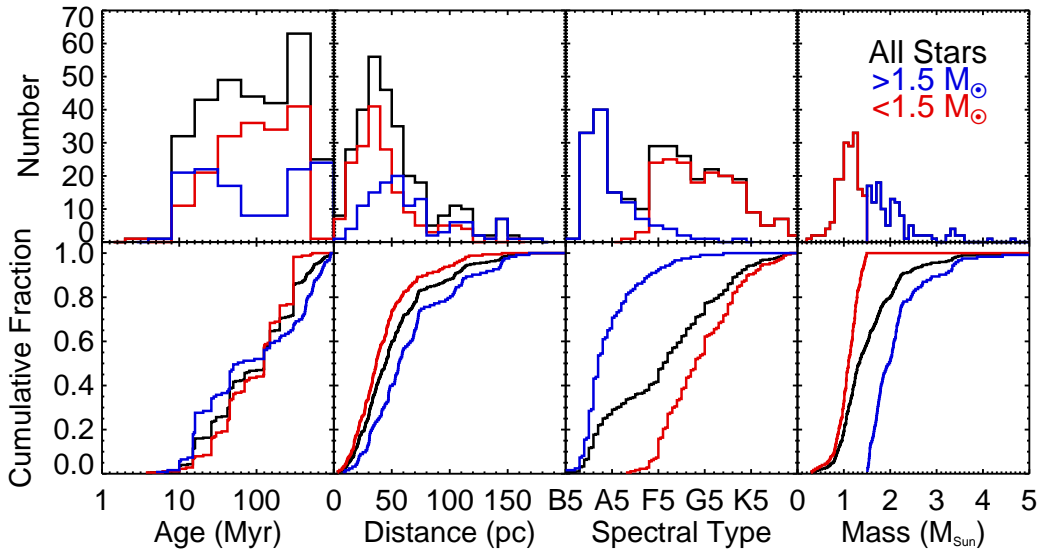}
\caption{Properties of the first 300 GPIES target stars.  While lower-mass stars ($<$1.5 M$_\odot$) are generally nearby (82\% are within 60 pc), higher-mass stars are made up of three groups: nearby moving group stars, nearby field stars, and Sco-Cen stars, with 47\% of higher-mass stars beyond 60 pc.  For these 300 stars, the distributions have median values of 125 Myr, 45.2 pc, F6, and 1.34 M$_\odot$. \label{fig:targ_demo}}
\end{figure}
A sample of young, nearby stars was constructed by combining members of kinematic associations \citep{zucerkman:2001,Zuckerman:2011bo,deZeeuw:1999fe} with X-ray selected FGKM stars within 100 pc that also satisfy the observing limits (m$_I\lesssim9.0$ mag, $\delta \leq 25\degr$). Candidate B and A stars were selected for youth using evolutionary tracks \citep{siess:2000}. We obtained echelle spectra for approximately 2000 candidate stars to further restrict the sample to the $\sim$800 most promising candidates by measuring lithium abundance and chromospheric activity indicators. Apparent binaries with angular separation $0.02$--$3.0\arcsec$ and $\Delta$mag $\leq5$ (including those discovered during the course of the campaign) were then removed because the presence of the companion degrades closed-loop performance. The target list was updated in late-2016 to replace newly-resolved binaries with stars from the original sample, and newly-identified moving group members. The final GPIES sample consisted of 602 nearby stars. Almost half of the final sample are members of kinematic associations; 152 are from nearby moving groups and 104 from the more distant Scorpius-Centaurus OB2 association. Of the remaining 346 stars not in known moving groups or associations, 92 were B and A-type stars selected based on their position on the color-magnitude diagram, and 254 were solar-type stars selected based on the strength of chromospheric activity indicators (e.g., Ca H\&K). A few stars in the sample were selected for other indicators of planetary systems such as debris disk structure.  For example, even though the age of Fomalhaut is $>$400 Myr \citep{mamajek:2013a}, the stellocentric offset of the outer belt \citep{kalas:2005} and eccentric orbit of Fomalhaut b could be the dynamical outcome of a yet-to-be-detected Fomalhaut c in the system \citep{Kalas:2013hp}.

The distributions of various target properties for the first 300 stars observed during the GPIES campaign are shown in Figure~\ref{fig:targ_demo}. Parallax measurements were obtained from the {\it Gaia} DR2 catalog for 251 stars \citep{GaiaCollaboration:2018io}. We used {\it Hipparcos} parallaxes from \citet{vanLeeuwen:2007dc} for the remaining 49 stars because these were either too bright for {\it Gaia} or had worse astrometric precision in the DR2. Spectral types were obtained from SIMBAD, compiled from a number of literature sources. The properties for each target are given in the observing log in Table~\ref{tbl:sample}.

The ages of the stars in the 300-star sample were estimated using a number of methods: 145 based on membership of a nearby kinematic moving group or association \citep{zucerkman:2001,Zuckerman:2011bo}, 109 based on activity indicators in their optical spectra or measured X-ray luminosity, and 48 based on their position on the color-magnitude diagram. For these 109 stars, the process to derive their ages from X-ray and spectroscopic indicators, as well as the new high-resolution spectra obtained for a subset, will be discussed in more detail in an upcoming paper. The ages of the kinematic groups are derived from isochrone fitting (e.g., \citealp{Naylor:2006iz}), lithium depletion boundary analysis (e.g., \citealp{binks:2014}), trace-back analysis (e.g., \citealp{Riedel:2017dg}) and eclipsing or resolved spectroscopic binaries (e.g., \citealp{Nielsen:2016ct}). Applying these various techniques to many coeval stars simultaneously provides tight constraints on the ages of these groups, and as such these ages are preferred.

For the stars not in a known kinematic group, we instead estimated their ages by the measured decrease in lithium abundance and chromospheric activity indicators for lower-mass stars, and the evolution across the color-magnitude diagram for higher-mass stars. We used a combination of lithium abundance, Ca II H/K and H$\alpha$ emission line strength, and X-ray luminosity to assign an age to the lower-mass stars (F-type and later) based on a comparison to similar measures of stars within open clusters and kinematic associations \citep{mamajek:2008,Soderblom:2010}. These methods cannot be applied to the more massive stars within the sample (B through early F-type). For example, lithium depletion only occurs efficiently in stars that have convective envelopes deep enough to transport lithium from the photosphere to regions where the temperature is sufficient for lithium burning to occur. As the depth of the convective envelope is inversely proportional to the mass of the star, this process only occurs efficiently in lower-mass stars \citep{Castro:2016gz}. Early-type stars also do not possess the magnetic field necessary to generate the chromospheric activity indicators observed in later-type stars. Instead, we rely on the rapid evolution of these more massive stars across the color-magnitude diagram to determine their ages. The position of early-type (B and A-type) stars on the color-magnitude diagram is primarily dictated by their age and mass. These stars rapidly evolve onto the zero-age main sequence (ZAMS) where they spend roughly one-third of their main sequence lifetime at their ZAMS location on the HR diagram, before the rising helium abundance in their cores causes them to become cooler and more luminous. Secondary effects include metallicity, rotation, binarity, and extinction, which may need to be considered when attempting to determine the age of a star from its position on the color-magnitude diagram.  Table~\ref{tbl:sample} presents our best estimate for the ages and masses of each target star, the median of the distribution if posteriors are available.  We are currently exploring the production of posteriors for the 109 stars that rely on X-ray, calcium, and lithium age indicators, and will present the results of this analysis and age and mass posteriors for the full sample in an upcoming paper.

\subsubsection{Higher-mass stars: ages and masses}
We combined the stellar evolutionary model MESA \citep{Paxton:2010jf}, used to predict how the fundamental properties of a star evolves with time, with the ATLAS9 model atmospheres \citep{Castelli:2004ti}, used to predict the emergent flux from the stellar photosphere, to estimate both the ages and masses of higher-mass stars and the masses of lower-mass stars based on their position on the color-magnitude diagram. We first generated a grid of evolutionary models spanning a range of masses ($0.6 \le M/{\rm M}_{\odot} \le 9.8$), metallicities ($-0.5 \le [M/H] \le 2.0$), and initial rotation rates ($0.0 \le \Omega/\Omega_c \le 0.7$ where $\Omega/\Omega_c$ denotes the rotation rate as a fraction of the critical rotation rate, see \citealp{Paxton:2013}), assuming a solar abundance of $X_{\odot}=0.7154$, $Y_{\odot}=0.2703$, and $Z_{\odot}=0.0142$ \citep{Asplund:2009eu}. These models were generated using the same parameters used to create the MESA Isochrones and Stellar Tracks (MIST) model grid \citep{Dotter:2016fa, Choi:2016kf}, and our models are consistent with their grid at the rotation rates they computed. Rotation is treated by initializing solid-body rotation at the ZAMS, and we exclude the pre-main sequence (PMS) portion of the evolutionary track given the discontinuity seen at the ZAMS for the rapidly-rotating stars. For higher-mass stars, the PMS is very brief ($\sim10$\,Myr), and it is far more likely that a star consistent with the location of a PMS A-type star on the color-magnitude diagram is actually an older star evolving away from the ZAMS. As with the MIST model grid, stars with $M<1.2$\,M$_{\odot}$ were only computed without rotation.

We collected optical photometry ($G_B$, $G$, $G_R$) from the {\it Gaia} catalogue \citep{GaiaCollaboration:2018io}, combining the catalog uncertainties with the systematic uncertainties reported in \citet{Evans:2018cj}. For the handful of stars too bright for {\it Gaia}, we used Tycho2 photometry ($B_T$, $V_T$; \citealp{Hog:2000wk}) instead. We fit these measurements to synthetic photometry derived from a combination of our evolutionary model and the ATLAS9 model atmospheres. We used the {\tt emcee} affine-invariant Markov Chain Monte Carlo (MCMC) sampler \citep{ForemanMackey:2013io} to sample the posterior distributions of star mass $M$, age $t$, metallicity $[M/H]$, initial rotation $\Omega/\Omega_c$, inclination of the rotation axis $i$, and parallax $\pi$. We used the following priors: 
\begin{itemize}
    \item uniform for $t$ and $\cos i$,
    \item a \citet{Kroupa:2001ki} initial mass function for $M$,
    \item a normal distribution for $[M/H]$ ($-0.05\pm0.11$; \citealp{nielsen:2013}),
    \item a mass-dependent Maxwellian distribution for the rotation rate $v$ \citep{Zorec:2012il},
    \item a combination of a normal distribution ({\it Gaia} or {\it Hipparcos} measurement and uncertainty) and an assumed uniform space density with an exponential drop-off in distance \citep{bailerjones2015} for $\pi$.
\end{itemize}
\noindent If the star is part of a known kinematic group we used a normal distribution described by the age and corresponding uncertainty of that group as the age prior. We also use $v\sin i$ measurements as an additional prior, when available. At each step in the MCMC chain, the rotation rate and radius of the star are combined to give equatorial velocity.  This is combined with the inclination angle, also an MCMC parameter, to give $v\sin i$ for the star at that step in the chain.  The prior on this derived quantity has a form of a gaussian centered on the measured value with a sigma of the reported error.  For known spectroscopic binaries, each stellar component of the system was included in the fit. Three parameters ($M$, $\Omega/\Omega_c$, $i$) were added for each additional star in the system, with a single age, metallicity, and parallax describing the system. The mass ratio measured from the orbit fit to the radial velocities of both components in the double-lined systems was used as an additional prior. For systems that either eclipsed or contained three stars, the  mass measurements for each star were used as priors instead. The masses reported in Table~\ref{tbl:sample} for these spectroscopic systems are the total system mass, rather than the mass of the brightest component.

At each step in the MCMC chain the evolutionary model grid was interpolated to determine the luminosity, equatorial radius, and current rotation rate of the star. A two-dimensional model of the star was constructed using the formalism described in \citet{Lovekin:2006ij} and \citet{EspinosaLara:2011dl} to account for the latitudinal dependence of the effective temperature and surface gravity, and for the angle between the rotation axis and the observer. We sampled the surface of the star equally in latitude and longitude ($90\times180$) to model the star. The emergent flux at each point within each bandpass was computed from the ATLAS9 model atmosphere given the the local effective temperature and surface gravity, the metallicity of the star, and the viewing angle between the surface normal and the observer. The total flux of the star in each bandpass was calculated by weighting the flux from each point by the projected surface area of each segment as viewed from Earth; segments not visible from Earth have zero projected surface area. The total flux was scaled by the distance to the star and compared to the measured photometry. The MCMC chains were advanced until the median and $1\sigma$ credible region derived from the last half of the chain were no longer evolving.

The ages we derive for these higher-mass stars are consistent with previous works using this method.  For example, we derive an age of $750^{+170}_{-190}$ for Fomalhaut, consistent with $520\pm100$ Myr from \citet{nielsen:2013} and $520^{+100}_{-60}$ from \citet{david:2015}.  \citet{mamajek:2012} finds an age for the system of $440\pm40$ Myr, based largely on an analysis of the age of the companion K4V star TW PsA (Fomalhaut B).  This is similar to the two previous analyses of Fomalhaut itself, and about $\sim$1.5 $\sigma$ lower than our estimate.  This age discrepancy is largely a result of different metallicity values used in these analyses, with the \citet{siess:2000} models used by \citet{nielsen:2013} assuming a solar metallicity of $Z_\odot=0.02$, \citet{david:2015} using the PARSEC model grid which assumes $Z_\odot=0.0152$, while \citet{mamajek:2012} used a $Z_\odot=0.017$ and the \citet{bertelli:2008} model grid.  We use a lower value of $Z_\odot=0.0142$ along with the model grid discussed above.  By increasing the metallicity from 0.0142 to 0.017 to match \citet{mamajek:2012}, we recover a more consistent age of $550\pm70$ Myr.  Similarly, increasing the metallicity further to $Z_\odot=0.02$ gives a still lower age, of $440\pm70$ Myr.  As we note above, the \citet{Asplund:2009eu} value of $Z_\odot=0.0142$ more closely matches modern estimates of the protosolar abundance.

For stars in young moving groups, this method tends to overestimate the ages of these higher-mass stars, due to the fact that there is essentially no movement across the HR diagram during the first third of a star's main sequence life.  For example, the absolute $V$ magnitude of a 1.8\,$M_\odot$ star will change less than 0.05 mags from the ZAMS ($\sim$15 Myr) until $\sim$240 Myr, given the models of \citet{siess:2000}.  Thus $\beta$ Pictoris, despite being the namesake of the $26\pm3$ Myr $\beta$ Pic moving group, has a poorly constrained age based on its color-magnitude diagram position alone, with a 1-$\sigma$ upper limit of $<345$ Myr, and $<510$ Myr at 2-$\sigma$.  This is similar to the age derived by \citet{nielsen:2013} of $110^{+20}_{-30}$ Myr, and $520^{+400}_{-170}$ Myr from \citet{david:2015}.

Thus, while there is variation between the different literature age determinations using this method, the precise set of models used and especially assumed protosolar metallicity affects the final shape of the age posterior.  Nevertheless, the ages used for the higher-mass stars in our sample not in moving groups are derived in a consistent manner, rather than relying on a heterogeneous compilation of literature estimates.

\subsubsection{Lower-mass stars: masses}
The above technique was used for stars bluer than $B-V = 0.35$ (earlier than F1).  We used a similar framework to estimate the masses of the lower-mass stars within the sample. Rotation of these stars can safely be ignored; observed $v\sin i$ values are significantly lower than for early-type stars \citep{Gallet:2013gy}, and as such there is a negligible dependence of the effective temperature and surface gravity on stellar latitude. For these stars, the {\tt emcee} sampler was used to sample the posterior distributions of star mass $M$, age $t$, metallicity $[M/H]$, and parallax $\pi$. As for the higher-mass stars, an additional mass term was included in the fit for each stellar companion within known spectroscopic binary systems. A prior on age was applied for these stars, either for the group age for members of moving groups or associations, or the age distribution derived by activity indicators.  At each step the effective temperature, surface gravity, and radius of the star was calculated by interpolating the evolutionary model. The flux from the star in each bandpass was then computed from an interpolation of the ATLAS9 model atmosphere grid, scaled by the square of the ratio of the radius of the star and the distance.  These lower-mass stars not in moving groups or associations are assigned a gaussian age uncertainty with $\sigma$ of 20\%.  We are in the process of deriving more realistic age uncertainties for individual FGKM stars based on calcium emission, X-ray emission, and lithium absorption, and will present updated age and mass posteriors in a future paper.  For this work, then, we assume a single value of age and mass for each star when deriving completeness.  For a future paper detailing the final description of the GPIES sample, however, we will incorporate full posteriors in age and mass.

Extinction is not treated for either the higher-mass or lower-mass stars due to the relative proximity of the stars within the Local Bubble. The overall sample is also relatively free of binaries by construction, although there are undoubtedly spectroscopic binaries that remain undiscovered in the sample. The effect of a binary companion would be to make a star appear more luminous and redder, and therefore typically appear older within our analysis. An overestimate of the age of the star would lead to an underestimate of the sensitivity of the observations of that star to planetary-mass companions, and an overestimate in the overall planet frequency in the final analysis. This bias would be most significant for unresolved binaries with a near equal flux ratio, the type of systems that are most amenable for detection via radial velocity. The effect of this bias on the analysis presented below is expected to be small given that the B and A-type stars with ages determined using this technique only constitute 16\% of the sample, and only a small fraction of them are expected to be undiscovered equal-mass spectroscopic binaries.

There are a total of 12 spectroscopic binaries remaining in the sample and their properties
and estimates of the component masses from the SED fit described previously are given in Table~\ref{tbl:sb}.  There are several physical binaries and multiples wider than $3\arcsec$ (the limit used in constructing the sample), for example GJ 3305 orbits 51 Eridani at $66.7\arcsec$ \citep{feigelson:2006}.  We expect these very wide systems to have less of a dynamical influence on the disk and planet formation.  A more detailed analysis of the effect binarity has on planet formation requires a larger, dedicated sample that is complete to both single stars and binaries, with early work on this front being carried out by the SPOTS survey \citep{asensio:2018}.  In this work, we consider very close SBs and wide-orbit binaries the same as single stars, though for SBs we adopt the combined mass of the system as the stellar host mass.

\begin{deluxetable}{ccccccl}
\tablewidth{0pt}
\tabletypesize{\normalsize}
\tablecaption{Known Spectroscopic Binaries\label{tbl:sb}}
\tablehead{\colhead{Name} & \colhead{Type} & \colhead{$P$} & \colhead{$q$} & \colhead{$M_1$} & \colhead{$M_2$}  & \colhead{Reference}\\
& & (days) & & (M$_{\odot}$) & (M$_{\odot}$)}
\startdata
CC Eri & SB2 & $1.56$ & $0.54\pm0.02$ & $0.54_{-0.03}^{+0.02}$ & $0.29\pm0.02$ & \citet{2000AA...359..159A}\\
HR 784 & SB1 & $37.09$ & \nodata & $1.21_{-0.04}^{+0.02}$ & $0.60_{-0.13}^{+0.11}$& \citet{2018MNRAS.475.1375G}\\
$\zeta$ Hor & SB2 & $12.93$ & $0.88\pm0.03$ & $1.82_{-0.13}^{+0.12}$ & $1.58_{-0.30}^{+0.04}$ & \citet{1964AnAp...27...11S}\\
HR 3395 & SB1 & $14.30$ & \nodata & $1.24_{-0.05}^{+0.05}$ & $0.84_{-0.10}^{+0.07}$ & \citet{2006ApJS..162..207A}\\
4 Sex & SB2 & $3.05$ & $0.945\pm0.002$ & $1.42_{-0.02}^{+0.01}$ & $1.35_{-0.01}^{+0.01}$ & \citet{2003AJ....125..825T}\\
HD 142315 & SB1 & $1.264$ & \nodata & $2.15_{-0.06}^{+0.08}$ & $2.01_{-0.08}^{+0.06}$ & \citet{1987ApJS...64..487L}\\
$\zeta$ TrA & SB1 & $12.98$ & \nodata & $1.13\pm0.01$ & $0.40_{-0.03}^{+0.05}$ & \citet{2004MNRAS.352..975S}\\
V824 Ara & SB2 & $1.68$ & $0.909\pm0.014$ & $1.18\pm0.01$ & $1.08\pm0.01$ & \citet{2000AA...360.1019S}\\
V4200 Sgr & SB2 & $46.82$ & $0.983\pm0.001$ & $0.88\pm0.01$ & $0.87\pm0.01$ & \citet{2017AJ....154..120F}\\
Peacock & SB1 & $11.75$ & \nodata & $5.31_{-0.27}^{+0.29}$ & $4.45_{-0.63}^{+0.36}$ & \citet{1936ApJ....84...85L}\\
$\iota$ Del & SB1 & $11.04$ & \nodata & $1.93_{-0.24}^{+0.10}$ & $0.87_{-0.21}^{+0.76}$ & \citet{1935PDAO....6..207H}\\
42 Cap & SB2 & $13.17$ & $0.727\pm0.005$ & $1.94\pm0.03$ & $1.41\pm0.02$ & \citet{1997AJ....114.2747F}\\
\enddata
\end{deluxetable}

\subsection{Data Acquisition}\label{sec:fmmf}
GPIES has been operating in priority visitor mode, where members of the GPIES team operate GPI to take campaign observations, since the start of the campaign in late 2014. An automated target-picker was used to select the best star to observe given the ages and distances of the remaining unobserved stars, and the current hour angle and environmental conditions \citep{McBride:2011}. Targets with known substellar companions were not prioritized. The instrumental setup was similar for each target. A typical sequence consists of 38 one-minute coronagraphic exposures taken with GPI's integral field spectrograph at a low spectral resolution ($\lambda/\Delta\lambda = 50$) in the H-band (1.5--1.8\micron). The Cassegrain rotator was fixed for all GPI observations causing astrophysical signals to rotate in the IFS images as the target being observed transits overhead. Observations were conducted under program codes GS-2014B-Q-500, GS-2014B-Q-501, GS-2015A-Q-500, GS-2015A-Q-501, GS-2015B-Q-500, and GS-2015B-Q-501. Observations in 2016 were conducted under the 2015B semester program codes. An observing log is given for the first 300 stars observed in the campaign---and any subsequent observations taken to confirm or reject candidate companions---in Table \ref{tbl:sample}.

\section{Data reduction and contrast curves}
GPIES data are processed automatically, with the reduction steps described in detail in \citet{Wang:2018}; we briefly describe them here. Each frame of integral field spectroscopy data from GPI is processed with the GPI Data Reduction Pipeline \citep[DRP;][]{Perrin:2014,Perrin:2016gm}. The GPI DRP produces dark frames and wavelength solutions for each lenslet using calibration data taken during the daytime. Immediately before each observing sequence, a snapshot argon arc lamp is taken and is processed by the GPI DRP to measure the instrument flexure between the observing sequence and the daytime calibration data. For each science frame, the GPI DRP subtracts the dark frame, corrects for bad pixels, compensates the wavelength solution for flexure, constructs 3D spectral data cubes, corrects for optical distortion, and measures the locations and fluxes of the satellite spots that are used for calibration.

A typical observing sequence consists of 38 one-minute exposures, each of which is reduced into a spectral datacube with 37 wavelength channels, resulting in a total of 1406 individual spectral slices, with some correlation between adjacent wavelength channels. The stellar PSF (i.e., speckles) is subtracted from each of these spectral slices. With GPI, we are able to utilize both angular differential imaging \citep[ADI;][]{Marois:2006df} and spectral differential imaging \citep[SDI;][]{Racine1999,Marois2000,Sparks2002} to disentangle the potential planets from the speckles. 
The speckle subtraction and the planet detection are performed using an open-source Python package \texttt{pyKLIP} \citep{Wang:2015th}, which includes an implementation of the Karhunen-Lo\`eve (K-L) Image Projection algorithm \citep[KLIP;][]{Soummer:2012ig,Pueyo:2015}. We follow the forward model matched filter approach described in \citet{Ruffio2017} to which a few improvements, which are described in the following, were made.

The first step of the reduction consists in removing the time-averaged uncorrected atmospheric speckles, which result in a smooth stellar halo steeply rising near the edge of the coronagraphic mask. When the wind is strong, the halo becomes lopsided and aligned with the direction of the wind, sometimes denoted wind-butterfly. Consequently, the halo cannot be efficiently subtracted with a classical spatial high-pass filter and a Gaussian kernel, which was the approached taken in \citet{Ruffio2017}. In this work, we first subtract the radial intensity profile in the image, which removes the symmetric component of the stellar halo. Then, the lopsided component is subtracted from its projection onto a set of principal components computed from a library of GPI images containing a strong wind-butterfly.

The second processing step consists in subtracting the speckles individually in small sectors of each image. The results presented in \citet{Ruffio2017} suffered from artifacts related to the edge of the sectors. In this work, we used tight overlapping sectors centered at each separation in the image therefore mitigating the edge-effects.

Finally, flux, standard deviation and Signal-to-Noise ratio (S/N) maps are computed according to \cite{Ruffio2017} using a matched filter accounting for the heteroskedasticity (i.e., non-uniformity) of the noise and the bias in the planet signal induced by the speckle subtraction \citep{Pueyo2016}. The matched filter assumes a specific spectrum for the planet; however, it was demonstrated that two templates (a cool T-type and a warm L-type spectral template) were sufficient to recover most spectral types \citep[See][]{Ruffio2017SPIE}. The entire survey was therefore reduced twice, once for each template.

The speckle subtraction has several free parameters that were optimized in \citet{Ruffio2017SPIE} for GPI. We used 30 K-L modes to model the stellar PSF. For each subtraction zone, the K-L modes were generated from the 150 most correlated reference images selected from the subset of the observing sequence in which a planet in that zone would move by at least 0.7 pixel in H band due to a combination of ADI and SDI

Planet detection thresholds are defined from the standard deviation map after calibration (in unit of planet-to-star flux ratio). The standard deviation map calibration is described in \citet{Ruffio2017}. It guarantees that the S/N map has a standard deviation of one and corrects for any algorithm related flux losses (i.e., algorithm throughput) using extensive planet injection and recovery. Then, the 1D standard deviation as a function of separation is computed as the root-mean-square of the standard deviation map in 4-pixel concentric annuli.  Detection thresholds are generally derived from a false positive rate chosen a priori. However, the estimation of planet occurrence rates, the goal of this work, requires that all the candidates above the detection threshold be vetted. The vetting process generally requires follow-up observations, which is a limited resource due to telescope time constraints. As a result, we define the detection threshold of $\sigma$ as the boundary used over the GPIES campaign for candidate follow-up. This 1D detection threshold is the contrast curve.
Unfortunately, while the stellar halo subtraction generally improved GPIES' sensitivity at intermediate separations, it introduced a large number of false positives at the very edge of the coronagraphic mask when using the T-type spectral template. We therefore increased the detection threshold to $8\sigma$ below $<0.3''$ in these reductions.  Full contrast curves for the entire GPIES campaign, including the subset discussed here, will be presented along with the final statistical analysis of the campaign in a future paper.

\section{Companion Detections}\label{companions}
\begin{deluxetable}{ccccccl}
\tablewidth{0pt}
\tabletypesize{\normalsize}
\tablecaption{Detected companion properties}
\tablehead{ \colhead{Name} & \colhead{Epoch} & \colhead{$\rho$\tablenotemark{1}} & \colhead{$a_{\rm proj}$\tablenotemark{1}} & \colhead{$\Delta H$} & \colhead{Mass\tablenotemark{2}} & \colhead{Astrometry reference} \\
& & (arc sec) & (au) & (mag) & (M$_{\rm Jup}$)  }
\startdata
51 Eri b & 20141218 & $0.451\pm0.002$ & $13.43\pm0.09$ & $14.4\pm0.2$ & $2.6\pm0.3$ & \citet{DeRosa:2015jl}\\
$\beta$ Pic b & 20151106 & $0.249\pm0.001$ & $4.84\pm0.02$ & $9.3\pm0.1$ & $12.9\pm0.5$ & \citet{Wang:2016gl}\\
HD 95086 b & 20160229 & $0.621\pm0.005$ & $53.65\pm0.46$ & $13.7\pm0.2$ & $2.6\pm0.4$ & \citet{rameau2016}\\
HR 8799 c & 20160919 & $0.944\pm0.001$ & $38.99\pm0.15$ & $12.0\pm0.1$ & $8.3\pm0.6$ & \citet{wang:2018b}\\
HR 8799 d & 20160919 & $0.674\pm0.001$ & $27.85\pm0.11$ & $12.0\pm0.1$ & $8.3\pm0.6$ & \citet{wang:2018b}\\
HR 8799 e & 20160919 & $0.385\pm0.002$ & $15.89\pm0.09$ & $11.6\pm0.1$ & $9.2\pm0.6$ & \citet{wang:2018b}\\
\hline
HD 984 B & 20150830 & $0.216\pm0.001$ & $9.90\pm0.04$ & $6.5\pm0.1$ & $48\pm3$ & \citet{JohnsonGroh:2017kh}\\
HR 2562 B & 20160125 & $0.619\pm0.003$ & $21.07\pm0.11$ & $11.7\pm0.1$ & $26^{+9}_{-13}$ & \citet{Konopacky:2016dk}\\
PZ Tel B & 20150730 & $0.502\pm0.001$ & $23.65\pm0.08$ & $5.3\pm0.1$ & $64\pm5$ & this work
\enddata
\tablenotetext{1}{Projected separation at the first detection in the GPIES campaign.}
\tablenotetext{2}{Mass derived from absolute $H$ magnitude.}
\label{tbl:companion_props}
\end{deluxetable}

\begin{figure}
\includegraphics[width=\columnwidth]{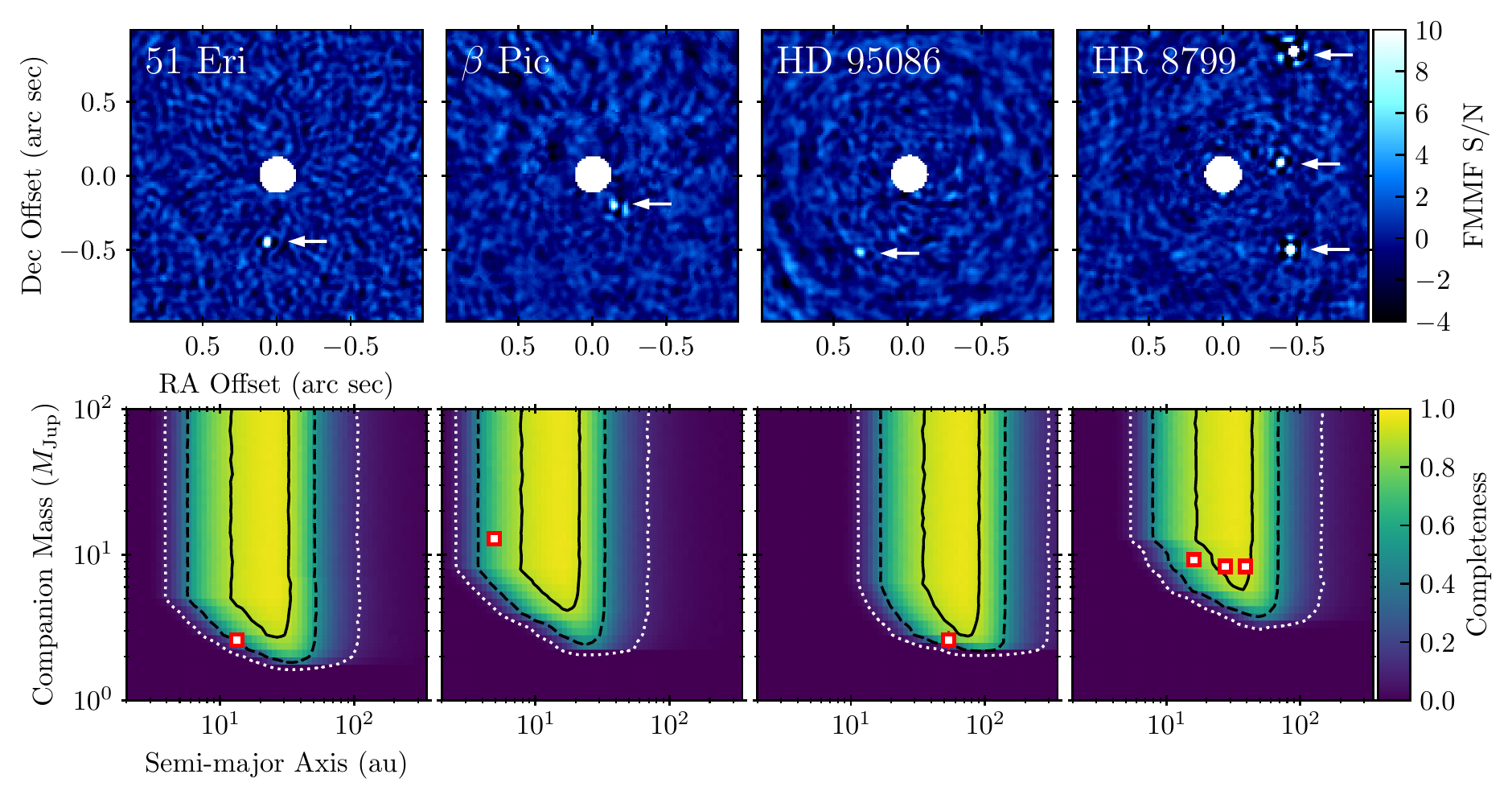}
\caption{(top row) FMMF S/N maps of the four planetary systems resolved within GPIES after image reduction and PSF subtraction. The 51 Eri S/N map was calculated using a T-type template, the others with an L-type template. (bottom row) Corresponding completeness maps for the four stars computed using the procedure described in Section \ref{complete_sec}. Contours are plotted for the 90\% (solid), 50\% (dashed), and 10\% (dotted) completeness levels. The location of each companion is based on the mass derived from the $H$-band photometry and the projected separation of the companion at the epoch of the campaign observation. \label{fig:gallery_planets}}
\end{figure}

Nine substellar companions around seven host stars have been detected within the first half of the GPIES survey, three brown dwarfs and six planetary-mass companions. The measured position, contrast, and derived mass for each companion is given in Table \ref{tbl:companion_props}. The first GPIES campaign image and corresponding sensitivity curves are given in Figure \ref{fig:gallery_planets} for the imaged planets, and Figure \ref{fig:gallery_bds} for the brown dwarfs.  51 Eridani b and HR 2562 were new discoveries from these images, while $\beta$ Pictoris b, HD 95086 b, HR 8799 cde, HD 984 B, and PZ Tel B were detections in our campaign images of known substellar companions.

\subsection{51 Eridani}
51~Eridani~b was discovered early in the GPIES campaign \citep{Macintosh:2015ew} around the F0IV $\beta$~Pictoris moving group member \citep{bell:2015} 51~Eridani (hereafter 51~Eri). The H-band spectrum measured in the discovery epoch exhibited strong methane and water absorption, consistent with both predictions for low-mass exoplanets around young stars, and also the observed spectra of low-temperature T-type brown dwarfs. Follow-up astrometric observations confirmed that 51~Eri~b was in a bound orbit around 51~Eri \citep{DeRosa:2015jl}, with a negligible probability of it being an interloping field T-dwarf in chance alignment with 51~Eri.

Subsequent spectroscopic observations with GPI \citep{Rajan:2017ur} and VLT/SPHERE \citep{Samland:2017vi}, combined with thermal-IR photometry obtained with Keck/NIRC2 (Macintosh+15, Rajan+17), have been used to characterize the atmospheric properties of the planet. The observed spectral energy distribution is consistent with a photospheric cloud fraction between 75-90\% \citep{Rajan:2017ur} and 100\% \citep{Samland:2017vi}, depending on the specifics of the cloud model used in the analysis. Both fits find a similar effective temperature ($T_{\rm eff}=$~600--760\,K), but a range of values for the surface gravity ($\log g=$~3.5--4.5\,[dex]). The estimated mass of the planet is strongly dependent on the assumptions made regarding its formation. Under an assumption of a high-entropy (so-called `hot-start') formation scenario \citep{marley07}, the luminosity of the planet corresponds to a mass of 1--2\,M$_{\rm Jup}$ \citep{Rajan:2017ur}. Uniquely for young directly-imaged companions, the luminosity of 51~Eri~b is also consistent with predictions of low-entropy, `cold-start' formation scenarios \citep{marley07,fortney08}, with a significantly higher model-dependent mass of 2--12\,M$_{\rm Jup}$. Measurements of the reflex motion induced by the orbiting planet, either via spectroscopy or astrometry, are required to differentiate between these two scenarios.

\subsection{$\beta$ Pictoris}
The exoplanet $\beta$~Pictoris~b ($\beta$~Pic~b) was discovered around the A6V namesake of the $\beta$~Pictoris moving group using VLT/NaCo in 2003, later confirmed to be a bound companion in follow-up observations taken with the same instrument in 2008 \citep{Lagrange:2009hq,Lagrange:2010}. The planet is orbiting interior to a large circumstellar disk \citep{smith:1984}, and has a measurable effect on the morphology of the inner disk \citep{Lagrange:2012,MillarBlanchaer:2015ha}.  $\beta$~Pic~b was observed several times with GPI; as a first light target during commissioning, as a part of an astrometric monitoring program to constrain the orbit of the system, and finally as part of the GPIES Campaign.  Due to its proximity and youth, the primary was a highly ranked GPIES target, and was selected by our automated target-picker early in the campaign, resulting in the detection of $\beta$ Pic b.  Milli-arcsecond astrometry obtained with GPI was critical for constraining the orbital parameters of the planet \citep{Wang:2016gl}. Observed variability in the light curve of the host star $\beta$~Pic was interpreted as being caused by a transit of either the planet or by material within its Hill sphere \citep{LecavelierDesEtangs:2009jt}. \citet{Wang:2016gl} used GPI astrometry to tightly constrain the inclination of the orbit of the planet and excluded a transit at the 10-$\sigma$ level, although a transit of the Hill sphere of the planet was still a possibility.  Precise timings on the transit of the planet's Hill sphere were made, allowing for a unique opportunity to probe the circumplanetary environment of a young exoplanet. 

The low contrast between $\beta$~Pic~b and its host star compared to other directly-imaged exoplanets makes it an ideal target for atmospheric characterization. \citet{Chilcote:2017fv} used a combination of GPI spectroscopy and literature photometry (compiled and calibrated by \citealp{morzinski:2015}) to measure the spectral energy distribution of the planet between 0.9 and 5\,\micron, wavelengths spanning the bulk of the emergent flux of the planet. The data were consistent with models that incorporated the condensation of dust into thick vertically-extended clouds within the photosphere, and matched the observed spectral energy distributions of isolated low-surface gravity brown dwarfs \citep{Chilcote:2017fv}.
Unlike 51 Eri b, the bolometric luminosity of $\beta$~Pic b is only consistent with the predictions of the `hot-start' luminosity models, and inconsistent with the \citet{fortney08} `cold-start' models. However, this is not a criticism
of core accretion; see, e.g., 
\citet{Berardo:2017,Owen:2016} for how core accretion can
lead to hot/warm starts, and also our Section \ref{ssec:theory}.
\citet{Chilcote:2017fv} derived a model-dependent mass of $12.7\pm0.3$\,M$_{\rm Jup}$, consistent with the model-independent mass derived from {\it Hipparcos} and {\it Gaia} astrometry of the host star of $11\pm2$\,M$_{\rm Jup}$ \citep{Snellen:2018} and 13$\pm$3 M$_{\rm Jup}$ \citep{dupuy:2019}.

\subsection{HD 95086}
HD 95086 b was discovered around the A8III member of the 15 Myr \citep{Pecaut:2016} Lower Centaurus Crux subgroup of the Scorpius-Centaurus OB2 association \citep{Rizzuto:2011gs} using VLT/NaCo in 2011 \citep{Rameau:2013dr,Rameau:2013ds}, one of a growing number of exoplanets that have been resolved around Sco-Cen members \citep{Bailey:2014,Chauvin:2017ev}. The planet was noted for its unusually red infrared colors that were ascribed to a photosphere dominated by thick clouds \citep{Meshkat:2013fz}. HD~95086 was observed as a part of the GPIES campaign to characterize the atmosphere of the planet, monitor its orbital motion, and to search for additional interior companions.  Like $\beta$ Pic, the target star was selected for observations for our planet-search program by our automated target-picker, resulting in a detection of the planet.  The low luminosity and very red infrared colors of the planet combined with its distance of 86\,pc make it a challenging target for spectroscopic observations, especially at shorter wavelengths. \citet{DeRosa:2016kh} reported the first spectroscopic measurement of the planet obtained using GPI, a spectrum covering the blue half of the $K$ band. This was combined with literature photometry at $J$ and $L^{\prime}$ to reveal a spectral energy distribution consistent with model atmospheres incorporating significant amounts of photospheric dust, a result confirmed in a later analysis incorporating SPHERE observations of the planet \citep{Chauvin:2018vm}.

Monitoring the orbital motion of the planet can provide insight into the architecture of the HD 95086 system. The presence of two large circumstellar dust rings was inferred from the spectral energy distribution of the star \citep{Moor:2013bg,Su:2017kl}; the outer ring was subsequently resolved with millimeter interferometric observations with ALMA \citep{Su:2017kl}. HD 95086 b lies between these two rings and, despite only sampling a small fraction of the full orbit of the planet with GPI and literature astrometry, \citet{rameau2016} were able to constrain the eccentricity of the orbit to be near-circular, under the assumption of co-planarity with the outer ring. While the dynamical impact of the planet on the outer ring is still uncertain \citep{Su:2017kl}, \citet{rameau2016} demonstrated that it is unlikely that the planet is responsible for clearing the gap in to the predicted outer radius of the inner disk, and posited this as indirect evidence of the existence of additional lower-mass companions that are below present detection limits, a finding also confirmed by \citet{Chauvin:2018vm}.

\subsection{HR 8799}
The HR 8799 system is currently the only example of a multiple planet system detected via direct imaging \citep{Marois:2008ei,Marois:2010gp,konopacky:2018}. The host star is a $\lambda$ Bootis star that exhibits $\gamma$ Doradus variability \citep{zerbi:1999,gray:1999}. It is a probable member of the $42^{+6}_{-4}$\,Myr \citep{bell:2015} Columba moving group \citep{Zuckerman:2011bo,malo:2013} and has a spectral type of either A5 and F0, depending on which set of spectral lines are used for typing \citep{Gray:2003fz}. The star is also host to two circumstellar dust rings, at $\sim$10 and $\sim100-300$ au, with the planets occupying the gap between them, and a large diffuse dust halo at $\sim$300-1000 au \citep{Su:2009ig}. While the inner ring is only inferred from the spectral energy distribution of the star, the outer ring has been resolved with ALMA and the SMA \citep{Booth:2016iy,Wilner:2018jy}.

The HR 8799 system has been observed multiple times with GPI; as an early science target during the commissioning of the instrument \citep{Ingraham:2014gx}, and during the campaign to both characterize the atmospheres of the known planets \citep{Greenbaum:2018hz} at multiple bands, and as part of the planet search in $H$ when chosen by the automated target-picker.  The campaign observations resulted in the detection of three of the four planets, HR 8799 cde, with HR 8799 b outside the field of view of GPI in our nominal search mode. Astrometry obtained with GPI was also used in conjunction with measurements obtained with Keck/NIRC2 to investigate the dynamical properties and long-term stability of the system \citep{wang:2018b}.  By utilizing N-body simulations, the combination of astrometry and dynamical limits placed tighter constraints on the orbital parameters and the masses of the individual planets, which when combined with evolutionary models give a mass for HR 8799 b of $5.8 \pm 0.5$ M$_{\rm Jup}$, and $7.2^{+0.6}_{-0.7}$ M$_{\rm Jup}$ for planets c, d, and e.

\subsection{Brown Dwarf Companions}
\begin{figure}
\includegraphics[width=\columnwidth]{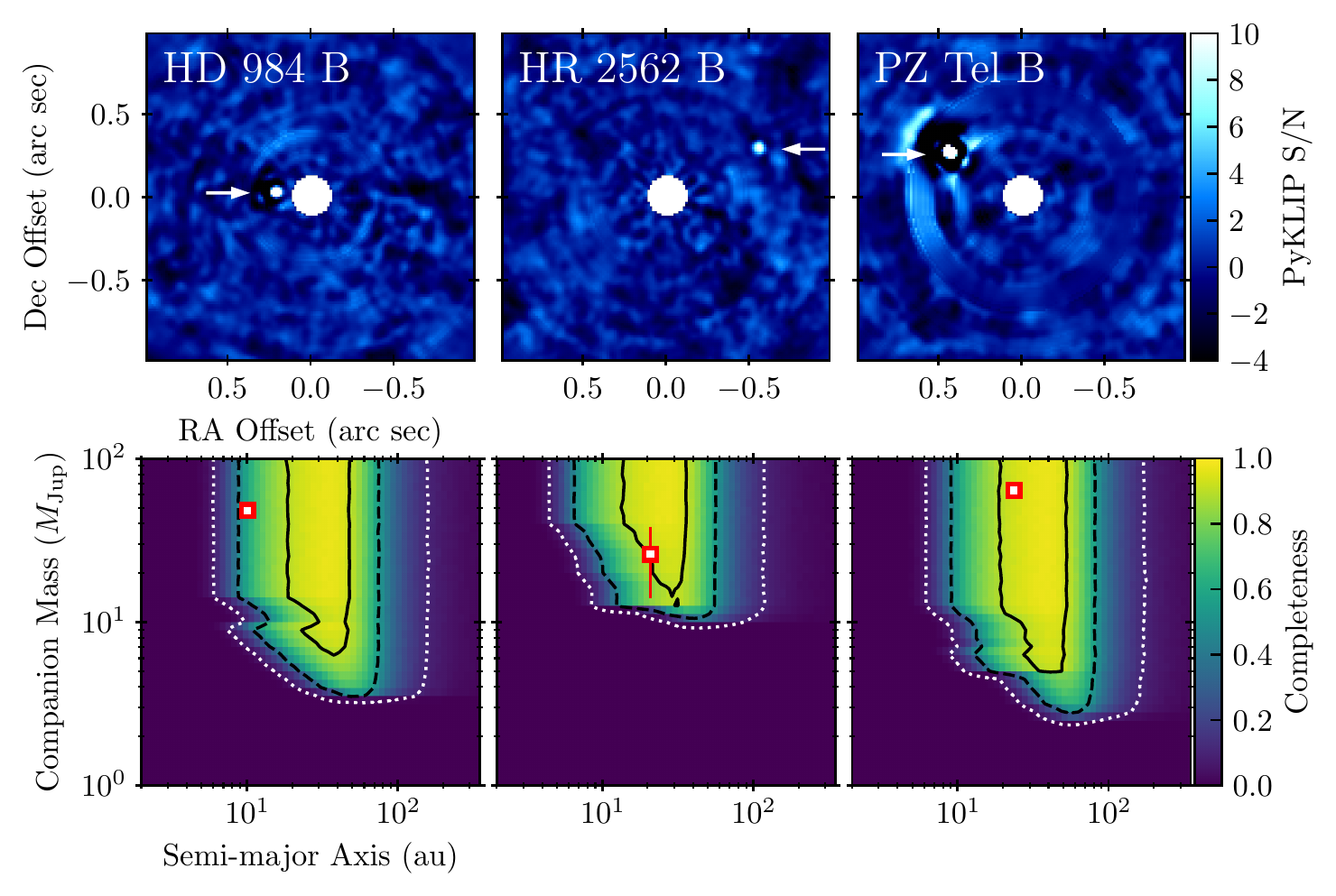}
\caption{(top row) PyKLIP S/N maps of the three systems with a brown dwarf companion resolved from the first 300 GPIES stars after image reduction and PSF subtraction. (bottom row) Corresponding completeness maps for the three stars computed using the procedure described in Section \ref{complete_sec}. Contours are plotted for the 90\% (solid), 50\% (dashed), and 10\% (dotted) completeness levels. The location of each companion is based on the mass derived from the $H$-band photometry and the projected separation of the companion at the epoch of the campaign observation. \label{fig:gallery_bds}}
\end{figure}
In addition to the four planetary systems discovered during the first half of the GPIES campaign, three brown dwarf companions were resolved; PZ Tel B, HD 984 B, and HR 2562 B. The companions to PZ Tel \citep{Mugrauer:2010cp,biller:2010} and HD 984 \citep{Meshkat:2015hd} were known before the GPIES observations of these stars, and were detected when their host stars were selected by our automated target-picker, while HR 2562 B was first discovered with GPI \citep{Konopacky:2016dk} as part of our planet search campaign.  PZ Tel and HD 984 are probable members of the $\beta$ Pictoris and Columba moving groups, respectively \citep{malo:2013}, while HR 2562 is thought to be $<$1 Gyr based on photometric \citep{Casagrande:2011ji} and spectroscopic \citep{Pace:2013cl} analyses. The three host stars have a later spectral type than the planet host stars discussed previously; G9IV (1.1 M$_{\odot}$) for PZ Tel, F7V (1.3 M$_{\odot}$) for HD 984, and F5V (1.3 M$_{\odot}$) for HR 2562.

 A brown dwarf companion was later detected around the campaign target HD 206893 (initially discovered with the VLT/SPHERE by \citealp{Milli:2016fs}), but these observations were obtained after the 300th star was observed and the companion is therefore not included in any of the analyses presented below for the first half of the GPIES campaign. 

\section{Survey Modeling and Inference}

Having observed 300 stars and detected six planets and three brown dwarfs, GPIES is one of the largest, deepest direct imaging surveys, with a high yield for substellar companions.  Using this rich dataset, we proceed to measure the overall occurrence rate of giant planets and trends in the underlying population.

\subsection{Survey Completeness to Date}\label{complete_sec}

We determine the completeness to planets for each of our observations using the Monte Carlo procedure described in \citet{Nielsen:2008}, \citet{Nielsen:2010}, and \citet{nielsen:2013}.  For each target star, completeness to substellar companions is determined over a grid that is uniform in log space in mass and semi-major axis.  At each grid point, $10^4$ simulated substellar companions, each with the same value of mass and semi-major axis, are randomly assigned orbital parameters: inclination angle from a $\sin(i)$ distribution, eccentricity from a linear distribution fit to wider-separation RV planets \citep{Nielsen:2008}, $P(e) = 2.1 - 2.2 e$ with $0 \leq e \leq 0.95$, and argument of periastron and epoch of periastron passage from a uniform distribution.  Since contrast curves are one dimensional, position angle of nodes is not simulated.  The projected separation (from the orbital parameters and mass and distance of the host star) and the $\Delta H$ (from the mass of the companion, the age and absolute magnitude of the host star, and luminosity models) for each companion are then computed.  A simulated companion that lies above the contrast curve is considered detectable, while one that lies below (or outside the field of view) is undetectable.  If more than one contrast curve is available for a single star, the simulated companions generated at the first epoch are advanced forward in their orbits to the next epoch, and compared to the next contrast curve; a companion is detectable if it is above at least one contrast curve.

As noted above, for the forward model matched filter (FMMF) contrast curves described in Section~\ref{sec:fmmf}, we use an 8$\sigma$ threshold inside of 0.3$''$, and a 6$\sigma$ threshold outside.  FMMF contrasts were produced for both a T-type and L-type template for each star \citep{Ruffio2017}.  The T-type reduction assumes significant methane absorption in the $H$ band, so that images in redder channels may be subtracted from images in bluer channels without subtracting the planet from itself.  To stay consistent, the same evolutionary models that produce $H$ band flux are used to give temperature as a function of mass and age, and planets hotter than 1100 K are assumed undectable using the T-type curve.  The L-type curve, which does not suffer from the possibility of self-subtraction, is used for simulated companions of all temperatures.  These FMMF contrast curves are defined out to 1.7$''$, though beyond 1.1$''$ some parts of the field go beyond the edge of the detector, and observations are not complete over all position angles.  We account for this with the same method as \cite{nielsen:2013}, by recording the fractional completeness as a function of separation, and using a uniform random variable to reject simulated planets that would not fall on the detector.

\begin{figure}[!ht]
\includegraphics[width=0.95\columnwidth,trim=0 3cm 0 8cm]{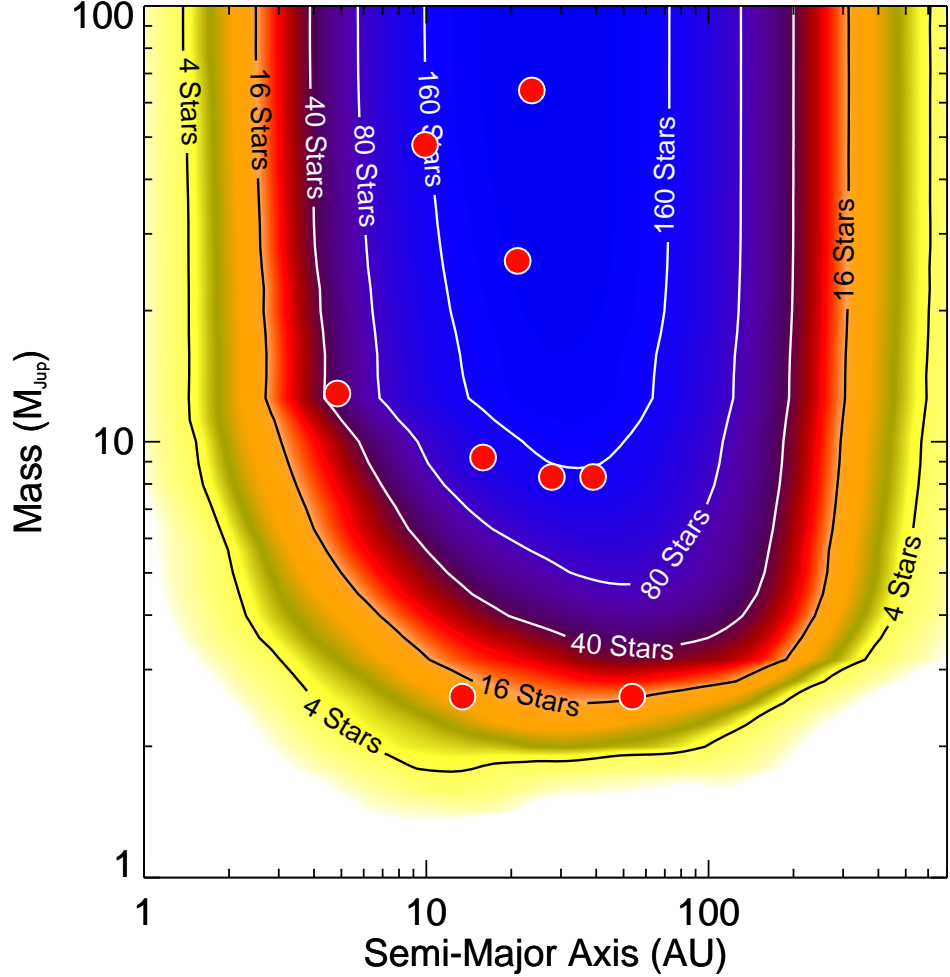}
\caption{Depth of search for the first 300 stars observed by GPIES, showing the number of stars to which the survey is complete for planets and brown dwarfs as a function of mass and semi-major axis.  Overplotted in red circles are the detected companions, plotted at the projected separation they were first imaged by GPIES. \label{fig:tongue_plot}}
\end{figure}

We utilize the CIFIST2011 BT-Settl atmosphere models \citep{baraffe:2015,allard:2014,Caffau:2011ik}\footnote{\url{https://phoenix.ens-lyon.fr/Grids/BT-Settl/CIFIST2011/COLORS/colmag.BT-Settl.server.2MASS.Vega}}, tied to the COND `hot start' evolutionary model grid \citep{baraffe03}, to give absolute $H$ magnitude and temperature as a function of mass and age.  From these, we produce completeness maps for each star.  These maps give the fractional completeness to companions as a function of mass and semi-major axis (completeness at a given value of mass and semi-major axis is not binary since orbital parameters will set whether a given companion is detectable for a certain contrast curve).  These completeness maps are shown for stars with detected companions in Figures~\ref{fig:gallery_planets} and \ref{fig:gallery_bds}.  Summing the maps across all target stars produces a map representing our depth of search (e.g. \citealt{lunine:2008}) for the survey, as given in Figure~\ref{fig:tongue_plot}.  The contours give the number of stars to which the survey is complete to companions of a given mass and semi-major axis.    Since target stars cover a large range of distances (3.2--176.7 pc), the limited field of view of GPIES ($\sim$0.15--1.1$''$, beyond which not all position angles lie on the detector) means that completeness plots for individual stars do not precisely line up.  Individual completeness plots never reach 100\% for a similar reason, as some simulated planets at each semi-major axis can fall within the IWA or outside the outer working angle, and so go undetectable.  As such the maximum value reached on the plot is 246 stars, though all 300 stars have some fractional sensitivity to substellar companions.  The detected companions are indicated by red dots, plotted at their inferred mass and projected separation at the first GPIES epoch, as given in Table~\ref{tbl:companion_props}.  In this figure, and in the subsequent analysis, we assume masses for each companion inferred from the BT-Settl models, the age of the system, and the $H$-band magnitude, even if a more robust mass is available from full SED fitting, to maintain consistency in the completeness calculation.  Most of these assumed masses are close to the published mass, with the notable exception of HD 95086, where the $H$-band magnitude and age suggest a mass of 2.6 $\pm$ 0.4 M$_{\mathrm{Jup}}$, compared to 4.4 $\pm$ 0.8 M$_{\mathrm{Jup}}$ inferred from the $Ks$-band magnitude and a bolometric correction \citep{DeRosa:2016kh}.  HD 95086 is a particularly unusual case, as its dusty atmosphere leads to a redder spectrum than most models predict.  We note that the analysis below does not change significantly whether a mass of 2.6 or 4.4 M$_\mathrm{Jup}$ is used.

We also investigated the effect of utilizing full age posteriors for each star, rather than using a single age for each target.  We considered the AB Dor moving group target star AN Sex, and computed the completeness plot for a single age of 149 Myr (as used in this work), for a disjoint gaussian, and for a symmetric gaussian.  The disjoint gaussian has a $\sigma$ of 19 Myr below 149 Myr, and a $\sigma$ of 51 Myr above, to match the \citet{bell:2015} age for the AB Dor moving group of 149$^{+51}_{-19}$ Myr.  The symmetric gaussian, computed for comparison, is given a $\sigma$ of 20 Myr (13\% age uncertainty).  As expected, assuming younger ages result in more detectable planets compared to older ages, however the effect on the completeness plot is marginal.  The 20\% completeness contour, for example, moves at most 7\% in planet mass, with a median shift of 1\%, for the disjoint gaussian.  For the symmetric gaussian, the effect is even smaller, maximum of 5\% and median 0.2\%.  In the symmetric case, the contours move toward smaller planets (becoming more sensitive when using a symmetric gaussian age posterior compared to a single age), while the completeness plot is less sensitive for the disjoint gaussian.  Similar results are found for the symmetric gaussian even if the value of $\sigma$ is doubled.  We thus expect a minor effect on our results by including age posteriors for each target star, but this effect will be explored more in depth in a future paper on the statistical constraints from the full survey, once full age posteriors are available for all of our stars.

Utilizing mass posteriors is expected to have an even smaller effect on the completeness plots.  Since simulated planets are generated as a function of semi-major axis, stellar mass has no effect on the completeness for a star observed at only a single epoch (273 out of our 300 stars), as planets are randomly assigned a mean anomaly, which does not require the period to be known.  For stars observed at multiple epochs, however, period is needed to determine the amount of orbital motion between the two observations, but given the slow orbital speeds of these wide-separation planets, and that the observations described here span less than two years, this is a minor effect as well, especially as we typically reach $\lesssim$5\% precision on stellar mass on our targets.

\subsection{A correlation between stellar mass and planet occurrence rate}\label{sec:stellarmass}
\begin{figure}[!ht]
\gridline{\fig{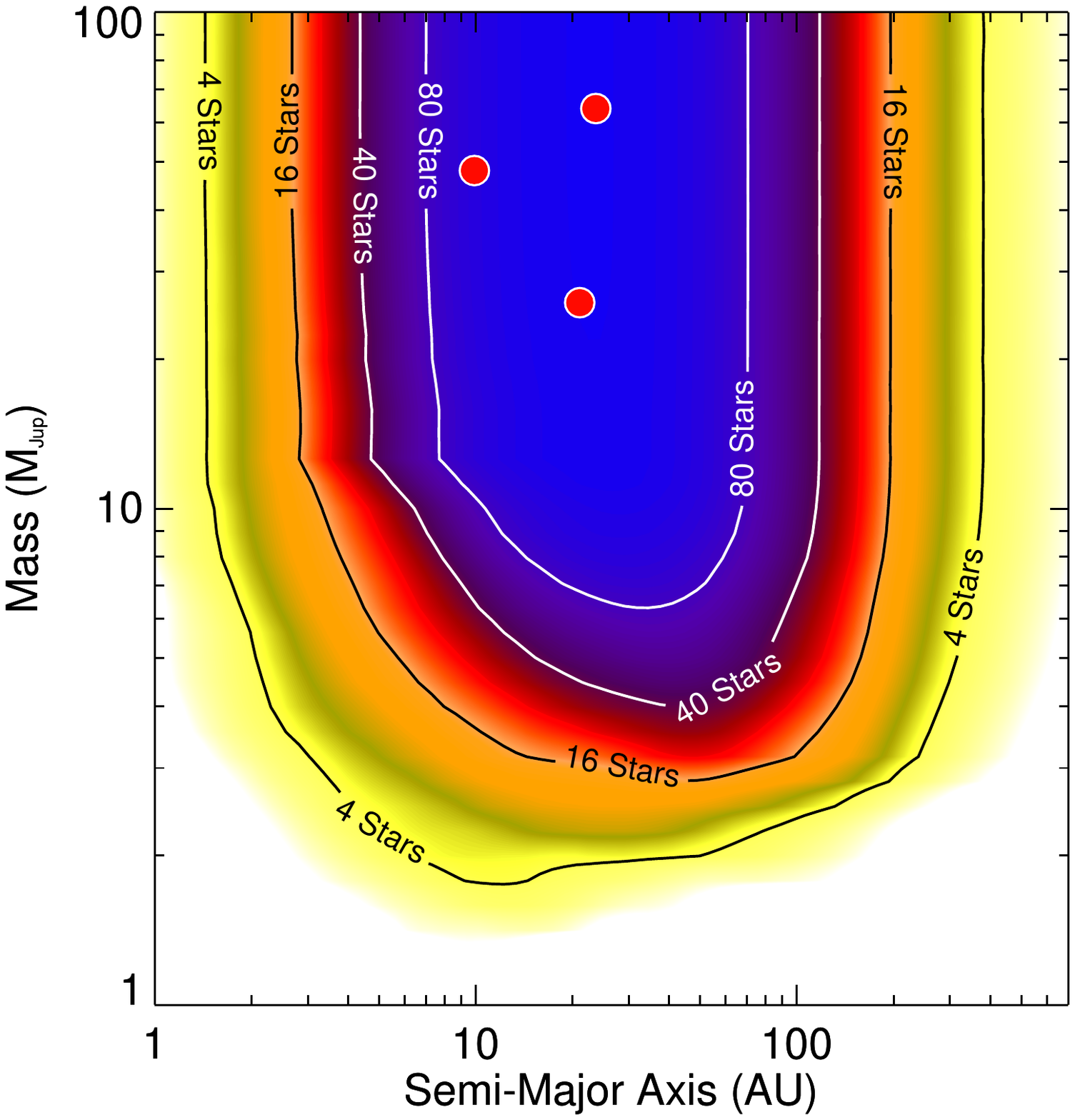}{0.5\columnwidth}{(a) M$<$1.5 M$_{\odot}$ (177 stars)}
         \fig{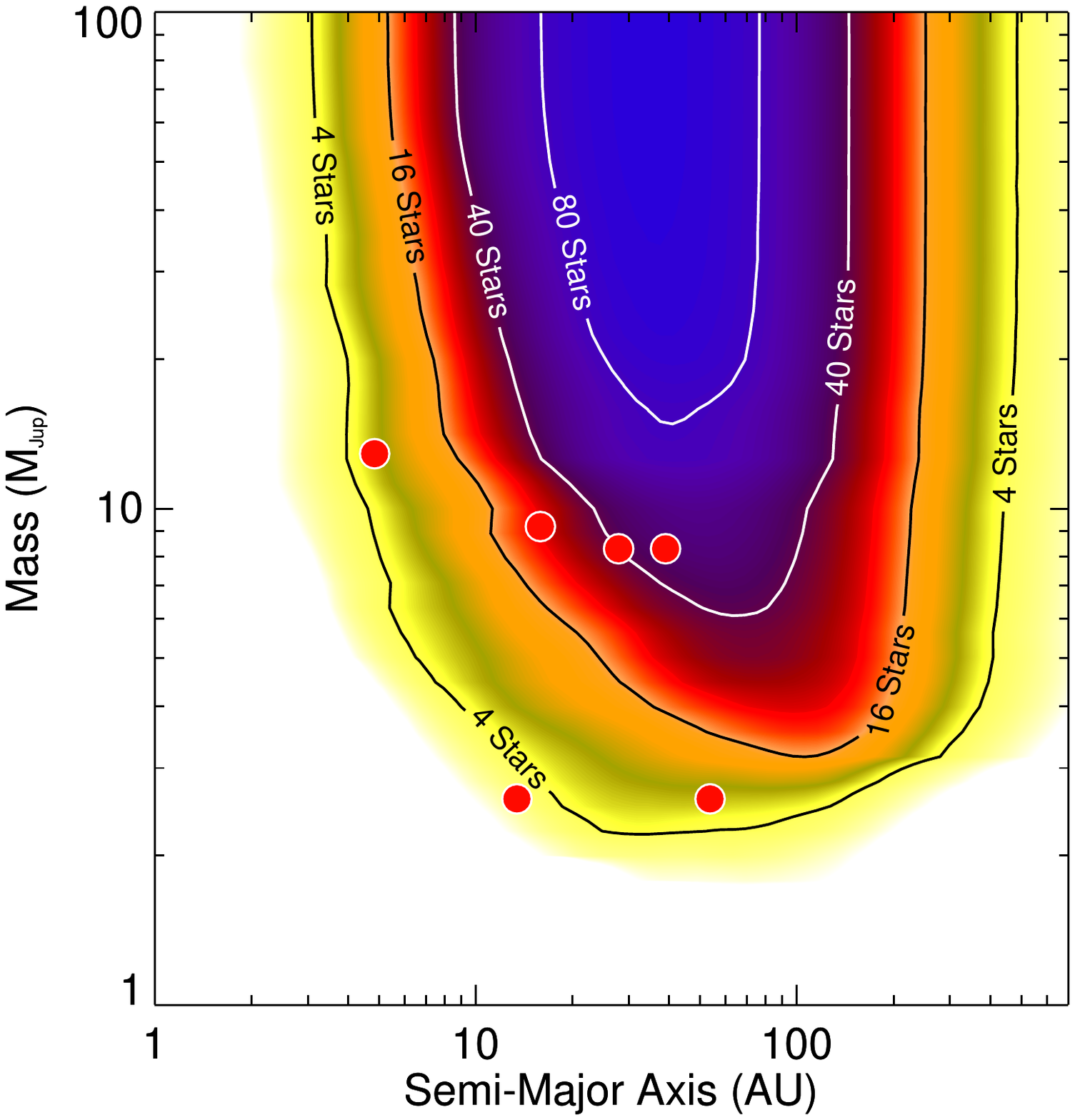}{0.5\columnwidth}{(b) M$\geq$1.5 M$_{\odot}$ (123 stars)}}
\caption{Depth of search for the sample divided by stellar mass.  All three brown dwarfs appear around lower-mass stars, which is not especially surprising given those stars comprise almost 60\% of the sample.  The fact that all six planets (and so all four planet-hosting stars) are among higher-mass stars, however, is more striking, and suggests a correlation between stellar mass and occurrence rate of intermediate period giant planets.\label{fig:twotongue}}
\end{figure}

A shared feature of the planetary companions described in Section~\ref{companions} is that all orbit the higher-mass stars in our sample: all host stars lie between 1.55 and 1.75 M$_\odot$ (Figure~\ref{fig:twotongue}).  As shown in Figure~\ref{fig:targ_demo}, there is no clear mass bias in our sample toward higher masses; in fact, 177 out of 300 stars (59\%) have masses below 1.5 M$_\odot$.  This trend is also the opposite from what we would expect for observational biases: lower mass stars are intrinsically fainter, and thus the same achievable contrast would correspond to a lower mass detectable planet.

We investigate the strength of this trend by comparing occurrence rates of wide-separation giant planets around higher-mass stars and lower-mass stars, here defined to have a mass determination above and below 1.5 M$_\odot$.  
We consider planets with semi-major axis between 3 and 100 au, and planet mass between 2 and 13 M$_{\rm Jup}$.
These limits are chosen to encapsulate a region of planet parameter space to which our observations are most sensitive and encompass the detected planets in the GPIES sample, which are between 2.6--12.9 M$_{\rm Jup}$ and a projected separation from 4.84--53.65 au (as shown in Figure~\ref{fig:tongue_plot}).  While the depth of search is computed for semi-major axis, the most direct measurement we have of each detected planet is projected separation, so we have chosen a broad enough range of semi-major axes so that planets at these projected separations are unlikely to fall outside this range.  Indeed, orbital monitoring of each of these planets \citep{DeRosa:2015jl,Wang:2016gl,rameau2016,wang:2018b} place them solidly in this range.  
We also begin by assuming planets are uniformly distributed over this range in log space in both semi-major axis and planet mass ($\frac{d^2N}{da \, dm} \propto a^{-1} m^{-1}$).  This is not too dissimilar to the power law distributions for small-separation ($a < 3.1$ au) giant planets found by \citet{cumming2008} of $\frac{d^2N}{da \, dm} \propto a^{-0.61} m^{-1.31}$, obtained by converting the period distribution of $\frac{dN}{dP} \propto P^{-0.74}$ to semi-major axis for solar-type stars.  For each star, then, we integrate the completeness to planets over this range with uniform weight given to each log bin in semi-major axis and mass.

\begin{figure}[ht]
\includegraphics[width=\columnwidth,trim=0 12cm 0 8cm]{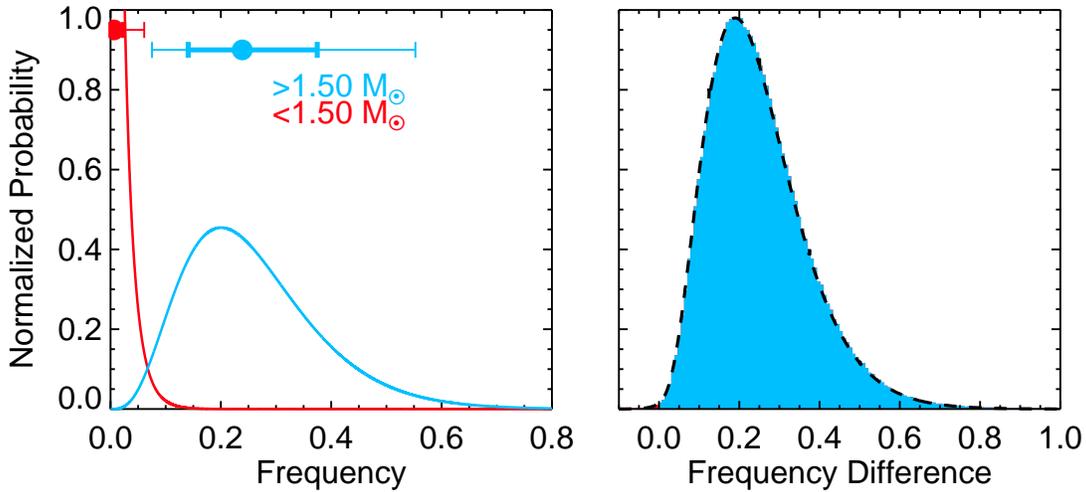}
\caption{(Left) The posterior probability of frequency of giant planet systems (3-100 au, 2-13 M$_{\rm Jup}$) for higher-mass stars (blue) and lower-mass stars (red).  Planets with these parameters are more common around higher-mass stars in this sample.  Filled circles represent the median of each distribution, and error bars give the 1 and 2 $\sigma$ confidence intervals.  (Right) Posterior for the difference between the two frequencies, with the lower-mass star occurrence rate subtracted from the higher-mass star occurrence rate.  The histogram (blue when the higher-mass star occurrence rate is larger, red otherwise) shows the results from Monte Carlo draws from the two posteriors, and the black dashed line gives the posterior computed using Equation~\ref{eq:diffprob}.  At 99.92\% confidence the occurrence rate is larger for higher-mass stars ($>1.5$ M$_\odot$) than lower-mass stars ($<1.5$ M$_\odot$). \label{fig:stellar_mass_prob}}
\end{figure}

Integrating over the entire range from 3--100 au and 2--13 M$_{\rm Jup}$ produces a completeness to massive giant planets of 17.5 stars among the higher-mass ($>1.5 M_\odot$) sample, and 43.3 among the lower-mass sample.  These two samples have 4 and 0 planetary systems detected, respectively, as HR 8799 counts as a single planetary system.  We infer the frequency of planetary systems for each subsample using a Bayesian approach with a Poisson likelihood ($L = \frac{e^{-\lambda} \lambda^k}{k!}$).  The expected number of planetary systems ($\lambda$) is the number of stars to which the subsample is complete multiplied by the frequency, and the measured number ($k$) is the number of detected planetary systems.  The prior is taken to be the Jeffrey's prior on the rate parameter of a Poisson distribution, $\mathrm{Prior}(\lambda) \propto \sqrt{\frac{1}{\lambda}}$.  We compute the posterior over a well-sampled regular grid of frequency for both sub-samples, plotting the results in Figure~\ref{fig:stellar_mass_prob}.

Given no planets were detected around lower-mass stars ($<1.5$ M$_\odot$), our measurement of the frequency for these stars represents an upper limit, $<$6.9\% at 95\% confidence.  With four detected systems, the fraction of higher-mass stars with giant planet systems is 24$^{+13}_{-10}$\% (68\% confidence interval).  To evaluate the confidence with which we can conclude that the higher-mass (HM) star frequency is larger than the lower-mass (LM) star frequency, we compute the posterior of the difference, $\delta = f_{HM} - f_{LM}$, given by 

\begin{equation}
P(\delta|\textrm{data}) \propto \int^{\infty}_0 P_{HM}(f|\textrm{data}) P_{LM}(f-\delta|\textrm{data}) df
\label{eq:diffprob}
\end{equation}

\noindent where $P_{HM}$ and $P_{LM}$ are the posteriors on the frequency ($f$) for higher- and lower-mass stars.  We confirm this expression with a Monte Carlo method, where a set of two frequencies are randomly generated from the two posteriors and differenced.  The two methods generated identical results, as shown in the right panel of Figure~\ref{fig:stellar_mass_prob}.  For 99.92\% of the samples, wide-separation giant planets around higher-mass stars are more common than around lower-mass stars, a $>$3$\sigma$ result.

We note that this result is not strongly dependent on our choice of limits in planet mass, semi-major axis, and stellar mass.  When increasing the upper semi-major axis limit from 100 au to 300 au, the significance of the result that wide-separation giant planets are more common around higher-mass stars drops marginally from 99.92\% to 99.85\%.  Since we are averaging over the depth of search plot in Figure~\ref{fig:tongue_plot}, expanding the range to areas of lower sensitivity reduces the effective number of stars to which the survey is sensitive, and for lower-mass stars this number drops from 43.3 to 36.2, and higher-mass stars drops from 17.5 to 17.4, when expanding the semi-major axis range from 100 to 300 au.  Increasing this range still further from 3--100 to 1-1000 au reduces the effective number of lower-mass stars to 24.9 and higher-mass stars to 12.0, and the significance in this case remains above 3$\sigma$ at 99.84\%.

Similarly, dropping the planet mass range considered from 2--13 M$_{\rm Jup}$ to 1--13 M$_{\rm Jup}$ has a negligible effect, with the significance remaining at 99.92\%.  Dropping the 12 SBs from the sample is also a minor effect, removing mainly higher-mass stars, and raising the significance slightly to 99.95\%.  Stellar mass has a more significant effect, since moving the boundary between higher-mass and lower-mass stars results in fewer lower-mass stars in the sample.  The significance drops to 99.70\% when the boundary is moved to 1.35 M$_\odot$ (an effective number of lower-mass stars of 37.9), and 96.99\% at 1.10 M$_\odot$ (24.3 stars).  Thus the larger frequency of wide-separation giant planets around higher-mass stars is relatively robust to specific choices of which range of parameter space we choose.

\subsection{Wide Separation Giant Planet Occurrence Rate around Higher-Mass Stars}\label{sec:bayes}

Following the example of \citet{cumming2008}, we consider a model of planet distributions defined by power laws in mass and semi-major axis.  We adopt a functional form of our model of:

\begin{equation}
\frac{d^2N}{dm \, da} = f \, C_1 \, m^\alpha \, a^\beta
\label{nomass_eq}
\end{equation}

\noindent where $m$ and $a$ are planet mass and semi-major axis.  We define this equation over a limited range, $[m_1 \leq m \leq m_2]$ and $[a_1 \leq a \leq a_2]$.  The frequency ($f$) is the occurrence rate of planets in this range, and so the normalization constant $C_1$ is then given by 

\begin{equation}
    C_1 = \left [\int_{m1}^{m2} \int_{a1}^{a2} m^\alpha \,  a^\beta \, dm \, da \right ]^{-1}
\end{equation}
\noindent Thus, if $\alpha \neq -1$ and $\beta \neq -1$:
\begin{equation}
    C_1 = \frac{\alpha+1}{m_2^{(\alpha+1)} - m_1^{(\alpha+1)}} \frac{\beta+1}{a_2^{(\beta+1)} - a_1^{(\beta+1)}}
\end{equation}
\noindent otherwise the two terms become $[\ln(a_2) - \ln(a_1)]^{-1}$ and $[\ln(m_2) - \ln(m_1)]^{-1}$.

In order to constrain the three free parameters of this model ($f$, $\alpha$, and $\beta$) we adopt a Bayesian approach utilizing Monte Carlo completeness calculations.  Bayes' Equation then becomes:

\begin{equation}
P(f, \alpha, \beta | \mathrm{data}) \propto P(\mathrm{data} | f, \alpha, \beta) \, P(f) \, P(\alpha) \, P(\beta)
\label{nomass_bayes_eq}
\end{equation}

\noindent and in standard Bayesian terminology, the first term is the posterior, the second the likelihood, and the final three terms are the priors.  We follow a similar method to \citet{kraus2008}, dividing the two-dimensional observational space (mass vs. projected separation) into a series of bins.  Similar Bayesian methods were also used by \citet{biller13}, \citet{wahhaj13}, and \citet{brandt14} to fit a power law model to imaged planets.  41 bins in mass are used, logarithmically spaced between 1 and 100 M$_{\rm Jup}$, and 81 bins in separation, logarithmically spaced between 1 and 1000 au.  In each bin we calculate the expected and actual number of planets in that bin; as a result, we choose a Poisson distribution for the likelihood in each bin, then find the product of probabilities across all the bins:

\begin{equation}
P(\mathrm{data} | f, \alpha, \beta) = \prod_i \prod_j \frac{e^{-E_{i,j}} \, E_{i,j}^{O_{i,j}}}  {O_{i,j}!}
\label{poisson_eq}
\end{equation}

\noindent where $i$ and $j$ are indices for bins in mass and projected separation, and $E_{i,j}$ and $O_{i,j}$ are the expected and observed number of planets in each bin, respectively.

Unlike \citet{kraus2008}, our model and likelihood use different parameters: Equation~\ref{nomass_eq} is in terms of mass and semi-major axis, while Equation~\ref{poisson_eq} is in terms of mass and projected separation, since directly-imaged planets are detected with a particular separation, but an (often unknown) semi-major axis.  Using Monte Carlo simulations, however, we are able to project our modeled planets from semi-major axis into separation.  As in Section~\ref{complete_sec}, we inject simulated planets at a grid of mass and semi-major axis, and for each grid point save the fraction of simulated planets that are above the contrast curve (detectable with our observations).  Additionally, we also sort the detectable planets into projected separation bins, tracking the projected separation of each detectable planet.  The completeness map for each star, then, is a three-dimensional array across mass, projected separation, and semi-major axis.  Marginalizing over the separation dimension, and summing over all stars, gives the depth of search shown in Figure~\ref{fig:tongue_plot}.  If we instead marginalize over semi-major axis, we derive a completeness map in mass and projected separation, the same parameters that define our observed planets.

For a given set of ($f$, $\alpha$, $\beta$), we follow this procedure for evaluating Equation~\ref{nomass_bayes_eq}:

\begin{enumerate}
    \item For each star, generate a three-dimensional completeness map using Monte Carlo simulations
    \item Sum the completeness maps over all stars
    \item Weight each point in the map using Equation~\ref{nomass_eq} and the values of ($f$, $\alpha$, $\beta$), giving the expected number of planets detected in each bin
    \item Marginalize the map over semi-major axis, producing a map of expected planets vs.  mass and separation
    \item Produce a corresponding observed map of detected planets vs. mass and separation
    \item Compute the likelihood from Equation~\ref{poisson_eq}.
    \item Multiply by priors
\end{enumerate}

\begin{figure}[!ht]
\includegraphics[width=\columnwidth,trim=0 3cm 0 7cm]{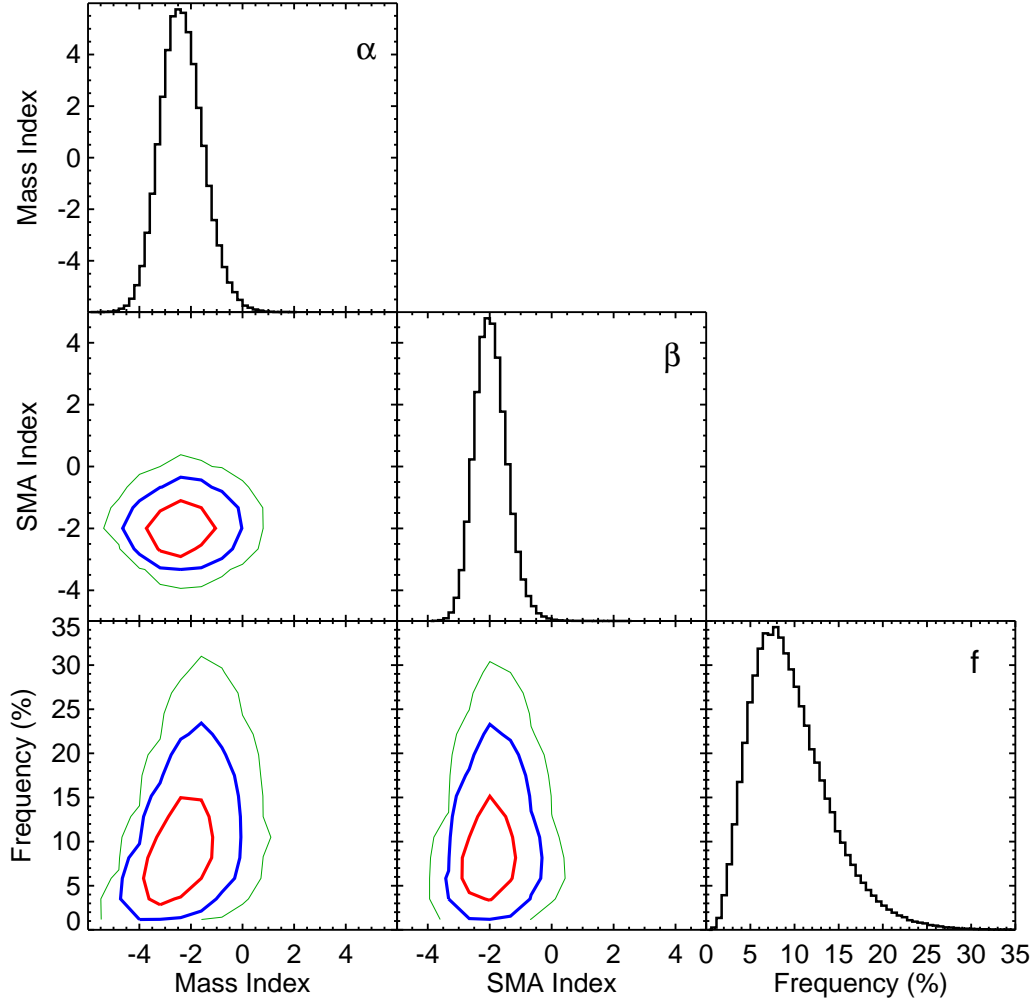}
\caption{The posteriors on our double power-law model of giant planet populations between 5--13 M$_{\rm Jup}$ and 10--100 au, for stars $>$1.5 M$_{\odot}$.  We find a relatively high occurrence rate of 9$^{+5}_{-4}$\% over these ranges, and negative power laws that point to a rise in planet frequency at smaller masses and smaller semi-major axis (SMA). Red, blue, and green contours give 1, 2, and 3~$\sigma$ contours on probability.  \label{fig:nomass_triangle}}
\end{figure}

We then utilize a Metropolis-Hastings MCMC procedure (as in \citealt{Nielsen:2014}) to explore this parameter space and generate posteriors on frequency and power law indices.  For priors we utilize uniform priors for $\alpha$ and $\beta$, and a Jeffrey's prior for $f$ appropriate to a Poisson distribution, $\mathrm{Prior}(f) \propto \sqrt{\frac{1}{f}}$.  Our results are not overly dependent on this choice of prior, we find similar posteriors with a prior that is uniform in $\ln{(f)}$.  We first restrict ourselves to a region of parameter space where we have both good completeness and our detections, $[1~$M$_{\rm Jup} \leq m \leq 13~$M$_{\rm Jup}]$ and $[1~{\rm au} \leq a \leq 100~{\rm au}]$, and M$_*>$ 1.5 M$_\odot$.  This range encompasses all six planets detected by the GPIES survey.  Given our low sensitivity to low-mass, short period planets, there is a strong degeneracy between frequency and power law indices, since very negative indices (that place large numbers of planets at small separations and small masses) result in larger frequencies.  Fits with large occurrence rates place most of these planets in the region of parameter space to which GPIES is least sensitive.  In order to report an occurrence rate with high confidence, we define occurrence over a region of high completeness, and so we quote not the frequency over the entire range, but instead a modified frequency over the truncated range $[5~$M$_{\rm Jup} \leq m \leq 13~$M$_{\rm Jup}]$ and $[10~{\rm au} \leq a \leq 100~{\rm au}]$.  Thus while 51 Eridani b, HD 95086 b, and $\beta$ Pic are excluded from this modified frequency, they fall within the broader range, and thus are all used to constrain the model.  The resulting posteriors are shown in Figure~\ref{fig:nomass_triangle}, and Table~\ref{tbl:allposteriors}.

\begin{deluxetable}{|c|c|c|c|c|}
\tabletypesize{\small}
\tablecaption{Constraints on substellar object populations from the GPIES 300-star sample, between 10-100 au.  The power-law model has a number of substellar companions per star of $f$ within the given mass and semi-major axis limits, with power law indexes of $\alpha$, $\beta$, and $\gamma$ for companion mass, semi-major axis, and stellar host mass, respectively.\label{tbl:allposteriors}}
\tablehead{\colhead{} & \colhead{$f$} & \colhead{$\alpha$} & \colhead{$\beta$} & \colhead{$\gamma$} }
\startdata
\multicolumn{5}{l}{\scriptsize{Planets around high-mass stars: 5 M$_{\rm Jup}$ $<$ $m$ $<$ 13 M$_{\rm Jup}$, $1.5 $M$_\odot < $M$_* < 5 $M$_\odot $}}\\
\hline
Median & 8.9\% & -2.37 & -1.99 & -- \\
68\% CI & 5.3 -- 13.9\% & -3.16 -- -1.51 & -2.47 -- -1.44 & -- \\
95\% CI & 2.9 -- 20.6\% & -3.90 -- -0.56 & -2.92 -- -0.81 & -- \\
\hline
\multicolumn{5}{l}{\scriptsize{Planets around all stars: 5 M$_{\rm Jup}$ $<$ $m$ $<$ 13 M$_{\rm Jup}$, $0.2 $M$_\odot < $M$_* < 5 $M$_\odot $ }}\\
\hline
Median & 3.5\% & -2.277 & -1.68 & 2.03 \\
68\% CI & 2.1 -- 5.4\% & -2.98 -- -1.48 & -2.18 -- -1.10 & 1.05 -- 3.07 \\
95\% CI & 1.1 -- 8.0\% & -3.65 -- -0.54 & -2.63 -- -0.38 & 0.09 -- 4.16 \\
\hline
\multicolumn{5}{l}{\scriptsize{Brown dwarfs around all stars: 13 M$_{\rm Jup}$ $<$ $m$ $<$ 80 M$_{\rm Jup}$, $0.2 $M$_\odot < $M$_* < 5 $M$_\odot $ }}\\
\hline
Median & 0.8\% & -0.47 & -0.65 & -0.85 \\
68\% CI & 0.3 -- 1.6\% & -1.53 -- 0.67 & -1.69 -- 0.89 & -2.33 -- 0.59 \\
95\% CI & 0.1 -- 2.8\% & -2.60 -- 1.96 & -2.52 -- 3.86 & -3.76 -- 2.01 \\
\hline
\multicolumn{5}{l}{\scriptsize{Substellar objects around all stars: 5 M$_{\rm Jup}$ $<$ $m$ $<$ 80 M$_{\rm Jup}$, $0.2 $M$_\odot < $M$_* < 5 $M$_\odot $ }}\\
\hline
Median & 4.7\% & -2.09 & -1.44 & 1.15 \\
68\% CI & 3.2 -- 6.7\% & -2.49 -- -1.71 & -1.92 -- -0.91 & 0.35 -- 1.96 \\
95\% CI & 2.0 -- 9.1\% & -2.89 -- -1.33 & -2.36 -- -0.28 & -0.48 -- 2.75 \\
\enddata
\end{deluxetable}
\normalsize

\begin{figure}[!ht]
\includegraphics[width=\columnwidth,trim=0 2.5cm 0 7cm]{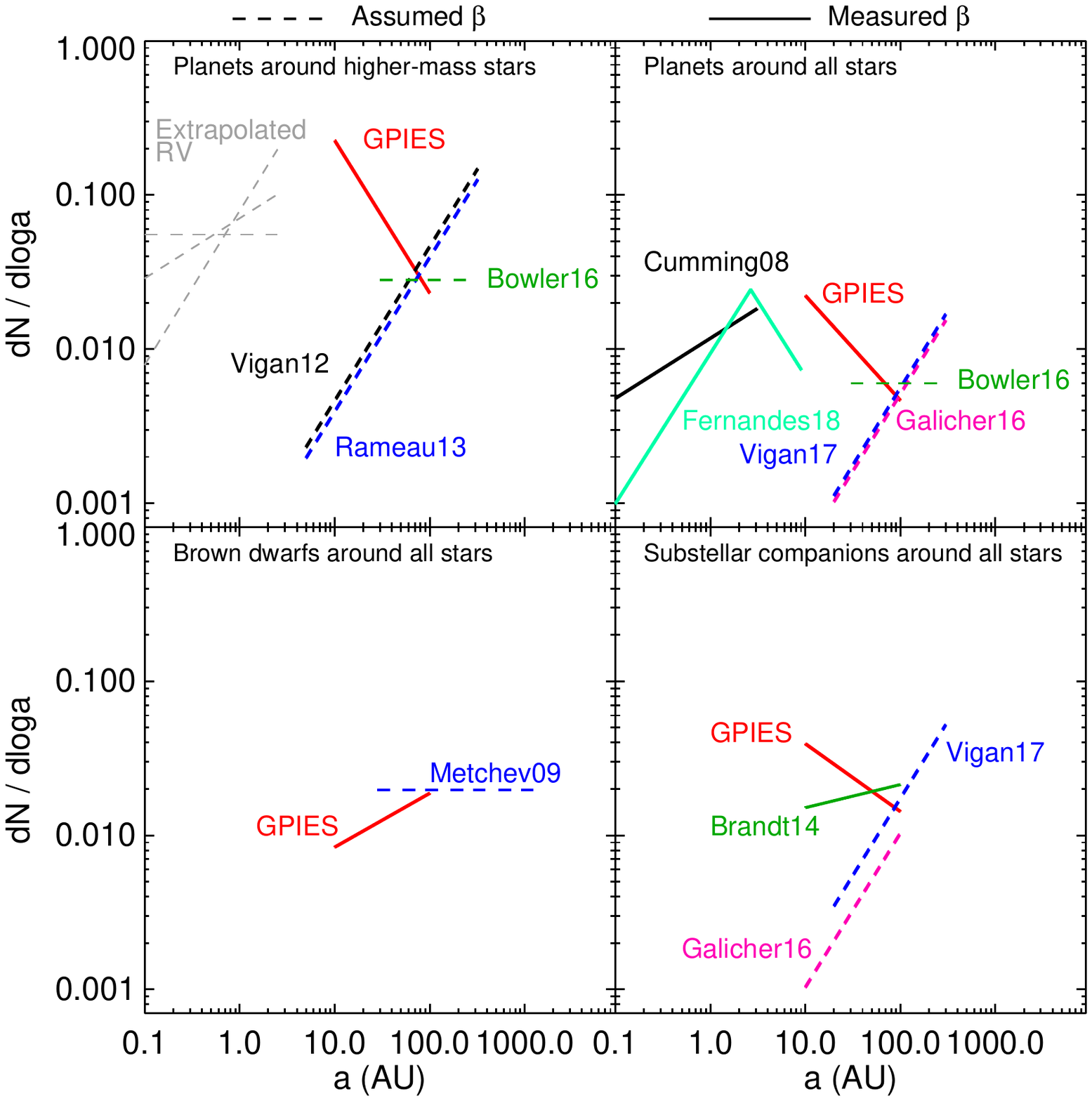}
\caption{A comparison of the occurrence rates of 5--13 M$_{\rm Jup}$ planets (top two panels), 13--80 M$_{\rm Jup}$ brown dwarfs (bottom left), and 5--80 M$_{\rm Jup}$ substellar companions (bottom right) as a function of semi-major axis for the GPIES results and other surveys.  The top left panel considers only higher-mass stars from GPIES, while all 300 stars are used in the other three panels with a stellar mass power law term (in these three panels the occurrence rate is given for 1 M$_\odot$).  The end points of each line represent the range of semi-major axis reported (10--100 au for GPIES), with the slope giving the value of $\beta$, the semi-major axis power-law index, and the normalization showing the planet occurrence rate.  We utilize the value of the companion mass power law ($\alpha$) in each survey to find the occurrence rate over the GPIES range in each panel, and in the last three panels we use our value of $\gamma$ (the power law index for stellar host mass) to convert the GPIES occurrence rate from 1.75 M$_\odot$ to 1.0 M$_\odot$, to more closely match analyses.  Solid lines denote fits to the value of $\beta$ (GPIES, \citealt{cumming2008}, \citealt{brandt14}, and \citealt{fernandes18}), while dashed lines are used for authors that assumed power law indexes ($\alpha$ and $\beta$ of $-1$ or 0) rather than fitting for them directly.  For clarity, we omit error bars and only display the median values for power law indexes and occurrence rates. \label{fig:compare_occurrence}}
\end{figure}

We find a relatively high giant planet occurrence rate: 9$_{-4}^{+5}$\%, for stars with mass $>1.5 M_\odot$, and planets in the range $5~M_{\rm Jup} \leq m \leq 13~M_{\rm Jup}$ and $10~{\rm au} \leq a \leq 100~{\rm au}$.  Given the very negative values for $\alpha$ and $\beta$, if we were to extrapolate beyond the region of high completeness to lower planet masses and smaller semi-major axes, we would infer a much larger occurrence rate with significantly larger uncertainties. \citet{Vigan:2012jm} combined the IDPS sample of 42 stars of AF spectral types with three literature stars, and assuming values of $\alpha = \beta = 0$, found a fraction of high-mass stars with at least one 3--14 M$_{\rm Jup}$ planet between 5--320 au of 8.7$^{+10.1}_{-2.8}$\%.  \citet{nielsen:2013} examined the 70 B and A stars observed by the Gemini NICI Planet-Finding Campaign, and with a null result for planets placed 2$\sigma$ upper limits on the fraction of high-mass stars with planets.  These limits have $<10$\% of high-mass stars having a 10--13 M$_{\rm Jup}$ planet with semi-major axis between 38--650 au, and $<20$\% with a 4--13 M$_{\rm Jup}$ planet between 59--460 au.  \citet{Rameau:2013} considered 37 AF stars, including detections of planets around HR 8799 and $\beta$ Pic, finding a planet fraction of 7.4$^{+3.6}_{-2.4}$\% from 5--320 au, between 3--14 M$_{\rm Jup}$, assuming $\alpha = \beta = 0$.  \citet{bowler:2016} considered multiple direct imaging surveys, performing a meta-analysis on 384 unique target stars.  This sample was then broken up by spectral type, into subsamples of M, FGK, and BA.  For the BA sample (roughly equivalent to our $>$1.5 M$_\odot$ sample), by assuming $\alpha = \beta = -1$, \citet{bowler:2016} find a planet fraction of 5--13 M$_{\rm Jup}$ 30--300 au planets of 2.8$^{+3.7}_{-2.3}$\%.

We compare our results to large imaging surveys with detected planets that explicitly measure occurrence rate in the top left panel of Figure~\ref{fig:compare_occurrence}, where we plot each survey as a line showing the semi-major axis distribution.  End points are taken to be the range of semi-major axes considered by each survey.  We utilize the value of $\alpha$ from each survey to find the expected occurrence rate over the GPIES mass range (5-13 M$_{\rm Jup}$).  In most cases authors assumed a value of $\beta$ (and $\alpha$) of $-1$ or 0, as we did in Section~\ref{sec:stellarmass}, rather than fitting for the power law index directly.  \citet{Vigan:2012jm} and \citet{Rameau:2013} find similar results, as expected given both had similar numbers of stars and detected planets.  The \citet{bowler:2016} meta-analysis included these surveys and others, for a total of 110 BA stars, and finds a lower occurrence rate, and assumes a distribution that is flat in log space, rather than one flat in linear space.  In all cases, the GPIES result is similar in fraction of planets at $\sim$100 au, but rises more steeply toward lower semi-major axis.  This is as we expect, given our finding of a more negative value of $\beta$ than $-1$ or 0, placing more planets close to the star.

We also plot possible extrapolations of RV occurrence rates for giant planets.  \citet{johnson:2010} reported a value of 20\% planet fraction for retired A-stars, for semi-major axis less than 2.5 au, and K$<$20 m/s (which they note corresponds to a lower mass limit of 1.12 M$_{\rm Jup}$ for a stellar host of 1.6 M$_\odot$).  \citet{johnson:2010} do not fit for $\alpha$, and do not give the upper mass for planets in their sample.  We assume both are similar to the \citet{cumming2008} sample for planets around FGK stars, and use the latter values for upper mass limit of 10 M$_{\rm Jup}$ and $\alpha = -1.31$, in order to convert the occurrence rate for all planets to 5--13 M$_{\rm Jup}$.  Similarly, while \citet{johnson:2010} give an upper limit of 2.5 au, they do not state the minimum semi-major axis of planets in their sample, though we speculate that this minimum value is 0.08 au, for HD 102956 b \citep{johnson:2010b}.  As there is no estimate for the value of $\beta$ for RV planets around higher-mass stars, we investigate three different values, flat in log space, the \citet{cumming2008} value, and flat in linear space ($\beta$ of -1, -0.61, and 0, respectively), and plot these as gray lines in Figure~\ref{fig:compare_occurrence}.

For integrated occurrence rate, the \citet{johnson:2010b} value of 20\% ($<$2.5 au) represents a rise from our value of 11.4\% (10--100 au), though for different mass ranges.  A more detailed study of the distributions of RV-detected giant planets around higher-mass stars is required to definitively determine the nature of the distribution of high mass planets within 10 au, and whether RV and imaged giant planets around higher-mass stars follow a continuous distribution, but with a break in the planet mass and semi-major axis power law indexes at $\sim$3 au.

\subsection{Occurrence Rate with a Stellar Mass Dependence}

We can instead fit the entire GPIES 300-star sample by including a term that allows the occurrence rate to vary as a function of stellar mass.  We thus introduce an additional term into Equation~\ref{nomass_eq}:

\begin{equation}
\frac{d^2N}{dm \, da} = f \, C_1 \, m^\alpha \, a^\beta \, \left ( \frac{M_*}{1.75 M_\odot} \right )^\gamma
\label{yesmass_eq}
\end{equation}

\noindent so that the overall occurrence rate $f$ now varies with stellar mass.  We choose $1.75 M_\odot$ as the normalization location given this is where most of our planet-hosting stars lie, and so occurrence rate is most constrained at this mass.  Similarly, Equation~\ref{nomass_bayes_eq} becomes:

\begin{equation}
P(f, \alpha, \beta, \gamma | \mathrm{data}) \propto P(\mathrm{data} | f, \alpha, \beta, \gamma) \, P(f) \, P(\alpha) \, P(\beta) \, P(\gamma)
\label{yesmass_bayes_eq}
\end{equation}

\noindent We solve this equation as before, but with one additional change: instead of summing the completeness map over all stars, we instead introduce an extra dimension in stellar mass, and bin the sample by stellar mass.  This produces a four-dimensional completeness map, as a function of companion mass, companion separation, companion semi-major axis, and stellar host mass.  Similarly, the map of expected planets is now over three dimensions, companion mass, companion separation, and stellar host mass.

\begin{figure}[!ht]
\includegraphics[width=\columnwidth,trim=0 3.5cm 0 8.5cm]{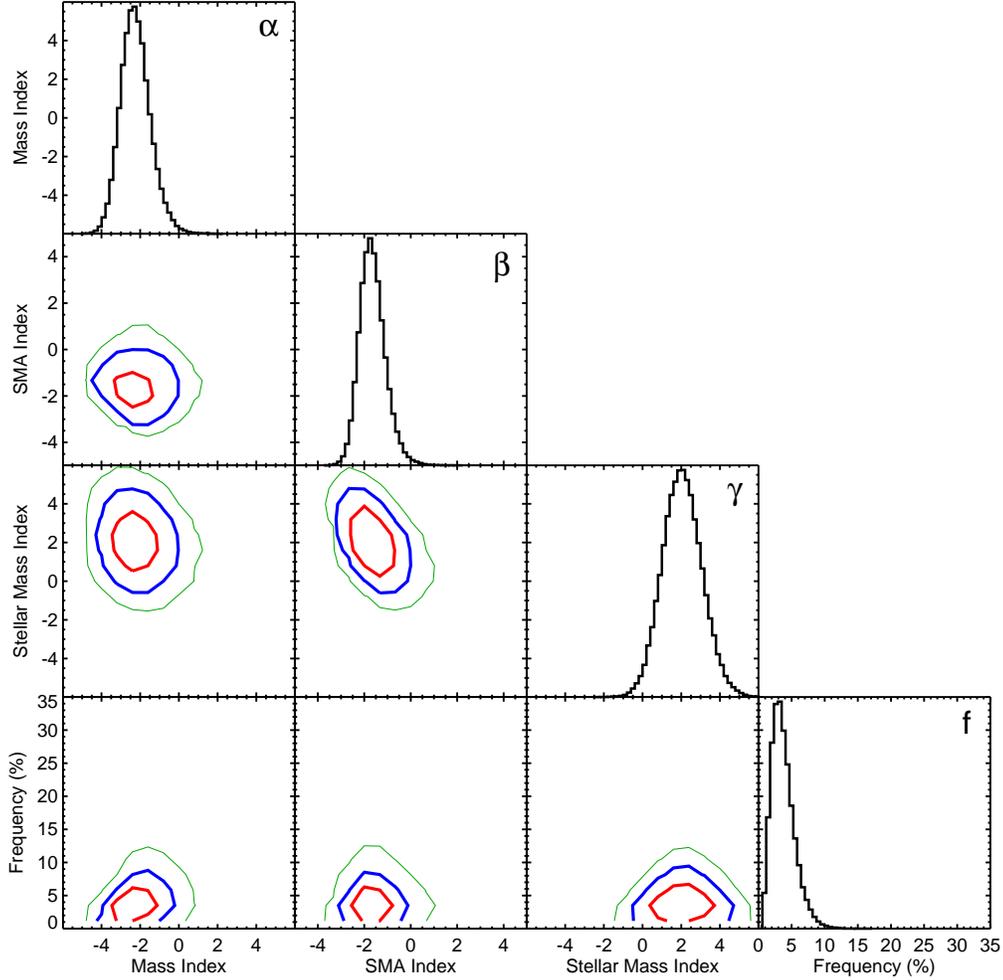}
\caption{A fit to the entire GPIES sample with a triple power-law model of wide-separation giant planet populations between 5--13 M$_{\rm Jup}$ and 10--100 au, with the occurrence rate ($f$) defined for 1.75 M$_\odot$ stars.  Given that all planets were detected around higher-mass stars, a large power-law index for stellar mass dependence is preferred. \label{fig:planets_gamma_triangle}}
\end{figure}

The results from the MCMC fit are plotted in Figure~\ref{fig:planets_gamma_triangle}.  As expected from the prevelance of detected planets around higher-mass stars, the distribution for $\gamma$ is peaked at relatively high values, $\gamma = 2.0 \pm 1.0$.  Interestingly, though likely coincidentally, this value of $\gamma$ is close to the negative of the Salpeter IMF power law index of -2.35 ($\frac{dN_*}{dM_*} \propto M_*^{-2.35}$, \citealt{salpeter:1955}), which would mean that for a volume-limited sample of young stars, while the number of giant planets per star would increase with increasing stellar mass, the number of giant planets per stellar mass bin would only change slightly, decreasing as the cube root of the stellar mass ($\frac{dN_p}{dM_*}\propto M_*^{-0.35}$).  When we compare this fit to our earlier fit to only higher-mass stars (Figure~\ref{fig:nomass_triangle}), the planet mass and semi-major axis power law indexes, $\alpha$ and $\beta$, are similar for both fits.  Occurrence rate, $f$, is significantly lower for the fit utilizing a stellar mass dependence.  While a more positive value of $\gamma$ ($\sim$4) better explains the additional 177 lower-mass stars without planets in our sample, such a distribution begins to overpredict the number of detectable planets around the highest mass stars in our sample ($\gtrsim$ 2 M$_\odot$).  Thus the best fit is found by combining a moderate value of $\gamma$ ($2.0 \pm 1.0$) and a lower overall occurrence rate.   While this fit incorporates the entire 300-star sample, we find the fit exclusively to the 123 higher-mass stars presented above to be more robust, as it centers around the detected planets.  Multiple models could account for the stellar mass dependence of giant planet occurrence seen in Figure~\ref{fig:twotongue}.  Here we have introduced a scaling of overall occurrence rate with stellar mass, but such a result could also be explained by a more negative power law fit to either companion mass or semi-major axis for lower-mass stars compared to $>1.5$M$_\odot$ stars, or some combination of varying power-law indices and occurrence rate as a function of stellar mass.  With no detections around our lower-mass star sample, we are unable to determine which model best represents these wider-separation giant planets.

\citet{Galicher:2016hg} described statistical results from the IDPS survey of 292 stars and two previous surveys, analyzing a sample of 356 stars with spectral types between B and M.  Assuming values of $\alpha = \beta = 0$, a planet fraction was inferred of 1.05$^{+2.80}_{-0.70}$\% between 0.5--14 M$_{\rm Jup}$ and 20--300 au, based on detections of planetary systems around HR 8799 and HIP 30034 (AB Pic).  An analysis of the VLT/NACO large program by \citet{vigan:2017} found similar results to IDPS with similar assumptions and analysis methods.  The survey observed 85 stars with NACO, which was then combined with previously published samples to reach a final sample of 199 FGK stars.  This final sample contained two detections of brown dwarfs, GSC 08047-00232 B and PZ Tel B, along with AB Pic b.  With $\alpha = \beta = 0$, the occurrence rate of 0.5--14 M$_{\rm Jup}$, 20--300 au planets was found to be between 0.85 and 3.65\% at 1$\sigma$, with a best value of 1.15\%.  From the \citet{bowler:2016} meta-analysis, assuming $\alpha = \beta = -1$, an occurrence rate of 0.6$^{+0.7}_{-0.5}$\% was derived for planets between 5--13 M$_{\rm Jup}$ and 30--300 au around stars between B and M spectral types.

The top right panel of Figure~\ref{fig:compare_occurrence} compares direct measurements of the planet occurrence rates to our findings for GPIES, where again our steeper slope places more planets at smaller semi-major axis.  We have used our median value of $\gamma$ to convert these occurrence rates from 1.75 to 1 M$_\odot$, to better match literature values.  As before, we use the values of $\alpha$ assumed from each reference to convert the given mass ranges to the GPIES range of 5--13 M$_{\rm Jup}$.  The occurrence rates estimates for different imaging surveys give about the same value at $\sim$100 au. \citet{cumming2008} and \citet{fernandes18} analyze RV surveys and find values of $\beta = -0.61 \pm 0.15$ and $\beta = -0.025^{+0.30}_{-0.22}$, respectively, at small separations.  While \citet{cumming2008} fit a single power law, \citet{fernandes18} find evidence for a broken power law, with a break at $\sim$3 au.  At these wider separations the \citet{fernandes18} value of $\beta = -1.975^{+0.22}_{-0.30}$ (from their symmetric {\tt epos} model) is a good match to our value of $\beta = -1.68^{+0.57}_{-0.50}$, though we note that while these RV surveys contain mainly FGK stars, all our planet detections are orbiting more massive stars, and so our GPIES results for 1 M$_\odot$ come from assuming a continuous distribution jointly constrained by non-detections on these lower-mass stars and six detections around higher-mass stars in the sample.

A number of analyses have also considered the giant planet occurrence rate at these separations for specific categories of target stars.  \citet{biller13} considered 80 stars that are members of moving groups that were observed by the Gemini NICI Planet-Finding Campaign, and used a Bayesian technique to fit a variant of Equation~\ref{nomass_eq}, where the upper semi-major axis cut-off was a free parameter.  This analysis included $\beta$ Pic b (which was not detected in the first NICI campaign observation of the star), and AB Pic b, but the small number of detections meant that limits could not be placed on the values of $\alpha$ and $\beta$, though a maximum in probability was reached for a planet fraction of 4\%, from 10--150 AU.  \citet{wahhaj13} performed a similar analysis on a combination of the NICI debris disk subsample and the debris disk hosts in the IDPS high-mass sample \citep{Vigan:2012jm}, with detections of $\beta$ Pic b and HR~8799~bcd.  This analysis found 2$\sigma$ limits that placed $\beta$ approximately between -2.5 and -1, and $\alpha$ between -1 and 2.5, with a peak value for the fraction of stars with planets of $\sim$10\%.  \citet{wahhaj13} also fit a model with a stellar mass dependence to the average planet multiplicity ($F_P$), which is a separate term from the fraction of stars with planets.  These results are suggestive of a rise of planet mulitiplicity with stellar mass, but with a large degree of uncertainty.  For their relation $F_P \propto M_*^\gamma$, the 1$\sigma$ results place $\gamma$ between 0.3 and 3.2, and a wider 2$\sigma$ range between -0.5 and 4.7, which encompasses the value of 0, corresponding to no stellar mass dependence.

\citet{bryan:2016} examined the occurrence rate of wide-separation giant planets in systems with a previously-known inner RV planet.  By combining long-period RV trends with (generally non-detections from) AO imaging, the possible mass and semi-major axis of the companions could be partially constrained.  Their high occurrence rate (62.1$^{+5.4}_{-5.7}$\% for substellar companions between 1-20 M$_{\rm Jup}$ between 5-100 au) is noticably higher than what we find with GPIES, suggesting that the presence of a close-in giant planet substantially boosts the likelihood of a wider substellar companion.  While they also fit values of $\alpha$ and $\beta$, \citet{bryan:2016} note that these posteriors are strongly dependent on the choice of mass and semi-major axis limits, as those limits truncate the probability distributions for each companion.

A recent paper by \citet{stone:2018} examined the constraints on occurrence rate from a sample of 98 stars observed in the LBT LEECH survey.  While a number of planets were observed in the characterization sample, the statistical sample had no planet detections.  For all stars, using the COND models \citet{Baraffe:2003bj}, they find a 2$\sigma$ upper limit of 25\% on the occurrence rate of planets between 4--14 M$_{\rm Jup}$ and 5--100 au, consistent with our results.

\subsubsection{Brown Dwarf Companions}

In addition to detecting six planets, the GPIES campaign has also detected three brown dwarfs, and so we apply our Bayesian framework to inferring the underlying population of brown dwarfs.  As before, we fit the four-parameter model of Equation~\ref{yesmass_eq} to brown dwarfs in the sample, restricting our fit to semi-major axis values between 1 and 100 au, and companion mass values between 13 and 100 M$_{\rm Jup}$.

\begin{figure}[!ht]
\includegraphics[width=\columnwidth,trim=0 3.0cm 0 8cm]{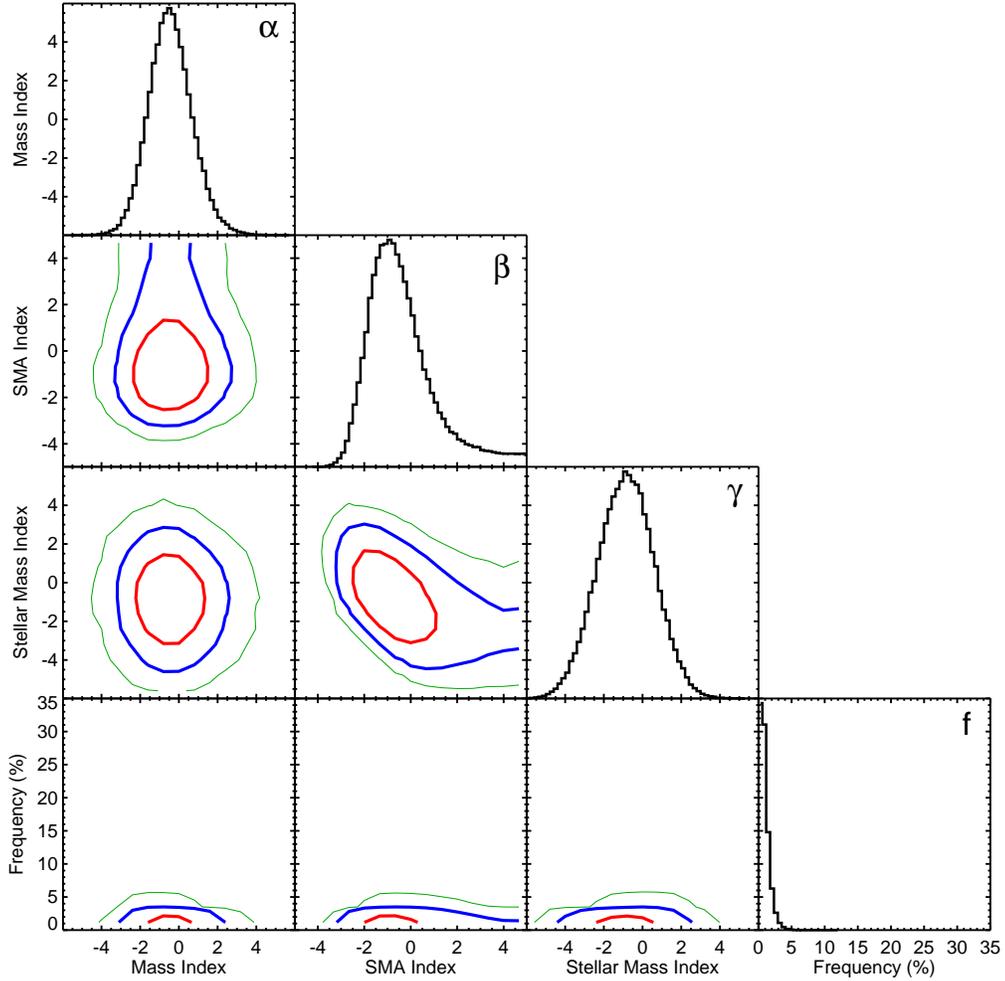}
\caption{Posteriors on the same parameterized model used to fit planets from the GPIES survey, but this time for brown dwarfs.  A significantly lower occurrence rate is observed, due to the higher sensitivity to these masses coupled with a detection rate lower by a factor of two. \label{fig:bd_triangle}}
\end{figure}

The depth of search, as shown by Figure~\ref{fig:tongue_plot}, is much more uniform over this mass range, and represents a much deeper sensitivity to these companions: the three brown dwarfs all lie within or near the 160-star contour.  Figure~\ref{fig:bd_triangle} displays the results from the brown dwarf analysis from the GPIES 300-star sample.

There is a tail to the posterior at large values of semi-major axis power law, which is truncated by the introduction of a tophat prior, uniform between values of -5 and 5, and 0 elsewhere.  This tail results from the limited number of brown dwarfs, and our model being defined in semi-major axis while our detections are made in separation.  A very positive power law index would place most brown dwarfs at the largest possible semi-major axis allowed, 100 au in our model.  Even if all brown dwarfs have a semi-major axis of 100 au, there is a non-zero probability of detecting three companions at projected separations between 9.9--23.7 au.

As for planets, we again truncate the range for reporting the occurrence rate, 10--100 au and 13--80 M$_{\rm Jup}$.  As expected given the deeper sensitivity and fewer detected objects, the brown dwarf occurrence rate is noticeably lower than for giant planets, 0.8$^{+0.8}_{-0.5}$\%.  

This occurrence rate is consistent with previous literature measurements of the brown dwarf companion frequency.  \citet{metchev09} conducted a deep survey of 100 stars (out of  a larger 266-star sample), and determined an occurrence rate of brown dwarfs (13--75 M$_{\rm Jup}$) orbiting between 28--1590 au of 3.2$^{+3.1}_{-1.7}$\%, assuming $\alpha = 0$ and $\beta = -1$.  The occurrence rate of a few percent for wide-separation brown dwarfs has been noted by a number of studies, including \citet{oppenheimer:2001}, \citet{brandt14}, \citet{uyama17}, and \citet{cheetham15}.

In the bottom left panel of Figure~\ref{fig:compare_occurrence}, we compare the semi-major axis distributions of GPIES and \citet{metchev09}.  Our result is for a lower occurrence rate than that of \citet{metchev09}, with a more negative power-law slope.  This decreasing of slope as we move closer to the star is consistent with results from RV surveys for an even smaller brown dwarf occurrence rate at closer separations, with a brown dwarf frequency $<$1\% within 5 year periods \citep{grether06}.

Additionally, we find no significant evidence for a stellar mass dependence of the brown dwarf occurrence rate.  We find a value for $\gamma$, the stellar mass power-law, of $-0.9 \pm 1.5$.  Thus a value of 0, representing no stellar mass dependence, lies within the 1$\sigma$ confidence interval.  A similar conclusion was reached by \citet{lannier16}, who computed companion frequency for stars from the MASSIVE survey of 58 M-stars (which detected two substellar objects) and the survey of 59 mostly-dusty AF stars of \citet{Rameau:2013}.  Focusing on companions with an intermediate mass ratio of 0.01-0.05 (corresponding to 10.5 to 52 M$_{\rm Jup}$ for a solar-type primary), \citet{lannier16} find no statistical difference between the higher mass and lower mass sample.  Similarly, the Gemini-NICI Planet-Finding Campaign surveyed young 70 B and A stars, finding three brown dwarfs \citep{nielsen:2013}, while \citet{bowler15} found four brown dwarfs from a survey of 78 single M stars, again suggesting no strong stellar mass dependence in the wide-separation brown dwarf occurrence rate.

\subsubsection{A Single Population of Substellar Companions at Wide Separations}

Finally, we consider the entire range of substellar objects, and attempt to fit planets and brown dwarfs with the same distribution.  GPIES has detected a total of 9 substellar objects, spanning a factor of 25 in companion mass, a factor of 11 in projected separation, and a factor of 1.6 in stellar mass.  Thus if a single power-law distribution can describe all these objects, the GPIES 300-star sample provides excellent constraints on the parameters of this distribution.

\begin{figure}[!ht]
\includegraphics[width=\columnwidth,trim=0 3cm 0 8cm]{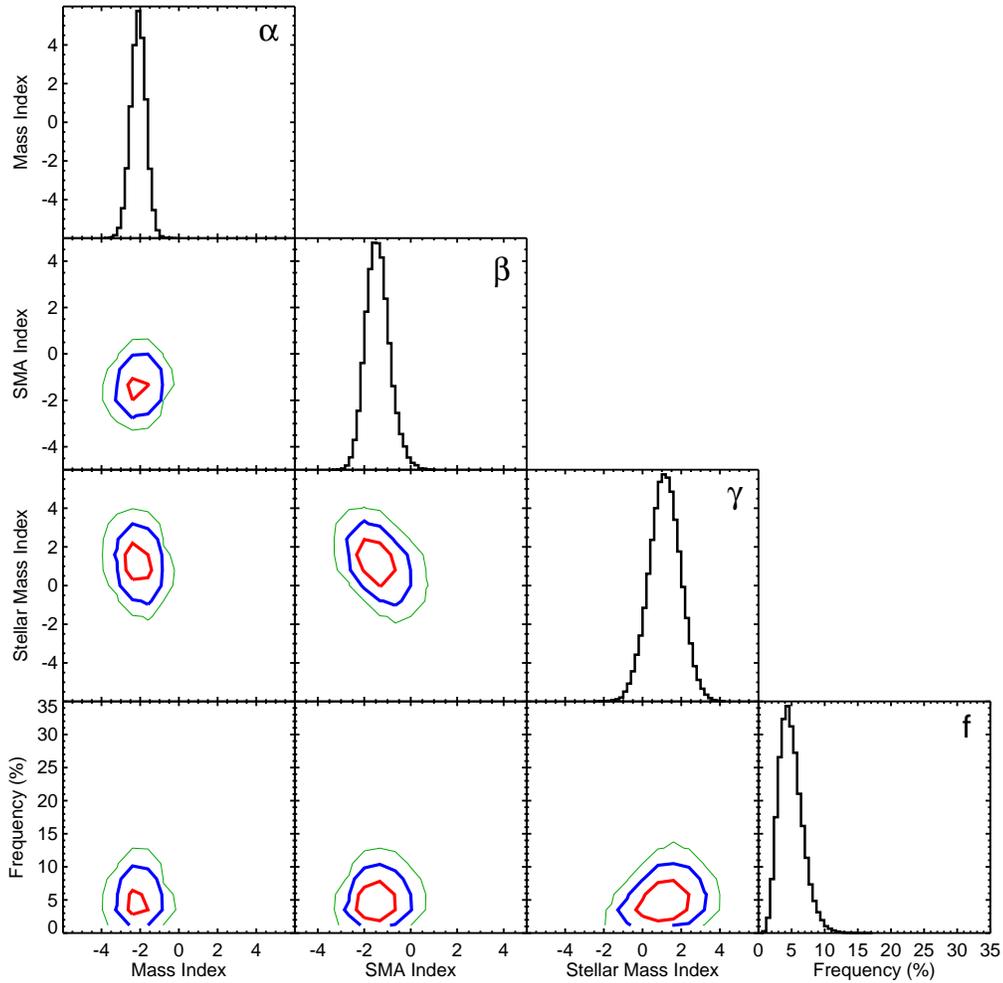}
\caption{A fit to the four-parameter model using all nine substellar companions detected in the GPIES 300 star sample. \label{fig:all_triangle}}
\end{figure}

Figure~\ref{fig:all_triangle} displays the results from fitting the entire GPIES substellar object yield with Equation~\ref{yesmass_eq}.  Compared to the other subsamples, the power law indices are most constrained with this fit, as expected from the larger number of objects to fit.  It is also not surprising that the medians of the posteriors of the three power law indices for the fit to all substellar objects are intermediate to posteriors for the fits to only planets and to only brown dwarfs.

A similar analysis was undertaken by \citet{brandt14} based on the SEEDS sample and previously published surveys.  Over a similar range to ours, 5--70 M$_{\rm Jup}$ and 10--100 au, \citet{brandt14} found values of $\alpha = -0.65 \pm 0.60$ and $\beta = -0.85 \pm 0.39$, compared to the values we derive here of $\alpha = -2.1 \pm 0.4$ and $\beta = -1.4 \pm 0.5$.  Thus we are finding significantly more negative power laws from our GPIES sample, placing more planets at smaller semi-major axis, and smaller masses.  \citet{brandt14} also found an overall occurrence rate of substellar companions in this regime of $1.7^{+1.1}_{-0.5}$\%. While we explicitly fit for $\gamma$, \citet{brandt14} do not, but instead adopted a fixed value of $\gamma = 1$.  The comparison is shown in the bottom right panel of Figure~\ref{fig:compare_occurrence}, with the two distributions consistent at $\sim$100 au, but the number of planets more steeply rising for GPIES when moving to smaller separations. \citet{vigan:2017} also computed the occurrence rates for substellar companions (0.5--75 M$_{\rm Jup}$, 20--300 au) of 2.1$^{+1.95}_{-0.60}$\%, assuming $\alpha = \beta = 0$.  With the same assumed distribution, \citet{galicher:2016} derived an occurrence rate of 0.35\%, with 95\% confidence between 0.10-1.95\%, for substellar companions between 5--70 M$_{\rm Jup}$ and 10--100 au.

The different values found of $\alpha$, $\beta$, and $f$ by GPIES and \citet{brandt14} can be accounted for by differences in the sensitivity and detected companions between the two samples.  \citet{brandt14} found one planetary mass companion (GJ 504 b), five brown dwarfs, and one intermediate object crossing the planet/brown dwarf boundary ($\kappa$ And b).  By contrast, GPIES detected more lower-mass companions, finding six planets and three brown dwarfs.  This leads to a more negative companion mass power law, predicting more lower mass companions.  In addition, the better IWA of GPI has led to three detections of substellar objects closer than 20 au (51 Eri b, $\beta$ Pic b, and HD 984 B), which in turn drives the semi-major axis power law more negative as well.  Finally, the larger number of detections, as well as power laws placing more planets at smaller masses (with lower sensitivity), are responsible for the larger occurrence rate we infer.

\subsection{Cold Start}\label{sec:cold}

In addition to the `hot start' evolutionary models of giant planets we have considered up until this point, there is an alternate set of `cold start' models.  In the cold start models \citep{marley07,fortney08}, as the planet forms most of the gravitational potential energy of the infalling gas is radiated away very early on ($\lesssim5$ Myr) in a shock, so that young planets have comparatively less internal energy to radiate away compared to their hot start counterparts, and so are cooler and less luminous at early ages.  At older ages ($\gtrsim300$ Myr) the cold start and hot start luminosity tracks converge.

The direct observational consequence of the cold start models is fainter, more difficult to detect planets than would be predicted from the hot start models.  Additionally, these models predict a pile-up in luminosity at the youngest ages, where planets between 2 and 10 M$_{\rm Jup}$ maintain luminosities between about 1-3$\times 10^{-6}$ L$_\odot$ for the first 30 Myr of their evolution  \citep{fortney08}.

\begin{figure}[!ht]
\gridline{\fig{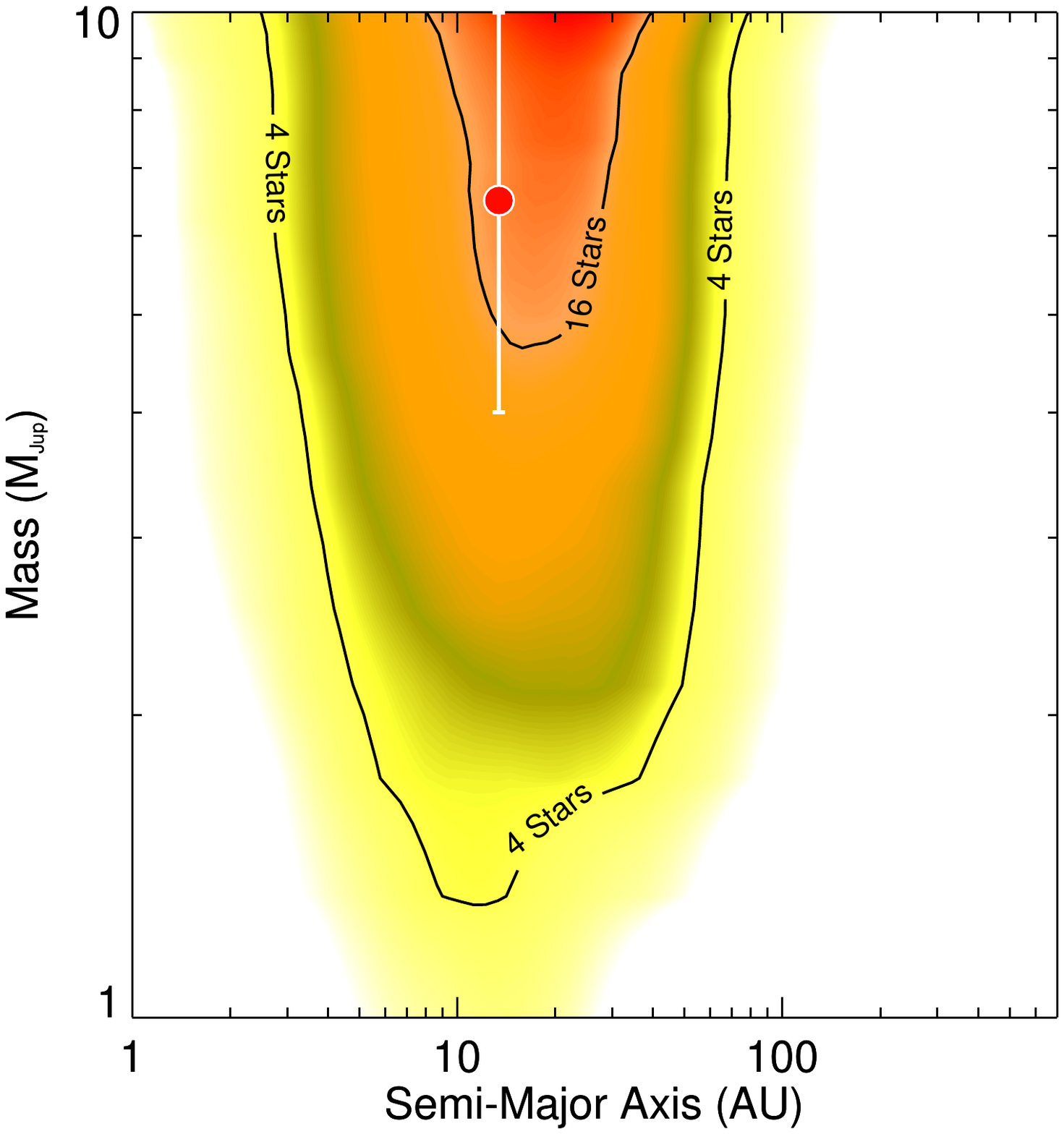}{0.5\columnwidth}{(a) \citet{fortney08}}
         \fig{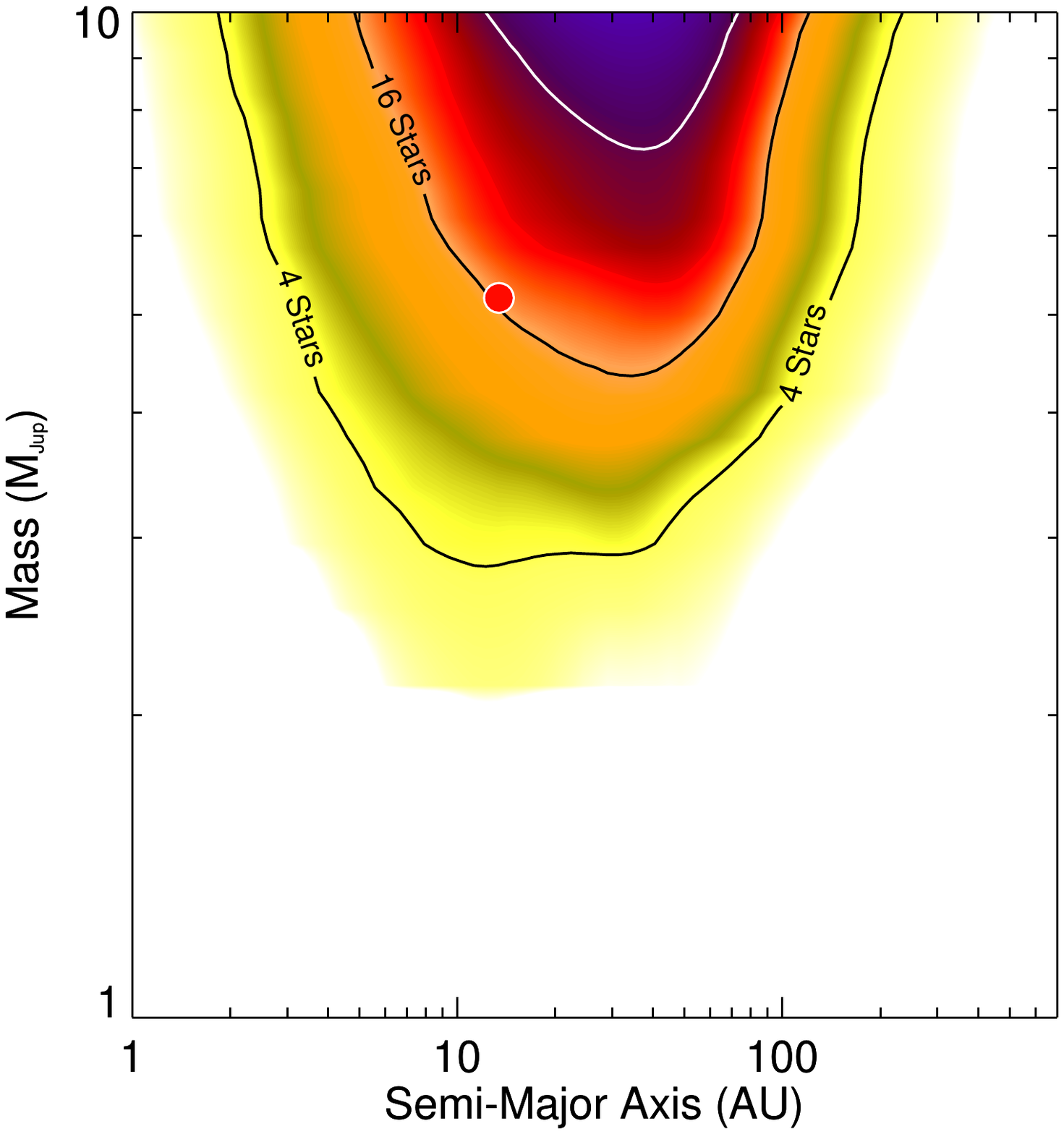}{0.5\columnwidth}{(b) Marley et al. (2019) in prep}}
\caption{Depth of search for cold start luminosity models.  Of the nine substellar objects detected by GPIES, only 51 Eri b has a luminosity low enough to potentially be a cold start planet.  The Sonora 2018 grid (right) predicts generally brighter giant planets, with the exception of a limited range of parameter space: between $\sim$10-50 Myr and $\lesssim$3 M$_{\rm Jup}$ the \citet{fortney08} models are slightly brighter.  As a result, the GPIES results are overall more sensitive to giant planets with the Sonora 2018 grid, but a handful of very young nearby stars result in the extension of the 4 star contour to the lowest masses for \citet{fortney08}. \label{fig:twocoldtongue}}
\end{figure}

We use cold start evolutionary models  \citep{fortney08}, and couple them with the AMES-COND atmospheric models \citep{allard01,baraffe03} by matching values of luminosity and temperature predicted by the cold start evolutionary grid with gridpoints in the AMES-COND models to extract $H$ magnitude.  We then produce a depth of search plot from the GPIES survey, given in the left panel of Figure~\ref{fig:twocoldtongue}.  There are currently no cold start models for brown dwarf masses, and of the six planets detected by GPIES, five have luminosities too large to have formed according to the cold start models.  Only 51 Eri b could potentially be a cold start planet.  However, due to the pile-up in luminosity of this model grid, the estimated mass can only be constrained to be between 4 and 10 M$_{\rm Jup}$ \citep{Macintosh:2015ew}.

As there is only one possible detection of a cold start planet, rather than attempt to fit a power law distribution, we instead solve for the 
occurrence rate of cold start giant planets with an assumed power law distribution.  Following the method described in Section~\ref{sec:stellarmass}, we adopt a uniform distribution in log space, $\alpha = \beta = -1$.  The average value over the depth of search between 2--13 M$_{\rm Jup}$, between 3-100 au shown in Figure~\ref{fig:twocoldtongue} is then 10.9 stars.  From Bayes' equation with a Poisson likelihood and a Jeffrey's prior, and one detection over this range, we find an occurrence rate of 15$^{+15}_{-9}$\%.  This is intermediate to the value we found for hot start giant planets around high-mass stars when assuming $\alpha = \beta = -1$, 24$^{+14}_{-10}$\% and the 2$\sigma$ upper limit for low-mass stars of $<6.9$\% (we caution that 51 Eridani contributed to these occurrence rates for both sets of models).  It is not yet certain which set of evolutionary models 51 Eri b follows, and so this cold start occurrence rate could be overestimated if it is indeed a hot start planet.  Additionally, these hot start and cold start models represent extremes in evolutionary models, there also exist ``warm start" models \citep{mordasini:2017,Spiegel:2012} that give mass as a function of the initial entropy of the planet (e.g. Figure 17 in \citealt{Rajan:2017ur}). Direct measurements of the mass of 51 Eri b from \textit{Gaia} astrometry of the host star could break the degeneracy between the different models.

\begin{figure}[!ht]
\includegraphics[width=0.9\columnwidth,trim=0 11.0cm 0 7.5cm]{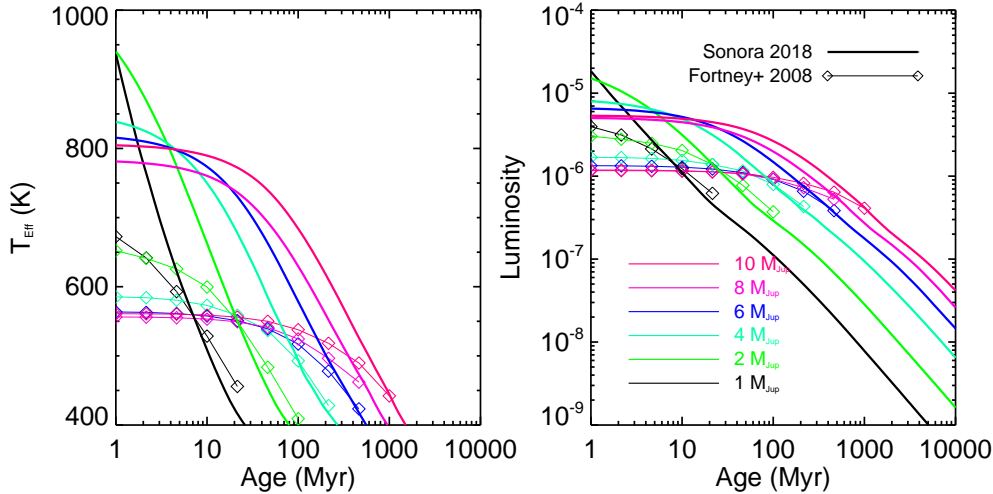}
\caption{Comparison of the two sets of cold start models we consider here, \citet{fortney08} and the Sonora 2018 grid (Marley et al. 2019 in prep).  The Sonora 2018 grid predicts generally higher temperatures and larger luminosities for giant planets at young ages. \label{fig:coldcomparison}}
\end{figure}

We also consider the new Sonora 2018 grid (Marley et al. 2019 in prep) with cloud-free atmospheres, and a new cold-start evolutionary grid.  This new set of cold start models predicts that young planets are generally hotter by $\sim$200 K and $\sim$5x more luminous than the \citet{fortney08} models,  as shown in Figure~\ref{fig:coldcomparison}.  Key differences between the two model sets include the assumed helium abundance (Y=0.24 and 0.28 for Marley et al. 2019 and \citealt{fortney08}, respectively), and the mass of the baryon in the equation of state, with \citet{fortney08} using $1\,\rm amu$ and Marley et al. (2019) using the mass of a hydrogen atom.

When using the Sonora 2018 grid, the mass of 51 Eri b is better defined, no longer lying in the overlap region of multiple mass tracks, but at a higher mass of 5.2 M$_{\rm Jup}$.  As before, no other detected GPIES planets are low luminosity enough to lie in the range of $H$ magnitudes predicted for 1--10 M$_{\rm Jup}$ planets.  The depth of search with the Sonora 2018 grid is given in the right panel of Figure~\ref{fig:twocoldtongue}.  The generally brighter planets results in deeper sensitivity to high-mass planets, though still not as deep as for the hot-start BT-Settl models.  For the lower-mass giant planets ($\lesssim$3 M$_{\rm Jup}$), however, the \citet{fortney08} models predict slightly brighter planets between $\sim$10--50 Myr.  Due to a small number of nearby stars with ages $\lesssim$50 Myr, then, the 4-star contour of the depth of search reaches 1.5 M$_{\rm Jup}$ for the \citet{fortney08} models, compared to 3 M$_{\rm Jup}$ for Sonora 2018.   Additionally, the Sonora 2018 grid has GPIES able to reach planets at larger semi-major axis, since a high luminosity planet means it can be detected around more distant target stars, so cold start planets can be detected around more of the GPIES sample.  Repeating the occurrence rate calculation above gives a slightly lower value, given the greater sensitivity, with 11$^{+11}_{-6}$\% of stars hosting a Sonora 2018 cold start planet between 2--13 M$_{\rm Jup}$ and 3--100 au.

Compared to hot start planets, potential cold start planets are very rare among the directly imaged exoplanets.  51 Eri b and GJ 504 b are the only two imaged planets have  low enough luminosity to be consistent with cold start models, and so how such a population depends on planet mass, semi-major axis, and stellar mass is not yet well understood.  Unlike previous generations of direct imaging exoplanet surveys, GPIES is reaching higher contrasts within 1", however surveys such as NICI \citep{nielsen:2013,wahhaj13,biller13,liu:2013} and IDPS \citep{Vigan:2012jm,Galicher:2016hg} reached higher contrasts than GPIES at $\gtrsim2''$, further from the glare of the central star.  For young, nearby moving group stars this corresponds to sensitivity to cold start planets to separations of $\gtrsim50$ au.  The relative dearth of detections could indicate a lower occurrence rate than hot start planets, or that cold start planets are found predominantly at a much smaller semi-major axis.  If 51 Eri b is indeed a cold start planet, and the occurrence rate of $\sim$10\% we infer here is accurate, then future surveys, either with an upgraded GPI \citep{chilcote18} or an upgraded SPHERE, or an ELT high-contrast imager, could allow us to better study these planets.

\section{Discussion}

At 300 stars, and with detections of six planets and three brown dwarfs, the early GPIES Campaign data is allowing us to probe the distribution of directly-imaged substellar companions at lower masses and closer separations than has been previously possible.  

\subsection{Do Wide Separation Brown Dwarfs and Giant Planets Follow the Same Distribution?}

The 13 M$_{\rm Jup}$ boundary that notionally separates giant planets from brown dwarfs represents the onset of deuterium burning, and a longstanding question has been whether there is a significant difference in formation mechanism between these two classes of objects (e.g., \citealt{schlaufman:2018}).  
Brown dwarfs and perhaps also giant planets of the highest masses are thought to represent
the low-mass outcome of the
process by which stars form, i.e., gravitational instability \citep{Kratter:2016,Forgan:2013} or turbulent fragmentation
(e.g., \citealp{Hopkins:2013}).
Objects of lower mass (Jupiters and below) may have formed
from the ``bottom up'' via core accretion, i.e., by coagulation of solids (e.g., \citealt{Goldreich:2004,Ormel:2017a}) and subsequent
accretion of gas (e.g., \citealt{Harris:1978,Mizuno:1978,Mizuno:1980,Pollack:1996jp}). 
Evidence for this dichotomy is seen for close-in companions observed by RV and transit surveys, where there exists a brown dwarf desert lacking companions near 30 $M_{\rm Jup}$ (e.g., \citealt{grether06,triaud:2017}). 
Such a deep minimum in the occurrence rate vs.~companion mass
distribution has not been seen at wider separations \citep{gizis01,metchev09}.  \citet{brandt14} found that
a single population can fit both brown dwarfs and giant planets at wide separations.

While turbulent fragmentation from a
molecular cloud---a process integral to the formation of 
stars---has been invoked to explain isolated planet-mass 
companions in systems with
primaries not too different in mass (e.g., 2MASSWJ 1207334-393254 and 2MASS
J11193254-1137466; \citealt{Chauvin:2004cy}; \citealt{best17}),
such a mechanism is not naturally applied to other directly
imaged planets, which show evidence for having formed
in a disk.
The orbit of $\beta$ Pic b is nearly coplanar with the star's
debris disk
\citep{Lagrange:2010,Nielsen:2014,Macintosh:2014js};
HD 95086 b appears to exhibit this same coplanarity with its star's
disk \citep{rameau2016};
and the HR 8799 planets are likely coplanar with each other
and with the surrounding debris disk 
\citep{wang:2018b}. Coplanarity points
to these planets forming in a disk, either by core accretion
or gravitational instability, rather than by turbulent 
fragmentation from a cloud.

\begin{figure}[!ht]
\includegraphics[width=0.9\columnwidth,trim=0 3.0cm 0 9cm]{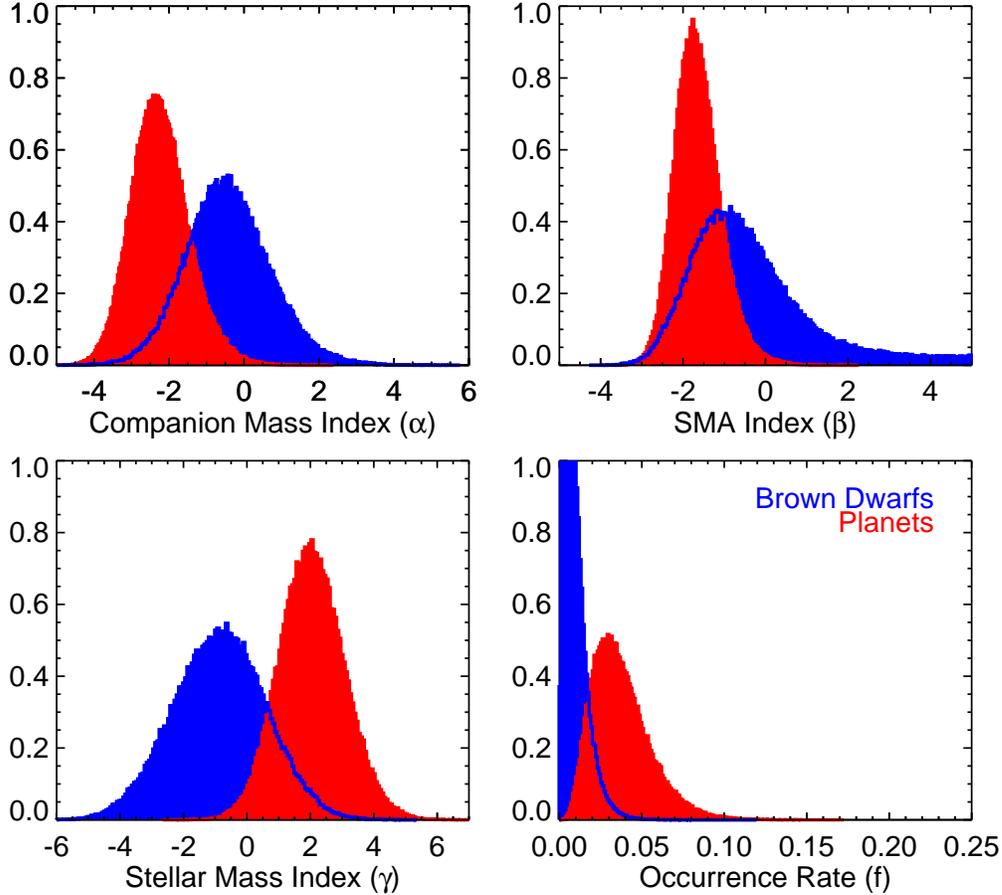}
\caption{Comparison of population parameter fits to planets and brown dwarfs, based on our GPIES sample. At 1-2$\sigma$ significance, brown dwarfs and giant planets appear to follow different underlying distributions between 10-100 au. \label{fig:bd_planet_compare}}
\end{figure}

In Section~\ref{sec:bayes} we fitted for the parameters to Equation~\ref{yesmass_eq} for three cases: using all substellar companions, only giant planets, and only brown dwarfs.  In Figure~\ref{fig:bd_planet_compare} we overplot the posteriors on the four parameters for brown dwarfs alone and planets alone.  Since the posteriors overlap, we cannot completely rule out the possibility that the two populations were drawn from the same underlying distribution. 
Nevertheless, when we follow the same procedure as in Section~\ref{sec:stellarmass}, we find the probabilities
that the values of $\alpha$ and $\beta$ are larger for brown dwarfs at 87.3\% and 74.3\%, while the probabilities that $\gamma$ and $f$ are smaller for brown dwarfs are 94.4\% and 95.1\%, respectively.
Thus, we find it likely that planets and brown dwarfs at 10-100 au separation follow different underlying distributions---planets obey a more bottom-heavy mass distribution, and are concentrated toward smaller
SMA---at a confidence level between 1 and 2$\sigma$.

The greatest degree of overlap in the posteriors is for $\beta$, while the other three parameters show a more significant difference for planets and brown dwarfs.  In particular, for wide separation giant planets this distribution is more likely to have lower companion masses and higher stellar host mass, with a higher overall occurrence rate compared to brown dwarfs.

\subsection{Comparing to Demographics of RV-Detected Planets}

The RV planet-search method represents one of the more robust ways to measure the underlying distributions of giant planets with periods $\lesssim$ 10 years.  While {\it Kepler}'s transiting planets greatly outnumber RV-detected planets, the RV method does not suffer as much bias toward longer periods, especially with precision RV surveys operating for multiple decades 
(e.g., \citealt{fischer:2014,bonfils:2013,astudillo:2017}).  \citet[][C08]{cumming2008} presented an analysis of planets orbiting a uniform sample of Keck data of FGK target stars, and fitted a double power law model to mass $m$ and period $P$ (similar in form to our Equation~\ref{nomass_eq}), finding that $d^2N/(dP \, dm) \propto P^{-0.74} m^{-1.31}$, with an overall planet occurrence rate of 10.5\% between 2--2000 days and
0.3--10 M$_{\rm Jup}$. This period distribution is equivalent to a semi-major axis ($a$) 
distribution of $dN/da \propto a^{-0.61}$ over a range of 0.031--3.1 au
for solar-mass stars. 
Past direct imaging surveys have sought to determine the extent to which this distribution holds for wider-separation planets \citep{Nielsen:2010,nielsen:2013,brandt14}.  While our occurrence rate $f$ gives the number of planets per star, C08 only consider the planet with the largest Doppler amplitude in each system, and therefore calculate the number of planetary \textit{systems} per star.  However, C08 note that this is not a strong bias, as only $\sim$10\% of their systems have multiple detected RV planets, and we proceed to use the C08 occurrence rate as an estimate of the number of planets per star.

Figure~\ref{fig:nomass_triangle} is inconsistent with C08's value of $\alpha$ (planet mass index) of -1.31 at the 1$\sigma$ level, and excludes their value of $\beta$ (semi-major axis index) of -0.61 at $\sim$2$\sigma$.  If we insist on the C08 values of $\alpha$, $\beta$, and $f$, and extend the distribution out to wider separations, we would predict a planet fraction of 4.7\% between 5--13 M$_{\rm Jup}$ and 10--100 au---this is also disfavored by GPIES at the 1$\sigma$ level.  A more significant difference, though, is in host stellar mass, since the \citet{cumming2008} sample consisted entirely of planets around FGK stars, while the GPIES planets are exclusively orbiting higher-mass stars.

\begin{figure}[!ht]
\includegraphics[width=0.9\columnwidth,trim=0 3cm 0 8cm]{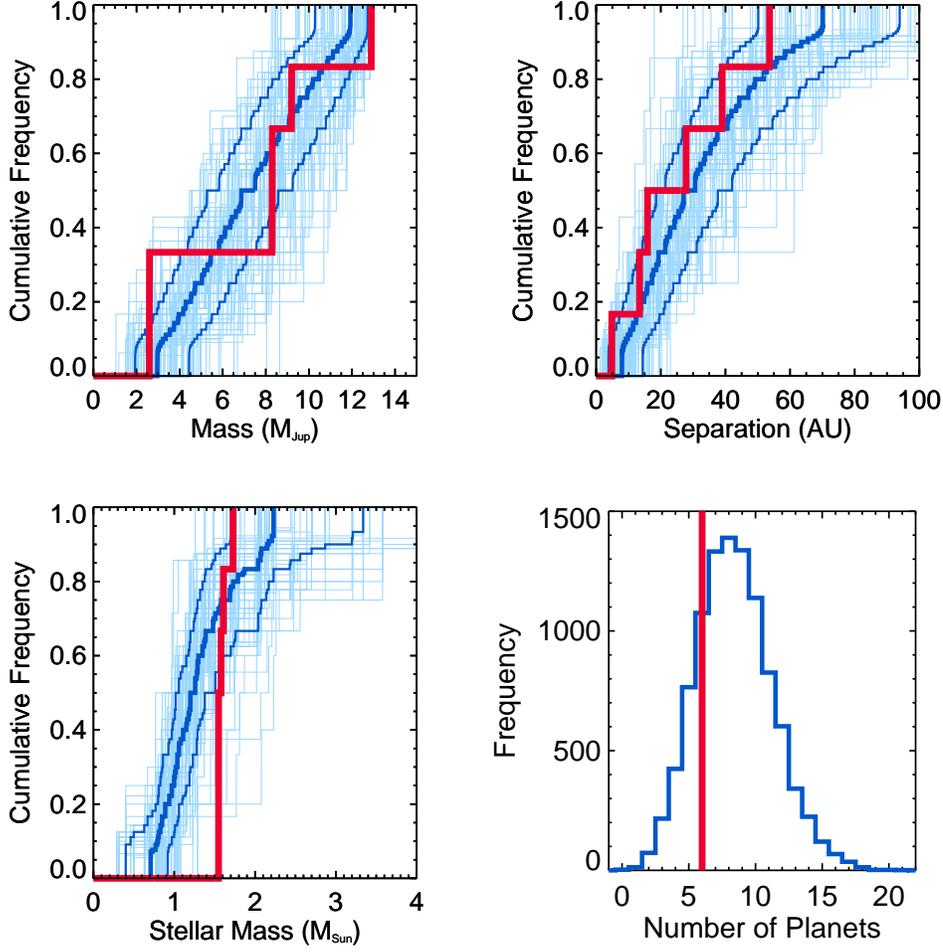}
\caption{Multiple mock GPIES surveys in a universe following the \citet{cumming2008} population of giant planets, extended out to 100 au, with no occurrence rate dependence on stellar host mass.  Cumulative histograms give the GPIES-detected planets (in red) and the mock surveys (in blue), for mass, separation, and stellar mass.  Thicker blue lines show the median and 1$\sigma$ confidence region for these cumulative distributions.  The final panel notes the predicted number of planets from these mock surveys, compared to the six planets that were actually detected.  While the \citet{cumming2008} distribution is consistent in number of planets and mass distribution, it somewhat overpredicts the separations of detected planets, and requires additional model parameters to account for the strong stellar mass dependence seen in the GPIES planet occurrence rate. \label{fig:cumming_comparison}}
\end{figure}

We further investigate the extent to which the C08 RV distribution is consistent with the GPIES results by setting up a Monte Carlo simulation where planets are drawn from the 
C08 power laws and occurrence rate, but extrapolated out to an upper semi-major axis cut-off of 100 au, to account for planets seen out to $\sim$70 au (HD 95086 b, HR 8799 b).  For now, we assume that the \citet{cumming2008} distribution fitted to FGK stars applies to stars of all masses, so that planet frequency is not a function of stellar mass ($\gamma=0$), and $\alpha$ and $\beta$ are the same for lower-mass and higher-mass stars.  We use the same method that generated the GPIES completeness maps to produce an ensemble of simulated planets drawn from a distribution using the 
C08 values of $\alpha$, $\beta$, and $f$ around each target star, and determine which planets are detectable given the contrast curve for that star \citep{Nielsen:2008,Nielsen:2010}.  We then record the mass, projected separation, and stellar host mass for each detectable planet from the ensemble.  Finally, we conduct 10,000 mock GPIES surveys, randomly assigning each star a chance of having a planet in the given mass and semi-major axis range, then using the per-star completeness to determine whether that planet would have been detected.

Figure~\ref{fig:cumming_comparison} compares these mock surveys to the results from GPIES.  Shown in red are cumulative histograms of the six planets detected by GPIES in planet mass, projected separation, and stellar host mass.  Overlaid are these same  parameters as measured from the mock surveys.  There is generally good agreement in planet mass, as expected from the aforementioned overlap between our measured value of $\alpha$ and that of \citet{cumming2008}.  Separation also appears consistent, though the GPIES distribution tends to slightly smaller values of separation than the median of the mock survey, again as we would expect from the GPIES-only fit for $\beta$ finding a more negative value than for the RV planets.  The largest departure, however, is clearly for stellar mass, where the median mock survey finds planets around stars over a range of stellar masses (by construction), whereas GPIES observes all planet hosts to be the higher-mass stars in the sample.  So while the overall occurrence rate of planets appears similar to predictions of the C08 distribution extended to 100 au, the breakdown by stellar mass is significantly different.

\begin{figure}[!ht]
\includegraphics[width=0.9\columnwidth,trim=0 11.5cm 0 8.5cm]{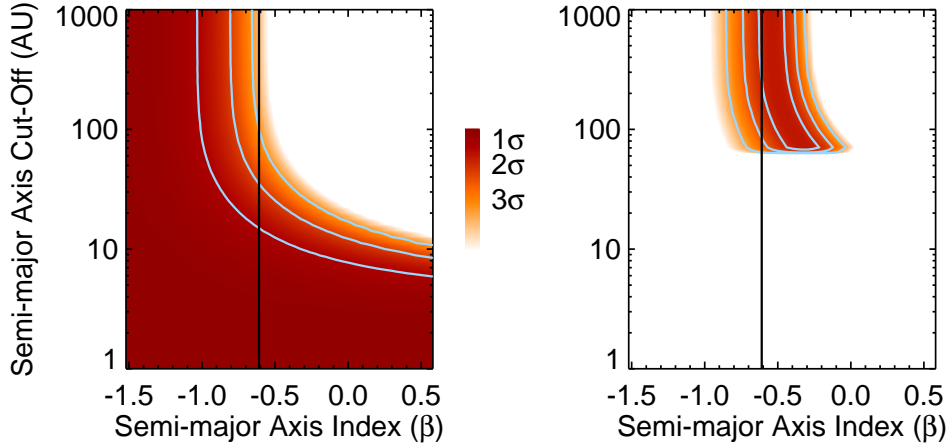}
\caption{A second way to compare the \citet{cumming2008} distributions
with GPIES data, this time distinguishing between stars below $1.5 M_\odot$ (left) and above (right).  We keep the mass power-law index ($\alpha$) and normalization ($f$) between 0.031-3.1 au fixed at the
C08 values, and allow the semi-major axis power-law index ($\beta$)
and the semi-major axis upper cut-off to find their most
probable values given the GPIES data. The color bar gives the probability of a given combination of $\beta$ and cut-off, with white areas being ruled out, and darker regions allowed. 
The value of $\beta$ exhibited by RV planets is -0.61 \citep{cumming2008}
and is shown by a vertical solid line. 
This value (and the C08 values of $\alpha$=-1.31 and $f$=10.5\% that are adopted
by construction) could apply to our sample of lower-mass stars if the semi-major axis distribution is cut off at less than 34 au (2$\sigma$)
or less than 15 au (1$\sigma$).
For higher-mass stars, when adopting the \citet{cumming2008} values of $\alpha$, $\beta$, and $f$, a value of semi-major axis cut-off of 264 au is consistent at 1$\sigma$,
while a cut-off of 85 au is needed to be
consistent at 2$\sigma$.  We emphasize that this analysis is distinct from the Bayesian method used in Figure~\ref{fig:nomass_triangle}, where the cut-off was fixed, and the values of $\alpha$, $\beta$, and $f$ were all allowed to float.  Additionally, while the MCMC analysis explicitly fitted for the observed separations of the planets, this method considers only  occurrence rate.  
\label{fig:cumming_comparison2}}
\end{figure}

To investigate the dependence of these results on the adopted
upper cut-off in semi-major axis, which is currently set to 100 au, and to better account for the
difference in stellar masses, 
we divide our sample into lower-mass stars ($<1.5$ M$_\odot$) and higher-mass stars $\geq$1.5 M$_{\odot}$), and conduct a second series of mock surveys for the two subsamples.  This time we keep the value of $\alpha$ and the total occurrence rate within 3.1 au constant using the \citet{cumming2008} values, and vary the value of the semi-major axis power law index ($\beta$) and the upper cut-off, which sets the maximum semi-major axis planets in the distribution may have.  For each combination of these two parameters, we compute the expected number of planets GPIES should have detected using our Monte Carlo simulations.  We then use the Poisson distribution to derive the probability that we achieved our actual number of detections in each subsample, displaying the results in Figure~\ref{fig:cumming_comparison2}.

For lower-mass stars ($<1.5$ M$_\odot$), with a null result for planets, we can only set upper limits for the two parameters, by taking the probability of detecting 0 planets given the expectation value.  For the \citet{cumming2008} value of $\beta$=-0.61, we find the upper cut-off must be less than 15 and 34 au
 at 1 and 2$\sigma$, respectively. Thus it is plausible that for solar-mass stars the \citet{cumming2008} distribution extends into the giant planet region of our own solar system, as our GPIES sensitivity does not extend inside 10 au.

For higher-mass stars ($>1.5$ M$_\odot$), since we have detected six planets, we can better constrain the values of $\beta$ and the upper cut-off within this modified \citet{cumming2008} distribution.  We again use the Poisson distribution to compute the probability of detecting six planets around the GPIES higher-mass stars at each point in the grid, then draw contours encircling 68\%, 95\%, and 99.7\% of the total probability.  We set a prior that imposes a lower limit on the semi-major axis cut-off of 70 au, corresponding to HR 8799 b \citep{Wang:2018}.  Though we only detected HR 8799 cde given the field of view of GPI, we assert that a distribution fitted to the inner three planets of the system should not exclude the outer planet b.  The \citet{cumming2008} value of $\beta$=-0.61 intersects the 1$\sigma$ contour at 264 au, and intersects the 2$\sigma$ contour at a semi-major axis cut-off of 85 au.  Such a large value for the upper cut-off is difficult to reconcile with the relative lack of detections of planets around higher-mass stars beyond $\sim$100 au by previous generations of imaging surveys.  An extrapolation of the C08 distribution would predict an equal number of planets between 10-80 au (where all six GPIES planets reside) and between 80--264 au.  The stated errors of C08 in the period distribution ($\frac{dN}{dlnP} \propto P^{0.26 \pm 0.1}$), if they are Gaussian, would correspond to errors in $\beta$ for semi-major axis of $-0.61 \pm 0.15$, and so a 1$\sigma$ value of $-0.46$ would be more consistent with our results, reaching the 1$\sigma$ value in the right panel of Figure~\ref{fig:cumming_comparison2} at 79 au. Adopting such a value for $\beta$ would make the upper limits for lower-mass stars more stringent, moving from 15 and 34 au at 1 and 2$\sigma$ to 12 and 23 au, respectively.  Previously we had fitted for all three parameters in Equation~\ref{nomass_eq} simultaneously---$\alpha$, $\beta$, $f$---while fixing the value of the upper cut-off to 100 au.  Here, as we fix $\alpha$ and $f$ to the \citet{cumming2008} values and let $\beta$ and the upper cut-off float, we reach a similar conclusion, namely that GPIES excludes C08's value of $\beta$ at $\sim$1$\sigma$, and that adjusting the value of the upper cut-off is not a straightforward path to removing this discrepancy.

The upper cut-off analysis above considers only the occurrence rate when evaluating the consistency of the extended \citet{cumming2008} distribution with our GPIES data.  In order to consider not just occurrence rate but also the properties of the detected planets, we repeat the mock surveys on only the 123 stars above $1.5 M_\odot$,
assuming an upper semi-major axis cut-off of 100 au,
and the C08 values of $\alpha$ and $\beta$.
Given this modified distribution and our completeness to planets around our 123 higher-mass stars, the mean number of planets found is only 2.6.  To more closely 
match our actual observed number of 6 planets, we increase the $<3.1$ au planet occurrence rate from the \citet{cumming2008} value by a factor of 2 (which for a 1.5 M$_\odot$ star would represent a value of $\gamma$ near our median value from Figure~\ref{fig:planets_gamma_triangle} of 2.03).  This moves the mean number of detections up to 5.2, with 42\% of the mock surveys now producing 6 or more detected planets.  As before, we assume there is no stellar mass dependence in planet occurrence within this subsample of higher-mass stars ($\gamma=0$).  

\begin{figure}[!ht]
\includegraphics[width=0.9\columnwidth,trim=0 3cm 0 8cm]{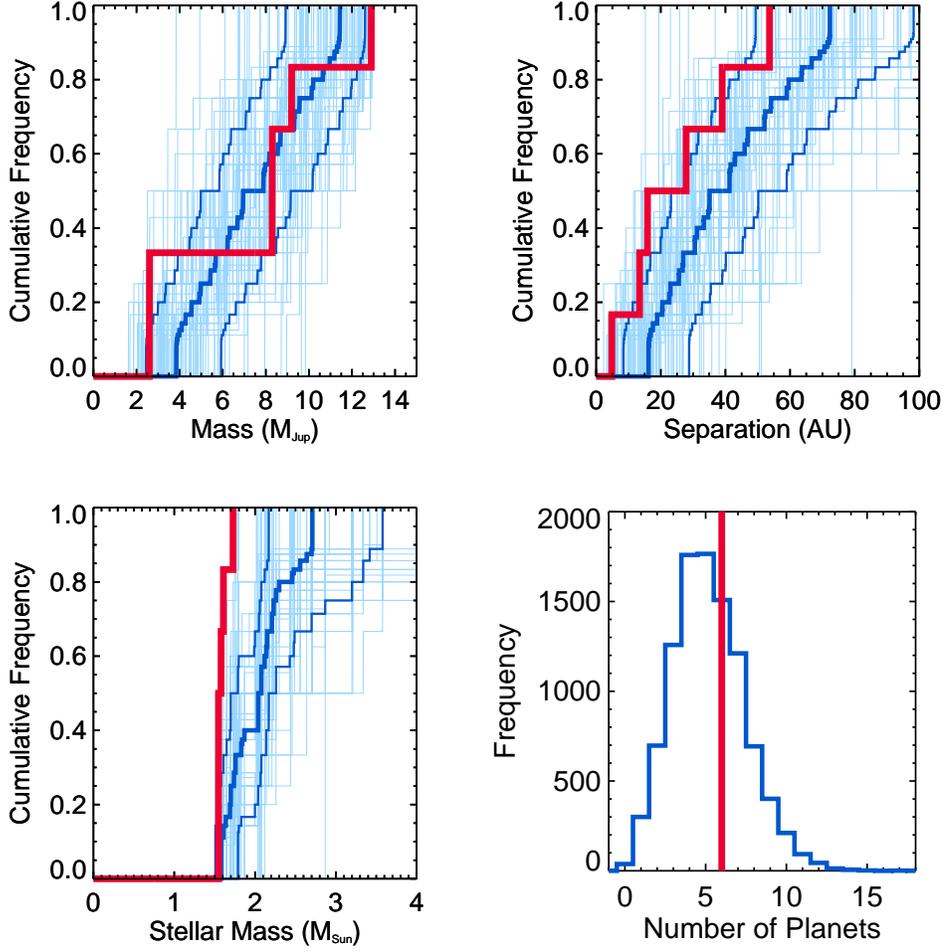}
\caption{Mock GPIES surveys, again based on the \citet{cumming2008} population of giant planets, extended out to 100 au, but only for the 123 higher-mass stars in the survey, and with planet occurrence rate $f$ boosted by a factor of 2 relative to the value
reported by C08.  With $f$ thus boosted, the number of actually detected planets is consistent with the mock surveys. The planet mass distributions also appear consistent. However, there is a significant discrepancy in the separation distribution, with GPIES observing planets at closer separations than the majority of the mock surveys.
\label{fig:cumming_comparison3}}
\end{figure}

Results from this third set of mock surveys restricted to higher-mass stars ($>1.5$ M$_\odot$) are plotted in Figure~\ref{fig:cumming_comparison3}.  While the planet mass distribution and overall occurrence rate appear consistent with the observed GPIES detections, there remain significant differences for the projected separation and stellar host mass distributions. Again, the \citet{cumming2008} separation distributions place more planets at wider separation than are observed.  We saw the same disagreement with the \citet{cumming2008} distribution in our full power-law fit (Figure~\ref{fig:nomass_triangle}), where we found a more negative value of $\beta$ from the GPIES sample.  Indeed, 80.2\% of mock surveys have a detected planet at a projected separation beyond 53.65 au, the value for HD 95086 b.  
We could try to reduce this discrepancy by boosting $f$ still further and decreasing
the upper cut-off in the semi-major axis distribution. 
But lowering the cut-off would exclude, without physical justification,
observed planets like HD 95086 b in our sample at 55 au, as well as the 
widest-separation planet in the HR 8799 system, planet b at 70 au.  Though HR 8799 b does not appear in our sample, the packed nature of the system \citep{wang:2018b} leads us to reject a model fit to the inner three planets that does not fit the fourth planet.

Figure \ref{fig:cumming_comparison3} also shows that despite our restriction
to $> 1.5$ M$_\odot$ stars, a mismatch persists in the stellar mass distributions:
the mock surveys predict planets to be detected around more 2 $M_\odot$ stars
than are actually observed. 
For such high-mass stars in the GPIES sample, there is no obvious bias where more massive stars are systematically older or more distant. The persistent discrepancy in stellar mass distributions likely arises because the C08 distribution places about as many
planets at large orbital distances as at small ones, and those
at large distances are detectable in the mock surveys of higher-mass
stars. In the actual GPI survey, 
as the absolute $H$ magnitude decreases by $\sim$1 mag from 1.5 to 2 M$_\odot$, planets like 51 Eri b, $\beta$ Pic b, and HR 8799 e that cluster at small orbital separations become undetectable around higher-mass stars.

In sum, we find that the semi-major axis distribution from C08 of 
RV-discovered giant planets is a poor fit to directly imaged giant planets at wider separations. 
For lower-mass stars ($<1.5$ M$_\odot$), while we cannot exclude the possibility that the \citet{cumming2008} RV population extends to orbital distances of 15--34 au, we do not have any data to support this model, either.  For higher-mass stars, the data
from GPIES place more planets at smaller semi-major axes than at larger ones;
GPIES prefers a semi-major axis power-law index of $\beta \simeq -2$,
which is discrepant from the C08 value of -0.61 at 2$\sigma$.

\citet{fernandes18} analyze RV and \textit{Kepler} giant planet occurrence rates, and observe a turnover in occurrence rate as a function of semi-major axis, with a peak at $\sim$3 au.  Their symmetric {\tt epos} fit gives a value of the planet mass power law $\alpha = -1.45 \pm 0.05$ (similar to the C08 value of $\alpha = -1.31 \pm 0.2$), and a broken power law fit to semi-major axis, with $\beta = -0.025^{+0.30}_{-0.22}$ for small periods and $\beta = -1.975^{+0.22}_{-0.30}$ for large periods, again converting a period power law index into semi-major axis.  At small semi-major axis, then, \citet{fernandes18} find a more positive value of power law index compared to the C08 value of $\beta = -0.61 \pm 0.15$, and a value of $\beta$ similar to our fit to planets around higher-mass stars ($\beta = -2.0^{+0.6}_{-0.5}$) and planets around stars with a stellar mass term ($\beta = -1.7^{+0.6}_{-0.5}$).  For the mass power law index, we find a more negative value of $\alpha = -2.3^{+0.8}_{-0.6}$, and while the 1$\sigma$ confidence intervals overlap, the offset could be a sign of a different planet mass distribution at larger separations, or between higher-mass and lower-mass stars.  \citet{fernandes18} find the location of the period break from their fit at $P = 1581^{+894}_{-392}$ days, which corresponds to a semi-major axis for solar-type stars of $2.66^{+0.93}_{-0.46}$ au.  The asymmetric {\tt epos} fit gives similar values for $\alpha$, $\beta$ within the break, and overall occurrence rate, as well as consistent values for $\beta$ beyond the break and the location of the break, though with significantly larger error bars: $\beta = -2.8^{+1.4}_{-1.9}$ and a break at $3.2^{+1.1}_{-1.4}$ au.  We find support for the conclusions of \citet{fernandes18}, that the power law index $\beta$ measured by C08 cannot continue to large values of semi-major axis, but must turn over within $\lesssim$10 au (modulo the difference in stellar mass between the planet hosts in the C08 and \citealt{fernandes18} samples and GPIES).

\begin{figure}[!ht]
\includegraphics[width=0.9\columnwidth,trim=0 3cm 0 7cm]{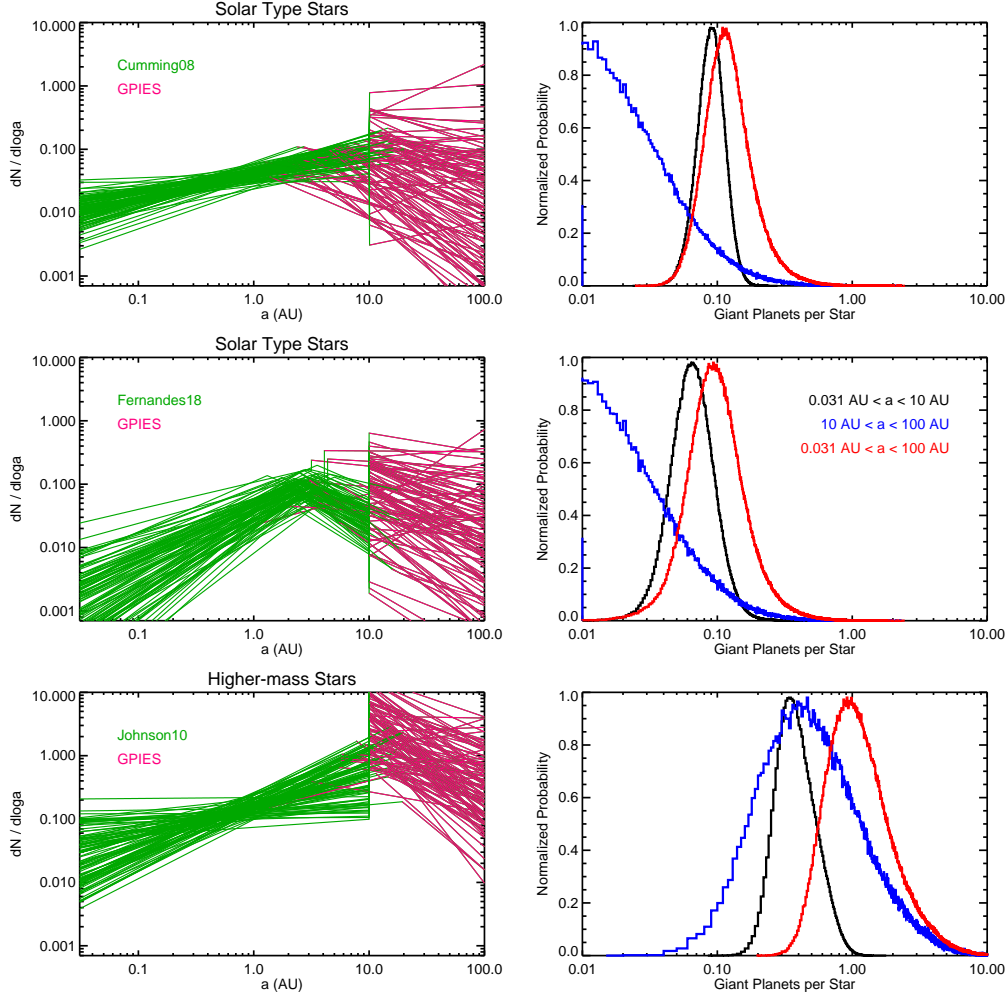}
\caption{Estimates of the total occurrence rate of giant planets between 1--13 M$_{\rm Jup}$ and 0.031--100 au, from combining RV occurrence rates and power laws for close-in giant planets with those presented in this work for wider-separation giant planets.  Occurrence rates from GPIES are extrapolated to 1 M$_{\odot}$ to compare to the C08 and \citet{fernandes18} results for solar-type stars, as well as extrapolated down to 1 M$_{\rm Jup}$, using the power laws from Table~\ref{tbl:allposteriors}.  From this analysis we estimate the fraction of giant planets within 100 au of solar-type stars to be 0.14$^{+0.10}_{-0.05}$ and 0.09$\pm 0.02$ for the C08 and \citet{fernandes18} distributions, while for higher-mass stars ($>1.5$ M$_\odot$) we find a significantly larger occurrence rate of 1.5$^{+1.7}_{-0.7}$.  As we have limited sensitivity to $\lesssim$3 M$_{\rm Jup}$ planets, future observations will be required to determine if the power laws seen by GPIES do indeed extend to these low planet masses, and thus more robustly determine the giant planet occurrence rate. 
\label{fig:extrapolate_occurrence}}
\end{figure}

Finally, we combine RV results for small-separation planets and GPIES results at larger separations to estimate the total number of planets within 100 au, as shown in Figure~\ref{fig:extrapolate_occurrence}.  In the top two panels, we couple GPIES and the C08 occurrence rate to calculate the number of planets as a function of semi-major axis between 1--13 M$_{\rm Jup}$ around 1 M$_\odot$ stars.  Draws are taken from the MCMC chain of our fit to giant planets including a stellar mass term, and combined with Monte Carlo draws of $\alpha$, $\beta$, and occurrence rate from C08, assuming each is Gaussian distributed and that the three parameters are independent, thus producing two power law distributions.  We attempt to make the combined distribution continuous by finding a point where the two power laws cross between 1--20 au; if the two do not intersect in this range, the handover point is taken to be 10 au, with a discontinuity in the final combined distribution.  Finally, we integrate the distribution to find the posterior on the total number of giant planets.  We find a total occurrence rate (number of planets per star) of 0.09$\pm$0.02 from $0.031-10$ au, and 0.04$^{+0.09}_{-0.03}$ from $10-100$ au, producing a total occurrence rate within 100 au of 0.14$^{+0.10}_{-0.05}$.  Repeating this analysis with the \citet{fernandes18} {\tt epos} symmetric distribution produces similar results, with occurrence rates of 0.07$^{+0.03}_{-0.02}$, 0.04$^{+0.09}_{-0.03}$, and 0.12$^{+0.10}_{-0.05}$ respectively.

As GPIES only detected planets around higher-mass stars, the posterior for giant planet occurrence approaches zero between 10--100 au, though the GPIES data do set a clear upper limit given the lack of detections of giant planets around these lower-mass stars.  By assuming this power-law model, we find a 2$\sigma$ upper limit on the occurrence rate of 1--13 M$_{\rm Jup}$ planets around 1 M$_\odot$ stars of 39.0\% and 38.1\% for the C08 and \citet{fernandes18} distributions, respectively.

For higher-mass stars, we repeat this analysis, this time using the \citet{johnson:2010} occurrence rate, stellar mass power law ($\gamma$) and errors, and setting the mass power law $\alpha$ equal to the C08 value and uncertainties, and allowing $\beta$ to follow a uniform distribution between -1 and 0.  We then draw from our MCMC fits to the high-mass star GPIES sample and continue as before.  Extrapolating the \citet{johnson:2010} results gives an occurrence rate within 10 au for 1.75 M$_\odot$ stars of 0.42$^{+0.20}_{-0.12}$.  Given the large occurrence rate determined by GPIES for 5--13 M$_{\rm Jup}$, and the very negative value of $\gamma$, the total occurrence rate for 1--13 M$_{\rm Jup}$ planets between 10--100 au orbiting higher-mass stars is even larger, 1.0$^{+1.7}_{-0.6}$, with a total occurrence rate within 100 au of 1.5$^{+1.7}_{-0.7}$ for these higher-mass stars.  Given the negative value of the planet mass power law index ($\alpha = -2.4^{+0.9}_{-0.8}$)  found from GPIES, many of these predicted planets have small masses.  With GPIES having limited sensitivity to $\lesssim 3$ M$_{\rm Jup}$ planets orbiting higher-mass stars, we cannot rule out a break in the power law distribution at lower planet masses, which would result in a smaller occurrence rate. 

\subsection{Connecting to Planet Formation Theory}\label{ssec:theory}
Our finding that giant planets
outnumber brown dwarfs at more than 10-to-1 around $>1.5 M_\odot$
stars argues against these planets having formed by
gravitational instability. Fragmentation of marginally 
Toomre-unstable disks
\citep{toomre:1964,goldreich:1965} is thought to produce a mass distribution 
weighted toward larger and not smaller masses. When fragments initially condense, 
they may already be
at least as massive as Jupiter (e.g., Figure 3 of
\citealp{Kratter:2016}; see also 
\citealt{Rafikov:2005}).
In a population synthesis study predicated 
on gravitational instability,
\citet{Forgan:2013} found the overwhelming majority of clumps 
that survived tidal disruption to lie above the deuterium burning 
limit. These initial fragments are expected to accrete still
more mass from their strongly
self-gravitating, dynamically active parent disks,
which themselves may still be accreting from infalling progenitor
clouds \citep{Kratter:2010gf}.

In addition, the requirement that disks rapidly cool
in order to fragment
\citep{Gammie:2001} 
predicts that collapse occurs at large stellocentric distances
where material is less optically thick, e.g., outside at least
$\sim$50 au in a model inspired by the HR 8799 system
(Figure 3 of \citealp{Kratter:2010gf}; 
see also \citealt{Matzner:2005} and \citealt{Rafikov:2009}).
The generic predictions of gravitational instability---a
top-heavy mass distribution extending to brown dwarfs and
stars, and large orbital separations (modulo orbital 
migration)---are not borne out by our
2--13 $M_{\rm Jup}$ planets located between 10--60 au. Instead
we infer mass distributions 
(using the fit to planets around higher-mass stars) 
that are bottom-heavy
($dN/dm \propto m^{\alpha}$ with 
$\alpha = -2.4 \pm 0.8$)
and separation distributions that favor small semi-major axes
($dN/da \propto a^{\beta}$ with $\beta = -2.0 \pm 0.5$).

Formation by gravitational instability remains an option for GPIES brown dwarfs, for which we infer mass and separation indices of $\alpha = -0.5 \pm 1.1$ (more top-heavy) and $\beta = -0.7^{+1.0}_{-1.5}$ (weighted more toward large distances).  The three brown dwarfs in our survey are located at orbital distances of 10--30 au, somewhat smaller than the values cited above for fragmentation. However, previous surveys with wider fields of view have found brown dwarfs at larger separations, e.g., HR 7329 B at 200 au \citep{lowrance:2000}, $\zeta$~Del~B at 910 au \citep{DeRosa:2014}, and HIP 79797 Ba/Bb, a pair of brown dwarfs in a 3 au orbit separated by 370 au from the primary star \citep{huelamo:2010,nielsen:2013}.  \citet{Moe:2018} present mechanisms (e.g., Lidov-Kozai cycles) for stellar/brown dwarf binary systems to decrease their separation over time.

Another observed property of brown dwarf companions that we have highlighted is their lack of preference for either higher-mass or lower-mass stars as host.  This, too, is consistent with formation within gravitationally unstable disks. If the disk gas surface density $\Sigma_{\rm g} \propto M_*$ and the sound speed $c_s \propto T^{1/2} \propto M_*^{1/4}$ (where we have taken the gas temperature $T \propto L_*^{1/4}$ and used the pre-main-sequence luminosity-mass relation $L_* \propto M_*^2$), then Toomre's $Q \propto 1/M_*^{1/4}$. This weak dependence on stellar mass is made even weaker when we fold in the large scatter in observed gas disk masses for given $M_*$ (e.g., \citealp{Andrews:2018}). Deeper analyses
that combine the $Q$ criterion with the requirement of
fast cooling reveal that disks gravitationally fragment
at orbital locations that are remarkably insensitive
to system parameters related to stellar mass 
\citep{Matzner:2005,Rafikov:2009}.

Our finding that giant planets at 10--100 au are more commonly hosted
by A stars than by lower-mass stars extends the trend
determined by prior demographic studies that closer-in gas giants
are more common around FGK stars than around M dwarfs (\citealp{Clanton:2014,Mulders:2015,ghezzi:2018}).
The correlation with stellar mass
presents a major constraint on theory. Having argued
above against planet formation by
gravitational instability (a `top-down' scenario),
we ask here whether the trend with stellar mass is consistent with
core accretion, a `bottom-up' scenario whereby giants nucleate from
sufficiently massive rocky cores. Note that the high luminosities of directly imaged planets, though
consistent with hot start models and inconsistent with
cold start models (Section~\ref{sec:cold}),
do not necessarily rule out core accretion. 
Cold and hot start models differ according to the planet's assumed
initial entropy, which can vary significantly within core accretion depending on the shock dynamics of runaway gas accretion (see, e.g., Figure 12 of \citealp{Berardo:2017}).

In a prescient analysis,
\citet{Laughlin:2004} anticipated fewer giant planets around
lower-mass stars,
arguing that core formation times would be longer around M dwarfs, and potentially
too long compared to gas disk dispersal times, because
their disks have longer orbital periods and are expected
to have fewer solids.
We update this reasoning within
the emerging paradigm of `pebble
accretion,' in which cores grow by accreting small solid particles
aerodynamically entrained in disk gas (see \citealp{Ormel:2017a} and
\citealp{Johansen:2017} for reviews). 
As particles drift radially inward through the gas disk
(e.g., \citealp{Weidenschilling:1977}), a fraction of them are accreted 
by the protocore. The final core mass is a monotonic increasing function
of $\int^{t_{\rm drift}} \Sigma_{\rm s} dt$:
the disk solid surface density $\Sigma_{\rm s}$
(local to the protocore) integrated over the time
it takes solids to drift past the protocore's orbit
and drain out from the disk (see equations 36, A7, and A9 of
\citealp{Lin:2018}).  There are 
dependencies of the final core mass on variables apart from this integral, but they 
are unlikely to change the qualitative conclusion  
that pebble accretion breeds more massive cores around more massive
stars.\footnote{In equations (A7) and (A9) of \citet{Lin:2018} pertaining to the `2D' limit
where the pebble accretion radius exceeds the solid
disk scale height, the extra variable dependencies only
strengthen the trend with stellar mass. In the case of their
equation 36 (`3D'),
the extra variables mute the trend but do not reverse it,
unless the particle Stokes number $St$ (aerodynamic stopping time
normalized to the orbit time) decreases more rapidly
than $1/M_*^{3/2}$. \citet{Pascucci:2016} support 
$St \propto 1/M_*^{1/2}$ for fragmentation-limited
particle growth \citep{Birnstiel:2012}; this relation
steepens to $1/M_*$ 
if particle sizes are fixed and gas disk masses
scale linearly with stellar mass.}
The surface density
$\Sigma_{\rm s}$ scales with the total solid disk mass, which should
increase with $M_*$ (a linear dependence is commonly assumed). 
The drift time $t_{\rm drift}$ also increases with $M_*$,
as inferred empirically from millimeter-wave dust continuum 
observations showing that disks are systematically
larger around higher-mass stars (\citealp{Pascucci:2016},
in particular their section 6.2; see also \citealp{Andrews:2018}).
Thus pebble/core accretion gets at least the sign right with respect
to the correlation of giant planet occurrence with host star mass, 
and it remains to work out more quantitative models that can satisfy the
constraints reported in our Table~\ref{tbl:allposteriors}.\footnote{Which is not to say that there are not outstanding problems intrinsic to the theory. 
\citet{Lin:2018} point out that pebble drift times in Jupiter-breeding scenarios
do not match observations. There may also be insufficient mass in pebbles (\citealt{Manara:2018}). Size distributions of accreting particles are crucial but uncertain; the origin of seed cores is tied to the perennial
problem of planetesimal formation. See \citet{Johansen:2017} for some current thinking.}

What about semi-major axis trends in the context of pebble 
accretion?  The preference reported here for giant planets to be located at
the lower end of the 10-100 au range makes sense insofar
as protocores at smaller orbital distances access larger
reservoirs of solids lying exterior to their orbits and drifting past.
Pebble accretion probabilities
also tend to be higher closer in because disk gas is denser there
and more efficiently drags pebbles onto cores: accretion
at small distances is more likely to be in the so-called `settling' regime,
with particle trajectories more strongly gravitationally focussed, 
as opposed to the less
efficient `hyperbolic' regime which obtains 
farther out
(\citealp{Ormel:2010}; \citealp{Lin:2018}, in particular their
Figure 15 which contrasts fast settling accretion at 5 au
with slow hyperbolic accretion at 40 au).

We can also try connecting the observed preference of directly
imaged giant planets for smaller semi-major axes---which we reiterate
applies to distances of 10--100 au---to
the observed preference of giant planets discovered by radial velocities (RV)  
for larger semi-major axes, which applies
to distances of 0.03--3 au (e.g., \citealp{cumming2008,fernandes18}).  Transit surveys report a similar rise in giant planets out to
$\sim$300 days (e.g., \citealp{dong:2013}; \citealp{Petigura:2018}).
Assuming these semi-major axis trends
can be stitched together---keeping in mind the caveat
that they derive from disjoint samples of higher-mass and lower-mass stars---we infer the giant planet occurrence rate
peaks between $\sim$3--10 au. 
Remarkably, a similar conclusion has been found by \citet{fernandes18} in a
combined RV-transit analysis of FGK stars: they find
evidence for a turn-over in the giant planet occurrence
rate at $\sim$2--3 AU.
A giant planet occurrence rate that peaks
at intermediate distances aligns at
least qualitatively with pebble/core accretion theory.
The aforementioned gains in growth with smaller orbital distance
are eventually reversed, at the shortest distances,
by other competing physical effects.
Cores stop accreting pebbles when they
gravitationally perturb 
the surrounding disk gas so strongly that the inward drift of pebbles
stalls \citep{Lambrechts:2014}. Such `pebble isolation' sets an upper mass limit that decreases as the orbital distance decreases \citep{Bitsch:2018,fung:2018}; cores do not grow past sub-Earth masses inside $\sim$0.1 AU (see Figure 12 of \citealp{Lin:2018}, who rule out in-situ formation of hot Jupiters by pebble accretion on this basis).
Another effect stifling growth at the shortest distances
is that disk gas becomes hotter and more opaque there and therefore
cools and accretes onto cores more slowly
\citep{Lee:2015,Lee:2016}. All the above theoretical reasons,
together with the many arguments pointing to inward migration of 
hot Jupiters by Kozai cycles/tidal friction and perhaps also
disk torques \citep{Dawson:2018},
lead us to posit a ``sweet spot'' for gas giant formation
at stellocentric distances that are neither too large nor too small. From Figures~\ref{fig:compare_occurrence} and \ref{fig:extrapolate_occurrence}, 
this sweet spot appears to be at $\sim$2--10 au.

\section{Conclusions}

From the first 300 stars observed out of the planned 600-star survey, reaching contrasts of 10$^6$ within 1$''$ radius, GPIES is one of the largest and deepest direct imaging surveys for exoplanets conducted to date.  Our analysis of the data shows that there is a clear stellar mass dependence on planet occurrence rate, with stars $> 1.5 M_\odot$ more likely to host giant planets (5--13 $M_{\rm Jup}$) at wide separations (semi-major axes 10--100 au) than lower-mass stars. Around higher-mass stars, the total occurrence rate of such planets is 9$^{+5}_{-4}$\%. When fitting our data for higher-mass stars ($>1.5$ M$_\odot$) with a power-law model of planet populations where $d^2N/(dm \,da) \propto m^\alpha a^{\beta}$, we find negative power law indices of $\alpha \simeq -2$ and $\beta \simeq -2$, values that favor smaller masses and smaller semi-major axes. All the statistical trends that we have uncovered---a bottom-heavy companion mass distribution, orbital distances concentrated more toward 10 au than 100 au, and a strong preference for higher-mass host stars---appear consistent with a ``bottom-up" formation scenario: core/pebble accretion.

By comparison to giant planets, brown dwarfs appear to exhibit different statistics: a more top-heavy mass distribution ($\alpha \simeq -0.47$), orbital distances that are somewhat more concentrated toward 100 au than 10 au ($\beta \simeq -0.65$), and no particular preference for host stellar mass. These are consistent with a ``top-down'' formation scenario: gravitational instability and fragmentation of circumstellar disks. Our finding of a brown dwarf occurrence rate on the order of 1\%, independent of stellar mass, is consistent with previous literature analyses of brown dwarfs from direct imaging surveys.

While planets around GPIES stars are more common around higher-mass stars ($>$1.5 M$_\odot$), it is interesting to note that the most massive planet host in the sample is $\beta$ Pictoris (1.73 M$_\odot$), yet there are 76 stars more massive than 1.8 M$_\odot$ in our 300-star sample.  Repeating the analysis of Section~\ref{sec:stellarmass}, but dividing the higher-mass sample between 1.5--1.8 M$_\odot$ and $>$1.8 M$_\odot$, we find at 99.7\% confidence that the intermediate stellar mass bin has more planets than the highest stellar mass bin.  But this result is dependent on our choice of bin boundary---when we extend the dividing line from 1.8 to 2.1 to 2.4 M$_\odot$, our confidence decreases from 99.7\% to 97.1\% to 87.1\%, respectively.  It is thus difficult to know if we are observing a real effect or whether we are fine-tuning our boundary based on a small number of detections.  For comparison, the limit we used here to separate higher-mass and lower-mass stars, 1.5 M$_\odot$, must be reduced to 1.04 M$_\odot$ (with only 67 stars out of 300 below this limit) before the confidence with which we can conclude that higher-mass stars have more planetary systems than lower-mass stars drops below 95\%.  If the frequency of wide-separation giant planet systems does indeed drop off beyond $\sim$2 M$_\odot$, this would suggest that the mechanism that formed the giant planets in the GPIES sample becomes less efficient at stellar masses above this value.  \citet{jones:2016} observe
reminiscent
behavior for close-in giant planets around giant stars, with a planet occurrence rate that increases out to $\sim$2.3 M$_\odot$, but then decreases at higher stellar masses.  \citet{jones:2016} model this as a Gaussian, though like us they have fewer 
targets in their sample at the high stellar mass end.  Additionally, their
RV targets have evolved off the main sequence, while ours are predominantly on the main sequence, and so it is unclear how to disentangle formation effects from post-main sequence evolution effects.  Nevertheless, further observations of additional higher mass stars ($\gtrsim 2$ M$_\odot$) can test whether this trend is real.

There are a limited number of bright, young, nearby stars available to high-contrast direct imagers like GPI and SPHERE.  As a result, at the current achievable contrasts there are unlikely to be a significant number of new detections of imaged giant planets following the completion of the GPIES and 
SPHERE/SHINE \citep{feldt:2017} campaigns.  An improvement in performance from instrument upgrades \citep{chilcote18}, however, could unlock mass/separation phase space around these target stars that is currently inaccessible.  For example, 51 Eridani was observed multiple times by planet-hunting surveys (e.g., \citealt{heinze:2010}, \citealt{Rameau:2013}, \citealt{biller13}), yet the planet 51 Eri b was not discovered until it was imaged with the more advanced GPI \citep{Macintosh:2015ew}.  Additional planets discovered with upgraded GPI and SPHERE would allow us to test the robustness of the trends with stellar mass and planetary mass uncovered in this paper.  The Gaia mission will identify astrometric signatures of planets with $\gtrsim 10$ year orbits that may be confirmed with direct imaging, which can also boost the number of companions in this separation range.

Alternatively, high contrast imaging at longer wavelengths offers the opportunity to expand the list of potential imaging targets to older stars ($\sim$300 Myr--1 Gyr).  While a 5 M$_{\rm Jup}$ planet is 100 times brighter at 30 Myr than 600 Myr at $H$-band, this ratio is closer to a factor of 10 at $M'$ (BT-Settl, \citealt{baraffe:2015}).  For a uniform star formation rate, there should be nine times more stars between 100 Myr and 1 Gyr than there are stars younger than 100 Myr.  JWST, GMT, TMT, and ELT will be well-suited to probe wide-separation giant planet populations around these intermediate-age stars at 5--10 $\mu$m.  Going forward, combining different detection methods to produce a full accounting of giant planet and brown dwarf populations from 0.01 to 1000 au will
provide still more constraints on formation mechanisms.

\acknowledgments
We thank the referee for helpful comments that improved the quality of this manuscript.  Based on observations obtained at the Gemini Observatory, which is operated by the Association of Universities for Research in Astronomy, Inc., under a cooperative agreement with the NSF on behalf of the Gemini partnership: the National Science Foundation (United States), the National Research Council (Canada), CONICYT (Chile), Ministerio de Ciencia, Tecnolog\'{i}a e Innovaci\'{o}n Productiva (Argentina), and Minist\'{e}rio da Ci\^{e}ncia, Tecnologia e Inova\c{c}\~{a}o (Brazil). This research has made use of the SIMBAD and VizieR databases, operated at CDS, Strasbourg, France. This work has made use of data from the European Space Agency (ESA) mission {\it Gaia} (\url{https://www.cosmos.esa.int/gaia}), processed by the {\it Gaia} Data Processing and Analysis Consortium (DPAC, \url{https://www.cosmos.esa.int/web/gaia/dpac/consortium}). Funding for the DPAC has been provided by national institutions, in particular the institutions participating in the {\it Gaia} Multilateral Agreement.
 This research used resources of the National Energy Research Scientific Computing Center, a DOE Office of Science User Facility supported by the Office of Science of the U.S. Department of Energy under Contract No. DE-AC02-05CH11231.
This work used the Extreme Science and Engineering Discovery Environment (XSEDE), which is supported by National Science Foundation grant number ACI-1548562. 

J.R., R.D. and D.L. acknowledge support from the Fonds de Recherche du Qu\'{e}bec. JRM's work was performed in part under contract with the California Institute of Technology (Caltech)/Jet Propulsion Laboratory (JPL) funded by NASA through the Sagan Fellowship Program executed by the NASA Exoplanet Science Institute. Support for MMB's work was provided by NASA through Hubble Fellowship grant \#51378.01-A awarded by the Space Telescope Science Institute, which is operated by the Association of Universities for Research in Astronomy, Inc., for NASA, under contract NAS5-26555. Supported by NSF grants AST-1411868 (E.L.N., K.B.F., B.M., and J.P.), AST-141378 (G.D.), and AST-1518332 (R.D.R., J.J.W., T.M.E., J.R.G., P.G.K.). Supported by NASA grants NNX14AJ80G (E.L.N., S.C.B., B.M., F.M., and M.P.), NNX15AC89G and NNX15AD95G (B.M., J.E.W., T.M.E., R.J.D.R., G.D., J.R.G., P.G.K.), NN15AB52l (D.S.), and  NNX16AD44G (K.M.M). K.W.D. is supported by an NRAO Student Observing Support Award SOSPA3-007. J.J.W. is supported by the Heising-Simons Foundation 51~Pegasi~b postdoctoral fellowship. E.C. thanks Courtney Dressing for helpful discussions. Portions of this work were performed under the auspices of the U.S. Department of Energy by Lawrence Livermore National Laboratory under Contract DE-AC52-07NA27344. This work benefited from NASA's Nexus for Exoplanet System Science (NExSS) research coordination network sponsored by NASA's Science Mission Directorate.  The Center for Exoplanets and Habitable Worlds is supported by the Pennsylvania State University, the Eberly College of Science, and the Pennsylvania Space Grant Consortium. 

\facility{Gemini:South (GPI)}
\software{Astropy \citep{astropy}, emcee \citep{ForemanMackey:2013io}, GPI DRP \citep{Perrin:2016gm}, IDL Astronomy Library \citep{idlastro}, pyKLIP \citep{Wang:2015th}}

\bibliographystyle{apj}   
\bibliography{ms} 

\begin{thebibliography}{}
\expandafter\ifx\csname natexlab\endcsname\relax\def\natexlab#1{#1}\fi

\bibitem[{{Abt}(2009)}]{2009ApJS..180..117A}
{Abt}, H.~A. 2009, \apjs, 180, 117

\bibitem[{{Abt} \& {Morrell}(1995)}]{1995ApJS...99..135A}
{Abt}, H.~A., \& {Morrell}, N.~I. 1995, \apjs, 99, 135

\bibitem[{{Abt} \& {Willmarth}(2006)}]{2006ApJS..162..207A}
{Abt}, H.~A., \& {Willmarth}, D. 2006, \apjs, 162, 207

\bibitem[{{Allard}(2014)}]{allard:2014}
{Allard}, F. 2014, in IAU Symposium, Vol. 299, Exploring the Formation and
  Evolution of Planetary Systems, ed. M.~{Booth}, B.~C. {Matthews}, \& J.~R.
  {Graham}, 271--272

\bibitem[{{Allard} {et~al.}(2001){Allard}, {Hauschildt}, {Alexander},
  {Tamanai}, \& {Schweitzer}}]{allard01}
{Allard}, F., {Hauschildt}, P.~H., {Alexander}, D.~R., {Tamanai}, A., \&
  {Schweitzer}, A. 2001, \apj, 556, 357

\bibitem[{{Amado} {et~al.}(2000){Amado}, {Doyle}, {Byrne}, {Cutispoto},
  {Kilkenny}, {Mathioudakis}, \& {Neff}}]{2000AA...359..159A}
{Amado}, P.~J., {Doyle}, J.~G., {Byrne}, P.~B., {et~al.} 2000, \aap, 359, 159

\bibitem[{{Andrews} {et~al.}(2018){Andrews}, {Terrell}, {Tripathi}, {Ansdell},
  {Williams}, \& {Wilner}}]{Andrews:2018}
{Andrews}, S.~M., {Terrell}, M., {Tripathi}, A., {et~al.} 2018, \apj, 865, 157

\bibitem[{{Asensio-Torres} {et~al.}(2018){Asensio-Torres}, {Janson},
  {Bonavita}, {Desidera}, {Thalmann}, {Kuzuhara}, {Henning}, {Marzari},
  {Meyer}, {Calissendorff}, \& {Uyama}}]{asensio:2018}
{Asensio-Torres}, R., {Janson}, M., {Bonavita}, M., {et~al.} 2018, \aap, 619,
  A43

\bibitem[{Asplund {et~al.}(2009)Asplund, Grevesse, Sauval, \&
  Scott}]{Asplund:2009eu}
Asplund, M., Grevesse, N., Sauval, A.~J., \& Scott, P. 2009, ARA{\&}A, 47, 481

\bibitem[{{Astropy Collaboration} {et~al.}(2013){Astropy Collaboration},
  {Robitaille}, {Tollerud}, {Greenfield}, {Droettboom}, {Bray}, {Aldcroft},
  {Davis}, {Ginsburg}, {Price-Whelan}, {Kerzendorf}, {Conley}, {Crighton},
  {Barbary}, {Muna}, {Ferguson}, {Grollier}, {Parikh}, {Nair}, {Unther},
  {Deil}, {Woillez}, {Conseil}, {Kramer}, {Turner}, {Singer}, {Fox}, {Weaver},
  {Zabalza}, {Edwards}, {Azalee Bostroem}, {Burke}, {Casey}, {Crawford},
  {Dencheva}, {Ely}, {Jenness}, {Labrie}, {Lim}, {Pierfederici}, {Pontzen},
  {Ptak}, {Refsdal}, {Servillat}, \& {Streicher}}]{astropy}
{Astropy Collaboration}, {Robitaille}, T.~P., {Tollerud}, E.~J., {et~al.} 2013,
  \aap, 558, A33

\bibitem[{{Astudillo-Defru} {et~al.}(2017){Astudillo-Defru}, {Forveille},
  {Bonfils}, {S{\'e}gransan}, {Bouchy}, {Delfosse}, {Lovis}, {Mayor}, {Murgas},
  {Pepe}, {Santos}, {Udry}, \& {W{\"u}nsche}}]{astudillo:2017}
{Astudillo-Defru}, N., {Forveille}, T., {Bonfils}, X., {et~al.} 2017, \aap,
  602, A88

\bibitem[{{Bailer-Jones}(2015)}]{bailerjones2015}
{Bailer-Jones}, C.~A.~L. 2015, \pasp, 127, 994

\bibitem[{{Bailey} {et~al.}(2014){Bailey}, {Meshkat}, {Reiter}, {Morzinski},
  {Males}, {Su}, {Hinz}, {Kenworthy}, {Stark}, {Mamajek}, {Briguglio}, {Close},
  {Follette}, {Puglisi}, {Rodigas}, {Weinberger}, \& {Xompero}}]{Bailey:2014}
{Bailey}, V., {Meshkat}, T., {Reiter}, M., {et~al.} 2014, \apjl, 780, L4

\bibitem[{{Bailey} {et~al.}(2016){Bailey}, {Poyneer}, {Macintosh}, {Savransky},
  {Wang}, {De Rosa}, {Follette}, {Ammons}, {Hayward}, {Ingraham}, {Maire},
  {Palmer}, {Perrin}, {Rajan}, {Rantakyr{\"o}}, {Thomas}, \&
  {V{\'e}ran}}]{bailey:2016}
{Bailey}, V.~P., {Poyneer}, L.~A., {Macintosh}, B.~A., {et~al.} 2016, in
  \procspie, Vol. 9909, Adaptive Optics Systems V, 99090V

\bibitem[{{Baraffe} {et~al.}(2003){Baraffe}, {Chabrier}, {Barman}, {Allard}, \&
  {Hauschildt}}]{baraffe03}
{Baraffe}, I., {Chabrier}, G., {Barman}, T.~S., {Allard}, F., \& {Hauschildt},
  P.~H. 2003, \aap, 402, 701

\bibitem[{Baraffe {et~al.}(2003)Baraffe, Chabrier, Barman, Allard, \&
  Hauschildt}]{Baraffe:2003bj}
Baraffe, I., Chabrier, G., Barman, T.~S., Allard, F., \& Hauschildt, P.~H.
  2003, A{\&}A, 402, 701

\bibitem[{{Baraffe} {et~al.}(2015){Baraffe}, {Homeier}, {Allard}, \&
  {Chabrier}}]{baraffe:2015}
{Baraffe}, I., {Homeier}, D., {Allard}, F., \& {Chabrier}, G. 2015, \aap, 577,
  A42

\bibitem[{{Barrado y Navascu{\'e}s} {et~al.}(2004){Barrado y Navascu{\'e}s},
  {Stauffer}, \& {Jayawardhana}}]{2004ApJ...614..386B}
{Barrado y Navascu{\'e}s}, D., {Stauffer}, J.~R., \& {Jayawardhana}, R. 2004,
  \apj, 614, 386

\bibitem[{{Bell} {et~al.}(2015){Bell}, {Mamajek}, \& {Naylor}}]{bell:2015}
{Bell}, C.~P.~M., {Mamajek}, E.~E., \& {Naylor}, T. 2015, \mnras, 454, 593

\bibitem[{{Berardo} {et~al.}(2017){Berardo}, {Cumming}, \&
  {Marleau}}]{Berardo:2017}
{Berardo}, D., {Cumming}, A., \& {Marleau}, G.-D. 2017, \apj, 834, 149

\bibitem[{{Bertelli} {et~al.}(2008){Bertelli}, {Girardi}, {Marigo}, \&
  {Nasi}}]{bertelli:2008}
{Bertelli}, G., {Girardi}, L., {Marigo}, P., \& {Nasi}, E. 2008, \aap, 484, 815

\bibitem[{{Best} {et~al.}(2017){Best}, {Liu}, {Dupuy}, \& {Magnier}}]{best17}
{Best}, W.~M.~J., {Liu}, M.~C., {Dupuy}, T.~J., \& {Magnier}, E.~A. 2017,
  \apjl, 843, L4

\bibitem[{{Beuzit} {et~al.}(2008){Beuzit}, {Feldt}, {Dohlen}, {Mouillet},
  {Puget}, {Wildi}, {Abe}, {Antichi}, {Baruffolo}, {Baudoz}, {Boccaletti},
  {Carbillet}, {Charton}, {Claudi}, {Downing}, {Fabron}, {Feautrier},
  {Fedrigo}, {Fusco}, {Gach}, {Gratton}, {Henning}, {Hubin}, {Joos}, {Kasper},
  {Langlois}, {Lenzen}, {Moutou}, {Pavlov}, {Petit}, {Pragt}, {Rabou}, {Rigal},
  {Roelfsema}, {Rousset}, {Saisse}, {Schmid}, {Stadler}, {Thalmann}, {Turatto},
  {Udry}, {Vakili}, \& {Waters}}]{Beuzit:2008}
{Beuzit}, J.-L., {Feldt}, M., {Dohlen}, K., {et~al.} 2008, in Society of
  Photo-Optical Instrumentation Engineers (SPIE) Conference Series, Vol. 7014,
  Society of Photo-Optical Instrumentation Engineers (SPIE) Conference Series

\bibitem[{{Biller} {et~al.}(2010){Biller}, {Liu}, {Wahhaj}, {Nielsen}, {Close},
  {Dupuy}, {Hayward}, {Burrows}, {Chun}, {Ftaclas}, {Clarke}, {Hartung},
  {Males}, {Reid}, {Shkolnik}, {Skemer}, {Tecza}, {Thatte}, {Alencar},
  {Artymowicz}, {Boss}, {de Gouveia Dal Pino}, {Gregorio-Hetem}, {Ida},
  {Kuchner}, {Lin}, \& {Toomey}}]{biller:2010}
{Biller}, B.~A., {Liu}, M.~C., {Wahhaj}, Z., {et~al.} 2010, \apjl, 720, L82

\bibitem[{{Biller} {et~al.}(2013){Biller}, {Liu}, {Wahhaj}, {Nielsen},
  {Hayward}, {Males}, {Skemer}, {Close}, {Chun}, {Ftaclas}, {Clarke}, {Thatte},
  {Shkolnik}, {Reid}, {Hartung}, {Boss}, {Lin}, {Alencar}, {de Gouveia Dal
  Pino}, {Gregorio-Hetem}, \& {Toomey}}]{biller13}
---. 2013, \apj, 777, 160

\bibitem[{{Binks} \& {Jeffries}(2014)}]{binks:2014}
{Binks}, A.~S., \& {Jeffries}, R.~D. 2014, \mnras, 438, L11

\bibitem[{{Birnstiel} {et~al.}(2012){Birnstiel}, {Klahr}, \&
  {Ercolano}}]{Birnstiel:2012}
{Birnstiel}, T., {Klahr}, H., \& {Ercolano}, B. 2012, \aap, 539, A148

\bibitem[{{Bitsch} {et~al.}(2018){Bitsch}, {Morbidelli}, {Johansen}, {Lega},
  {Lambrechts}, \& {Crida}}]{Bitsch:2018}
{Bitsch}, B., {Morbidelli}, A., {Johansen}, A., {et~al.} 2018, \aap, 612, A30

\bibitem[{{Bonfils} {et~al.}(2013){Bonfils}, {Delfosse}, {Udry}, {Forveille},
  {Mayor}, {Perrier}, {Bouchy}, {Gillon}, {Lovis}, {Pepe}, {Queloz}, {Santos},
  {S{\'e}gransan}, \& {Bertaux}}]{bonfils:2013}
{Bonfils}, X., {Delfosse}, X., {Udry}, S., {et~al.} 2013, \aap, 549, A109

\bibitem[{Booth {et~al.}(2016)Booth, Jord{\'a}n, Casassus, Hales, Dent,
  Faramaz, Matr{\`a}, Barkats, Brahm, \& Cuadra}]{Booth:2016iy}
Booth, M., Jord{\'a}n, A., Casassus, S., {et~al.} 2016, MNRAS, 460, L10

\bibitem[{{Bowler}(2016)}]{bowler:2016}
{Bowler}, B.~P. 2016, \pasp, 128, 102001

\bibitem[{{Bowler} {et~al.}(2015){Bowler}, {Liu}, {Shkolnik}, \&
  {Tamura}}]{bowler15}
{Bowler}, B.~P., {Liu}, M.~C., {Shkolnik}, E.~L., \& {Tamura}, M. 2015, \apjs,
  216, 7

\bibitem[{{Brandt} {et~al.}(2014){Brandt}, {McElwain}, {Turner}, {Mede},
  {Spiegel}, {Kuzuhara}, {Schlieder}, {Wisniewski}, {Abe}, {Biller},
  {Brandner}, {Carson}, {Currie}, {Egner}, {Feldt}, {Golota}, {Goto}, {Grady},
  {Guyon}, {Hashimoto}, {Hayano}, {Hayashi}, {Hayashi}, {Henning}, {Hodapp},
  {Inutsuka}, {Ishii}, {Iye}, {Janson}, {Kandori}, {Knapp}, {Kudo}, {Kusakabe},
  {Kwon}, {Matsuo}, {Miyama}, {Morino}, {Moro-Mart{\'{\i}}n}, {Nishimura},
  {Pyo}, {Serabyn}, {Suto}, {Suzuki}, {Takami}, {Takato}, {Terada}, {Thalmann},
  {Tomono}, {Watanabe}, {Yamada}, {Takami}, {Usuda}, \& {Tamura}}]{brandt14}
{Brandt}, T.~D., {McElwain}, M.~W., {Turner}, E.~L., {et~al.} 2014, \apj, 794,
  159

\bibitem[{{Bryan} {et~al.}(2016){Bryan}, {Knutson}, {Howard}, {Ngo}, {Batygin},
  {Crepp}, {Fulton}, {Hinkley}, {Isaacson}, {Johnson}, {Marcy}, \&
  {Wright}}]{bryan:2016}
{Bryan}, M.~L., {Knutson}, H.~A., {Howard}, A.~W., {et~al.} 2016, \apj, 821, 89

\bibitem[{Caffau {et~al.}(2011)Caffau, Ludwig, Steffen, Freytag, \&
  Bonifacio}]{Caffau:2011ik}
Caffau, E., Ludwig, H.~G., Steffen, M., Freytag, B., \& Bonifacio, P. 2011,
  Solar Physics, 268, 255

\bibitem[{{Campante} {et~al.}(2017){Campante}, {Veras}, {North}, {Miglio},
  {Morel}, {Johnson}, {Chaplin}, {Davies}, {Huber}, {Kuszlewicz}, {Lund},
  {Cooke}, {Elsworth}, {Rodrigues}, \& {Vanderburg}}]{campante:2017}
{Campante}, T.~L., {Veras}, D., {North}, T.~S.~H., {et~al.} 2017, \mnras, 469,
  1360

\bibitem[{{Cannon} \& {Pickering}(1993)}]{1993yCat.3135....0C}
{Cannon}, A.~J., \& {Pickering}, E.~C. 1993, VizieR Online Data Catalog, 3135

\bibitem[{Casagrande {et~al.}(2011)Casagrande, Sch{\"o}nrich, Asplund, Cassisi,
  Ram{\'\i}rez, Mel{\'e}ndez, Bensby, \& Feltzing}]{Casagrande:2011ji}
Casagrande, L., Sch{\"o}nrich, R., Asplund, M., {et~al.} 2011, A{\&}A, 530,
  A138

\bibitem[{Castelli \& Kurucz(2004)}]{Castelli:2004ti}
Castelli, F., \& Kurucz, R.~L. 2004, eprint arXiv:astro-ph/0405087

\bibitem[{Castro {et~al.}(2016)Castro, Duarte, Pace, \&
  do~Nascimento~Jr.}]{Castro:2016gz}
Castro, M., Duarte, T., Pace, G., \& do~Nascimento~Jr., J.~D. 2016, A{\&}A,
  590, A94

\bibitem[{Chauvin {et~al.}(2004)Chauvin, Lagrange, Dumas, Zuckerman, Mouillet,
  Song, Beuzit, \& Lowrance}]{Chauvin:2004cy}
Chauvin, G., Lagrange, A.-M., Dumas, C., {et~al.} 2004, A{\&}A, 425, L29

\bibitem[{Chauvin {et~al.}(2017)Chauvin, Desidera, Lagrange, Vigan, Gratton,
  Langlois, Bonnefoy, Beuzit, Feldt, Mouillet, Meyer, Cheetham, Biller,
  Boccaletti, D'Orazi, Galicher, Hagelberg, Maire, Mesa, Olofsson, Samland,
  Schmidt, Sissa, Bonavita, Charnay, Cudel, Daemgen, Delorme, Janin-Potiron,
  Janson, Keppler, Le~Coroller, Ligi, Marleau, Messina, Molli{\`e}re,
  Mordasini, M{\"u}ller, Peretti, Perrot, Rodet, Rouan, Zurlo, Dominik,
  Henning, Menard, Schmid, Turatto, Udry, Vakili, Abe, Antichi, Baruffolo,
  Baudoz, Baudrand, Blanchard, Bazzon, Buey, Carbillet, Carle, Charton,
  Cascone, Claudi, Costille, Deboulbe, De~Caprio, Dohlen, Fantinel, Feautrier,
  Fusco, Gigan, Giro, Gisler, Gluck, Hubin, Hugot, Jaquet, Kasper, Madec,
  Magnard, Martinez, Maurel, Le~Mignant, M{\"o}ller-Nilsson, Llored, Moulin,
  Orign{\'e}, Pavlov, Perret, Petit, Pragt, Puget, Rabou, Ramos, Rigal, Rochat,
  Roelfsema, Rousset, Roux, Salasnich, Sauvage, Sevin, Soenke, Stadler, Suarez,
  Weber, Wildi, Antoniucci, Augereau, Baudino, Brandner, Engler, Girard, Gry,
  Kral, Kopytova, Lagadec, Milli, Moutou, Schlieder, Szul{\'a}gyi, Thalmann, \&
  Wahhaj}]{Chauvin:2017ev}
Chauvin, G., Desidera, S., Lagrange, A.-M., {et~al.} 2017, A{\&}A, 605, L9

\bibitem[{Chauvin {et~al.}(2018)Chauvin, Gratton, Bonnefoy, Lagrange, de~Boer,
  Vigan, Beust, Lazzoni, Boccaletti, Galicher, Desidera, Delorme, Keppler,
  Lannier, Maire, Mesa, Meunier, Kral, Henning, Menard, Mo{\'o}r, Avenhaus,
  Bazzon, Janson, Beuzit, Bhowmik, Bonavita, Borgniet, Brandner, Cheetham,
  Cudel, Feldt, Fontanive, Ginski, Hagelberg, Janin-Potiron, Lagadec, Langlois,
  Le~Coroller, Messina, Meyer, Mouillet, Peretti, Perrot, Rodet, Samland,
  Sissa, Olofsson, Salter, Schmidt, Zurlo, Milli, van Boekel, Quanz, Wilson,
  Feautrier, Le~Mignant, Perret, Ramos, \& Rochat}]{Chauvin:2018vm}
Chauvin, G., Gratton, R., Bonnefoy, M., {et~al.} 2018, eprint arXiv:1801.05850,
  1801.05850

\bibitem[{{Cheetham} {et~al.}(2015){Cheetham}, {Kraus}, {Ireland}, {Cieza},
  {Rizzuto}, \& {Tuthill}}]{cheetham15}
{Cheetham}, A.~C., {Kraus}, A.~L., {Ireland}, M.~J., {et~al.} 2015, \apj, 813,
  83

\bibitem[{Chilcote {et~al.}(2017)Chilcote, Pueyo, De~Rosa, Vargas, Macintosh,
  Bailey, Barman, Bauman, Bruzzone, Bulger, Burrows, Cardwell, Chen, Cotten,
  Dillon, Doyon, Draper, Duch{\^e}ne, Dunn, Erikson, Fitzgerald, Follette,
  Gavel, Goodsell, Graham, Greenbaum, Hartung, Hibon, Hung, Ingraham, Kalas,
  Konopacky, Larkin, Maire, Marchis, Marley, Marois, Metchev, Millar-Blanchaer,
  Morzinski, Nielsen, Norton, Oppenheimer, Palmer, Patience, Perrin, Poyneer,
  Rajan, Rameau, Rantakyr{\"o}, Sadakuni, Saddlemyer, Savransky, Schneider,
  Serio, Sivaramakrishnan, Song, Soummer, Thomas, Wallace, Wang, Ward-Duong,
  Wiktorowicz, \& Wolff}]{Chilcote:2017fv}
Chilcote, J., Pueyo, L., De~Rosa, R.~J., {et~al.} 2017, AJ, 153, 182

\bibitem[{{Chilcote} {et~al.}(2012){Chilcote}, {Larkin}, {Maire}, {Perrin},
  {Fitzgerald}, {Doyon}, {Thibault}, {Bauman}, {Macintosh}, {Graham}, \&
  {Saddlemyer}}]{chilcote:2012}
{Chilcote}, J.~K., {Larkin}, J.~E., {Maire}, J., {et~al.} 2012, in \procspie,
  Vol. 8446, Ground-based and Airborne Instrumentation for Astronomy IV, 84468W

\bibitem[{{Chilcote} {et~al.}(2018){Chilcote}, {Bailey}, {De Rosa},
  {Macintosh}, {Nielsen}, {Norton}, {Millar-Blanchaer}, {Graham}, {Marois},
  {Pueyo}, {Rameau}, {Savransky}, \& {Veran}}]{chilcote18}
{Chilcote}, J.~K., {Bailey}, V.~P., {De Rosa}, R., {et~al.} 2018, ArXiv
  e-prints, arXiv:1807.07145

\bibitem[{Choi {et~al.}(2016)Choi, Dotter, Conroy, Cantiello, Paxton, \&
  Johnson}]{Choi:2016kf}
Choi, J., Dotter, A., Conroy, C., {et~al.} 2016, Astrophys. J., 823, 102

\bibitem[{{Clanton} \& {Gaudi}(2014)}]{Clanton:2014}
{Clanton}, C., \& {Gaudi}, B.~S. 2014, \apj, 791, 91

\bibitem[{{Corbally}(1984)}]{1984ApJS...55..657C}
{Corbally}, C.~J. 1984, \apjs, 55, 657

\bibitem[{{Cumming} {et~al.}(2008){Cumming}, {Butler}, {Marcy}, {Vogt},
  {Wright}, \& {Fischer}}]{cumming2008}
{Cumming}, A., {Butler}, R.~P., {Marcy}, G.~W., {et~al.} 2008, \pasp, 120, 531

\bibitem[{{David} \& {Hillenbrand}(2015)}]{david:2015}
{David}, T.~J., \& {Hillenbrand}, L.~A. 2015, \apj, 804, 146

\bibitem[{{Davison} {et~al.}(2015){Davison}, {White}, {Henry}, {Riedel}, {Jao},
  {Bailey}, {Quinn}, {Cantrell}, {Subasavage}, \&
  {Winters}}]{2015AJ....149..106D}
{Davison}, C.~L., {White}, R.~J., {Henry}, T.~J., {et~al.} 2015, \aj, 149, 106

\bibitem[{{Dawson} \& {Johnson}(2018)}]{Dawson:2018}
{Dawson}, R.~I., \& {Johnson}, J.~A. 2018, \araa, 56, 175

\bibitem[{{De Rosa} {et~al.}(2014){De Rosa}, {Patience}, {Ward-Duong}, {Vigan},
  {Marois}, {Song}, {Macintosh}, {Graham}, {Doyon}, {Bessell}, {Lai},
  {McCarthy}, \& {Kulesa}}]{DeRosa:2014}
{De Rosa}, R.~J., {Patience}, J., {Ward-Duong}, K., {et~al.} 2014, \mnras, 445,
  3694

\bibitem[{De~Rosa {et~al.}(2015)De~Rosa, Nielsen, Blunt, Graham, Konopacky,
  Marois, Pueyo, Rameau, Ryan, Wang, Bailey, Chontos, Fabrycky, Follette,
  Macintosh, Marchis, Ammons, Arriaga, Chilcote, Cotten, Doyon, Duch{\^e}ne,
  Esposito, Fitzgerald, Gerard, Goodsell, Greenbaum, Hibon, Ingraham,
  Johnson-Groh, Kalas, Lafreni{\`e}re, Maire, Metchev, Millar-Blanchaer,
  Morzinski, Oppenheimer, Patel, Patience, Perrin, Rajan, Rantakyr{\"o},
  Ruffio, Schneider, Sivaramakrishnan, Song, Tran, Vasisht, Ward-Duong, \&
  Wolff}]{DeRosa:2015jl}
De~Rosa, R.~J., Nielsen, E.~L., Blunt, S.~C., {et~al.} 2015, ApJL, 814, L3

\bibitem[{De~Rosa {et~al.}(2016)De~Rosa, Rameau, Patience, Graham, Doyon,
  Lafreni{\`e}re, Macintosh, Pueyo, Rajan, Wang, Ward-Duong, Hung, Maire,
  Nielsen, Ammons, Bulger, Cardwell, Chilcote, Galvez, Gerard, Goodsell,
  Hartung, Hibon, Ingraham, Johnson-Groh, Kalas, Konopacky, Marchis, Marois,
  Metchev, Morzinski, Oppenheimer, Perrin, Rantakyr{\"o}, Savransky, \&
  Thomas}]{DeRosa:2016kh}
De~Rosa, R.~J., Rameau, J., Patience, J., {et~al.} 2016, ApJ, 824, 121

\bibitem[{de~Zeeuw {et~al.}(1999)de~Zeeuw, Hoogerwerf, de~Bruijne, Brown, \&
  Blaauw}]{deZeeuw:1999fe}
de~Zeeuw, P.~T., Hoogerwerf, R., de~Bruijne, J. H.~J., Brown, A. G.~A., \&
  Blaauw, A. 1999, AJ, 117, 354

\bibitem[{{Dong} \& {Zhu}(2013)}]{dong:2013}
{Dong}, S., \& {Zhu}, Z. 2013, \apj, 778, 53

\bibitem[{Dotter(2016)}]{Dotter:2016fa}
Dotter, A. 2016, ApJS, 222, 8

\bibitem[{{Dupuy} {et~al.}(2019){Dupuy}, {Brandt}, {Kratter}, \&
  {Bowler}}]{dupuy:2019}
{Dupuy}, T.~J., {Brandt}, T.~D., {Kratter}, K.~M., \& {Bowler}, B.~P. 2019,
  \apjl, 871, L4

\bibitem[{Espinosa~Lara \& Rieutord(2011)}]{EspinosaLara:2011dl}
Espinosa~Lara, F., \& Rieutord, M. 2011, A{\&}A, 533, A43

\bibitem[{Evans {et~al.}(2018)Evans, Riello, Angeli, Carrasco, Montegriffo,
  Fabricius, Jordi, Palaversa, Diener, Busso, Cacciari, Leeuwen, Burgess,
  Davidson, Harrison, Hodgkin, Pancino, Richards, Altavilla,
  Balaguer-N{\'u}{\~n}ez, Barstow, Bellazzini, Brown, Castellani, Cocozza,
  Luise, Delgado, Ducourant, Galleti, Gilmore, Giuffrida, Holl, Kewley,
  Koposov, Marinoni, Marrese, Osborne, Piersimoni, Portell, Pulone, Ragaini,
  Sanna, Terrett, Walton, Wevers, \& Wyrzykowski}]{Evans:2018cj}
Evans, D.~W., Riello, M., Angeli, F.~D., {et~al.} 2018, A{\&}A, 616, A4

\bibitem[{{Feigelson} {et~al.}(2006){Feigelson}, {Lawson}, {Stark}, {Townsley},
  \& {Garmire}}]{feigelson:2006}
{Feigelson}, E.~D., {Lawson}, W.~A., {Stark}, M., {Townsley}, L., \& {Garmire},
  G.~P. 2006, \aj, 131, 1730

\bibitem[{{Fekel}(1997)}]{1997AJ....114.2747F}
{Fekel}, F.~C. 1997, \aj, 114, 2747

\bibitem[{{Fekel} {et~al.}(2017){Fekel}, {Henry}, \&
  {Tomkin}}]{2017AJ....154..120F}
{Fekel}, F.~C., {Henry}, G.~W., \& {Tomkin}, J. 2017, \aj, 154, 120

\bibitem[{{Feldt} {et~al.}(2017){Feldt}, {Olofsson}, {Boccaletti}, {Maire},
  {Milli}, {Vigan}, {Langlois}, {Henning}, {Moor}, {Bonnefoy}, {Wahhaj},
  {Desidera}, {Gratton}, {K{\'o}sp{\'a}l}, {Abraham}, {Menard}, {Chauvin},
  {Lagrange}, {Mesa}, {Salter}, {Buenzli}, {Lannier}, {Perrot}, {Peretti}, \&
  {Sissa}}]{feldt:2017}
{Feldt}, M., {Olofsson}, J., {Boccaletti}, A., {et~al.} 2017, \aap, 601, A7

\bibitem[{{Fernandes} {et~al.}(2018){Fernandes}, {Mulders}, {Pascucci},
  {Mordasini}, \& {Emsenhuber}}]{fernandes18}
{Fernandes}, R.~B., {Mulders}, G.~D., {Pascucci}, I., {Mordasini}, C., \&
  {Emsenhuber}, A. 2018, arXiv e-prints, arXiv:1812.05569

\bibitem[{{Fischer} {et~al.}(2014){Fischer}, {Marcy}, \&
  {Spronck}}]{fischer:2014}
{Fischer}, D.~A., {Marcy}, G.~W., \& {Spronck}, J.~F.~P. 2014, \apjs, 210, 5

\bibitem[{{Fischer} \& {Valenti}(2005)}]{fischer:2005}
{Fischer}, D.~A., \& {Valenti}, J. 2005, \apj, 622, 1102

\bibitem[{Foreman-Mackey {et~al.}(2013)Foreman-Mackey, Hogg, Lang, \&
  Goodman}]{ForemanMackey:2013io}
Foreman-Mackey, D., Hogg, D.~W., Lang, D., \& Goodman, J. 2013, PASP, 125, 306

\bibitem[{{Forgan} \& {Rice}(2013)}]{Forgan:2013}
{Forgan}, D., \& {Rice}, K. 2013, \mnras, 432, 3168

\bibitem[{{Fortney} {et~al.}(2008){Fortney}, {Marley}, {Saumon}, \&
  {Lodders}}]{fortney08}
{Fortney}, J.~J., {Marley}, M.~S., {Saumon}, D., \& {Lodders}, K. 2008, \apj,
  683, 1104

\bibitem[{{Fung} \& {Lee}(2018)}]{fung:2018}
{Fung}, J., \& {Lee}, E.~J. 2018, \apj, 859, 126

\bibitem[{{Gaia Collaboration} {et~al.}(2018){Gaia Collaboration}, Brown,
  Vallenari, Prusti, de~Bruijne, Babusiaux, Bailer-Jones, Biermann, Evans,
  Eyer, Jansen, Jordi, Klioner, Lammers, Lindegren, Luri, Mignard, Panem,
  Pourbaix, Randich, Sartoretti, Siddiqui, Soubiran, van Leeuwen, Walton,
  Arenou, Bastian, Cropper, Drimmel, Katz, Lattanzi, Bakker, Cacciari,
  Casta{\~n}eda, Chaoul, Cheek, De~Angeli, Fabricius, Guerra, Holl, Masana,
  Messineo, Mowlavi, Nienartowicz, Panuzzo, Portell, Riello, Seabroke, Tanga,
  Th{\'e}venin, Gracia-Abril, Comoretto, Garcia-Reinaldos, Teyssier, Altmann,
  Andrae, Audard, Bellas-Velidis, Benson, Berthier, Blomme, Burgess, Busso,
  Carry, Cellino, Clementini, Clotet, Creevey, Davidson, De~Ridder, Delchambre,
  Dell{\textquoteright}Oro, Ducourant, Fern{\'a}ndez-Hern{\'a}ndez, Fouesneau,
  Fr{\'e}mat, Galluccio, Garc{\'\i}a-Torres, Gonz{\'a}lez-N{\'u}{\~n}ez,
  Gonz{\'a}lez-Vidal, Gosset, Guy, Halbwachs, Hambly, Harrison, Hern{\'a}ndez,
  Hestroffer, Hodgkin, Hutton, Jasniewicz, Jean-Antoine-Piccolo, Jordan, Korn,
  Krone-Martins, Lanzafame, Lebzelter, L{\"o}ffler, Manteiga, Marrese,
  Mart{\'\i}n-Fleitas, Moitinho, Mora, Muinonen, Osinde, Pancino, Pauwels,
  Petit, Recio-Blanco, Richards, Rimoldini, Robin, Sarro, Siopis, Smith,
  Sozzetti, S{\"u}veges, Torra, van Reeven, Abbas, Abreu~Aramburu, Accart,
  Aerts, Altavilla, {\'A}lvarez, Alvarez, Alves, Anderson, Andrei,
  Anglada~Varela, Antiche, Antoja, Arcay, Astraatmadja, Bach, Baker,
  Balaguer-N{\'u}{\~n}ez, Balm, Barache, Barata, Barbato, Barblan, Barklem,
  Barrado, Barros, Barstow, Bartholom{\'e}~Mu{\~n}oz, Bassilana, Becciani,
  Bellazzini, Berihuete, Bertone, Bianchi, Bienaym{\'e}, Blanco-Cuaresma, Boch,
  Boeche, Bombrun, Borrachero, Bossini, Bouquillon, Bourda, Bragaglia,
  Bramante, Breddels, Bressan, Brouillet, Br{\"u}semeister, Brugaletta,
  Bucciarelli, Burlacu, Busonero, Butkevich, Buzzi, Caffau, Cancelliere,
  Cannizzaro, Cantat-Gaudin, Carballo, Carlucci, Carrasco, Casamiquela,
  Castellani, Castro-Ginard, Charlot, Chemin, Chiavassa, Cocozza, Costigan,
  Cowell, Crifo, Crosta, Crowley, Cuypers, Dafonte, Damerdji, Dapergolas,
  David, David, de~Laverny, De~Luise, De~March, de~Martino, de~Souza,
  de~Torres, Debosscher, del Pozo, Delbo, Delgado, Delgado, Di~Matteo, Diakite,
  Diener, Distefano, Dolding, Drazinos, Dur{\'a}n, Edvardsson, Enke, Eriksson,
  Esquej, Eynard~Bontemps, Fabre, Fabrizio, Faigler, Falc{\~a}o,
  Farr{\`a}s~Casas, Federici, Fedorets, Fernique, Figueras, Filippi, Findeisen,
  Fonti, Fraile, Fraser, Fr{\'e}zouls, Gai, Galleti, Garabato,
  Garc{\'\i}a-Sedano, Garofalo, Garralda, Gavel, Gavras, Gerssen, Geyer,
  Giacobbe, Gilmore, Girona, Giuffrida, Glass, Gomes, Granvik, Gueguen,
  Guerrier, Guiraud, Guti{\'e}rrez-S{\'a}nchez, Haigron, Hatzidimitriou,
  Hauser, Haywood, Heiter, Helmi, Heu, Hilger, Hobbs, Hofmann, Holland, Huckle,
  Hypki, Icardi, Jan{\ss}en, Jevardat~de Fombelle, Jonker, Juh{\'a}sz, Julbe,
  Karampelas, Kewley, Klar, Kochoska, Kohley, Kolenberg, Kontizas, Kontizas,
  Koposov, \& Kordopatis}]{GaiaCollaboration:2018io}
{Gaia Collaboration}, Brown, A. G.~A., Vallenari, A., {et~al.} 2018, A{\&}A,
  616, A1

\bibitem[{Galicher {et~al.}(2016)Galicher, Marois, Macintosh, Zuckerman,
  Barman, Konopacky, Song, Patience, Lafreni{\`e}re, Doyon, \&
  Nielsen}]{Galicher:2016hg}
Galicher, R., Marois, C., Macintosh, B., {et~al.} 2016, A{\&}A, 594, A63

\bibitem[{{Galicher} {et~al.}(2016){Galicher}, {Marois}, {Macintosh},
  {Zuckerman}, {Barman}, {Konopacky}, {Song}, {Patience}, {Lafreni{\`e}re},
  {Doyon}, \& {Nielsen}}]{galicher:2016}
{Galicher}, R., {Marois}, C., {Macintosh}, B., {et~al.} 2016, \aap, 594, A63

\bibitem[{Gallet \& Bouvier(2013)}]{Gallet:2013gy}
Gallet, F., \& Bouvier, J. 2013, A{\&}A, 556, A36

\bibitem[{{Gammie}(2001)}]{Gammie:2001}
{Gammie}, C.~F. 2001, \apj, 553, 174

\bibitem[{{Garrison} \& {Gray}(1994)}]{1994AJ....107.1556G}
{Garrison}, R.~F., \& {Gray}, R.~O. 1994, \aj, 107, 1556

\bibitem[{{Ghezzi} {et~al.}(2018){Ghezzi}, {Montet}, \&
  {Johnson}}]{ghezzi:2018}
{Ghezzi}, L., {Montet}, B.~T., \& {Johnson}, J.~A. 2018, \apj, 860, 109

\bibitem[{{Gizis} {et~al.}(2001){Gizis}, {Kirkpatrick}, {Burgasser}, {Reid},
  {Monet}, {Liebert}, \& {Wilson}}]{gizis01}
{Gizis}, J.~E., {Kirkpatrick}, J.~D., {Burgasser}, A., {et~al.} 2001, \apjl,
  551, L163

\bibitem[{{Goldreich} {et~al.}(2004){Goldreich}, {Lithwick}, \&
  {Sari}}]{Goldreich:2004}
{Goldreich}, P., {Lithwick}, Y., \& {Sari}, R. 2004, \araa, 42, 549

\bibitem[{{Goldreich} \& {Lynden-Bell}(1965)}]{goldreich:1965}
{Goldreich}, P., \& {Lynden-Bell}, D. 1965, \mnras, 130, 97

\bibitem[{{Gonzalez}(1997)}]{gonzalez:1997}
{Gonzalez}, G. 1997, \mnras, 285, 403

\bibitem[{{Gorynya} \& {Tokovinin}(2018)}]{2018MNRAS.475.1375G}
{Gorynya}, N.~A., \& {Tokovinin}, A. 2018, \mnras, 475, 1375

\bibitem[{{Gray} {et~al.}(2006){Gray}, {Corbally}, {Garrison}, {McFadden},
  {Bubar}, {McGahee}, {O'Donoghue}, \& {Knox}}]{2006AJ....132..161G}
{Gray}, R.~O., {Corbally}, C.~J., {Garrison}, R.~F., {et~al.} 2006, \aj, 132,
  161

\bibitem[{Gray {et~al.}(2003)Gray, Corbally, Garrison, McFadden, \&
  Robinson}]{Gray:2003fz}
Gray, R.~O., Corbally, C.~J., Garrison, R.~F., McFadden, M.~T., \& Robinson,
  P.~E. 2003, AJ, 126, 2048

\bibitem[{{Gray} {et~al.}(2003){Gray}, {Corbally}, {Garrison}, {McFadden}, \&
  {Robinson}}]{2003AJ....126.2048G}
{Gray}, R.~O., {Corbally}, C.~J., {Garrison}, R.~F., {McFadden}, M.~T., \&
  {Robinson}, P.~E. 2003, \aj, 126, 2048

\bibitem[{{Gray} \& {Garrison}(1987)}]{1987ApJS...65..581G}
{Gray}, R.~O., \& {Garrison}, R.~F. 1987, \apjs, 65, 581

\bibitem[{{Gray} \& {Garrison}(1989)}]{1989ApJS...70..623G}
---. 1989, \apjs, 70, 623

\bibitem[{{Gray} \& {Kaye}(1999)}]{gray:1999}
{Gray}, R.~O., \& {Kaye}, A.~B. 1999, \aj, 118, 2993

\bibitem[{Greenbaum {et~al.}(2018)Greenbaum, Pueyo, Ruffio, Wang, Rosa,
  Aguilar, Rameau, Barman, Marois, Marley, Konopacky, Rajan, Macintosh,
  Ansdell, Arriaga, Bailey, Bulger, Burrows, Chilcote, Cotten, Doyon,
  Duch{\^e}ne, Fitzgerald, Follette, Gerard, Goodsell, Graham, Hibon, Hung,
  Ingraham, Kalas, Larkin, Maire, Marchis, Metchev, Millar-Blanchaer, Nielsen,
  Norton, Oppenheimer, Palmer, Patience, Perrin, Poyneer, Rantakyr{\"o},
  Savransky, Schneider, Sivaramakrishnan, Song, Soummer, Thomas, Wallace,
  Ward-Duong, Wiktorowicz, \& Wolff}]{Greenbaum:2018hz}
Greenbaum, A.~Z., Pueyo, L., Ruffio, J.-B., {et~al.} 2018, VizieR Online Data
  Catalog, 155, 226

\bibitem[{{Grether} \& {Lineweaver}(2006)}]{grether06}
{Grether}, D., \& {Lineweaver}, C.~H. 2006, \apj, 640, 1051

\bibitem[{{Harlan}(1974)}]{1974AJ.....79..682H}
{Harlan}, E.~A. 1974, \aj, 79, 682

\bibitem[{{Harlan} \& {Taylor}(1970)}]{1970AJ.....75..507H}
{Harlan}, E.~A., \& {Taylor}, D.~C. 1970, \aj, 75, 507

\bibitem[{{Harper}(1935)}]{1935PDAO....6..207H}
{Harper}, W.~H. 1935, Publications of the Dominion Astrophysical Observatory
  Victoria, 6, 207

\bibitem[{{Harris}(1978)}]{Harris:1978}
{Harris}, A.~W. 1978, in Lunar and Planetary Science Conference, Vol.~9, Lunar
  and Planetary Science Conference, 459--461

\bibitem[{{Heinze} {et~al.}(2010){Heinze}, {Hinz}, {Sivanandam}, {Kenworthy},
  {Meyer}, \& {Miller}}]{heinze:2010}
{Heinze}, A.~N., {Hinz}, P.~M., {Sivanandam}, S., {et~al.} 2010, \apj, 714,
  1551

\bibitem[{{Hoffleit} {et~al.}(1970){Hoffleit}, {Eckert}, {L{\"u}}, \&
  {Paranya}}]{1970TOYal..30....1H}
{Hoffleit}, D., {Eckert}, D., {L{\"u}}, P., \& {Paranya}, K. 1970, Transactions
  of the Astronomical Observatory of Yale University, 30, 1

\bibitem[{H{\o}g {et~al.}(2000)H{\o}g, Fabricius, Makarov, Urban, Corbin,
  Wycoff, Bastian, Schwekendiek, \& Wicenec}]{Hog:2000wk}
H{\o}g, E., Fabricius, C., Makarov, V.~V., {et~al.} 2000, A{\&}A, 355, L27

\bibitem[{{Hopkins}(2013)}]{Hopkins:2013}
{Hopkins}, P.~F. 2013, \mnras, 430, 1653

\bibitem[{{Houk}(1978)}]{1978mcts.book.....H}
{Houk}, N. 1978, {Michigan catalogue of two-dimensional spectral types for the
  HD stars}

\bibitem[{Houk(1982)}]{Houk:1982vl}
Houk, N. 1982, Michigan Catalogue of Two-dimensional Spectral Types for the HD
  stars. Volume 3. Declinations -40 to -26., by Houk, N.. Ann Arbor, MI(USA):
  Department of Astronomy, University of Michigan, 12 + 390 p.

\bibitem[{{Houk} \& {Cowley}(1975)}]{1975mcts.book.....H}
{Houk}, N., \& {Cowley}, A.~P. 1975, {University of Michigan Catalogue of
  two-dimensional spectral types for the HD stars. Volume I. Declinations -90
  to -53.}

\bibitem[{Houk \& Smith-Moore(1988)}]{Houk:1988wv}
Houk, N., \& Smith-Moore, M. 1988, Michigan Catalogue of Two-dimensional
  Spectral Types for the HD Stars. Volume 4, Declinations -26.0 to -12.0.. N.
  Houk, M. Smith-Moore.Department of Astronomy, University of Michigan, Ann
  Arbor, MI 48109-1090, USA. 14+505 pp. Price US 25.00 (USA, Canada), US 28.00
  (Foreign) (1988).

\bibitem[{{Houk} \& {Swift}(1999)}]{1999MSS...C05....0H}
{Houk}, N., \& {Swift}, C. 1999, in Michigan Spectral Survey, Ann Arbor, Dep.
  Astron., Univ. Michigan, Vol. 5, p. 0 (1999), Vol.~5, 0

\bibitem[{{Hu{\'e}lamo} {et~al.}(2010){Hu{\'e}lamo}, {N{\"u}rnberger},
  {Ivanov}, {Chauvin}, {Carraro}, {Sterzik}, {Melo}, {Bonnefoy}, {Hartung},
  {Haubois}, \& {Foellmi}}]{huelamo:2010}
{Hu{\'e}lamo}, N., {N{\"u}rnberger}, D.~E.~A., {Ivanov}, V.~D., {et~al.} 2010,
  \aap, 521, L54

\bibitem[{Ingraham {et~al.}(2014)Ingraham, Marley, Saumon, Marois, Macintosh,
  Barman, Bauman, Burrows, Chilcote, De~Rosa, Dillon, Doyon, Dunn, Erikson,
  Fitzgerald, Gavel, Goodsell, Graham, Hartung, Hibon, Kalas, Konopacky,
  Larkin, Maire, Marchis, McBride, Millar-Blanchaer, Morzinski, Norton,
  Oppenheimer, Palmer, Patience, Perrin, Poyneer, Pueyo, Rantakyro, Sadakuni,
  Saddlemyer, Savransky, Soummer, Sivaramakrishnan, Song, Thomas, Wallace,
  Wiktorowicz, \& Wolff}]{Ingraham:2014gx}
Ingraham, P., Marley, M.~S., Saumon, D., {et~al.} 2014, ApJ, 794, L15

\bibitem[{{Johansen} \& {Lambrechts}(2017)}]{Johansen:2017}
{Johansen}, A., \& {Lambrechts}, M. 2017, Annual Review of Earth and Planetary
  Sciences, 45, 359

\bibitem[{{Johnson} {et~al.}(2010{\natexlab{a}}){Johnson}, {Aller}, {Howard},
  \& {Crepp}}]{johnson:2010}
{Johnson}, J.~A., {Aller}, K.~M., {Howard}, A.~W., \& {Crepp}, J.~R.
  2010{\natexlab{a}}, \pasp, 122, 905

\bibitem[{{Johnson} {et~al.}(2007){Johnson}, {Butler}, {Marcy}, {Fischer},
  {Vogt}, {Wright}, \& {Peek}}]{johnson:2007}
{Johnson}, J.~A., {Butler}, R.~P., {Marcy}, G.~W., {et~al.} 2007, \apj, 670,
  833

\bibitem[{{Johnson} \& {Wright}(2013)}]{johnson:2013}
{Johnson}, J.~A., \& {Wright}, J.~T. 2013, ArXiv e-prints, arXiv:1307.3441

\bibitem[{{Johnson} {et~al.}(2010{\natexlab{b}}){Johnson}, {Bowler}, {Howard},
  {Henry}, {Marcy}, {Isaacson}, {Brewer}, {Fischer}, {Morton}, \&
  {Crepp}}]{johnson:2010b}
{Johnson}, J.~A., {Bowler}, B.~P., {Howard}, A.~W., {et~al.}
  2010{\natexlab{b}}, \apjl, 721, L153

\bibitem[{{Johnson} {et~al.}(2014){Johnson}, {Huber}, {Boyajian}, {Brewer},
  {White}, {von Braun}, {Maestro}, {Stello}, \& {Barclay}}]{johnson:2014}
{Johnson}, J.~A., {Huber}, D., {Boyajian}, T., {et~al.} 2014, \apj, 794, 15

\bibitem[{Johnson-Groh {et~al.}(2017)Johnson-Groh, Marois, De~Rosa, Nielsen,
  Rameau, Blunt, Vargas, Ammons, Bailey, Barman, Bulger, Chilcote, Cotten,
  Doyon, Duch{\^e}ne, Fitzgerald, Follette, Goodsell, Graham, Greenbaum, Hibon,
  Hung, Ingraham, Kalas, Konopacky, Larkin, Macintosh, Maire, Marchis, Marley,
  Metchev, Millar-Blanchaer, Oppenheimer, Palmer, Patience, Perrin, Poyneer,
  Pueyo, Rajan, Rantakyr{\"o}, Savransky, Schneider, Sivaramakrishnan, Song,
  Soummer, Thomas, Vega, Wallace, Wang, Ward-Duong, Wiktorowicz, \&
  Wolff}]{JohnsonGroh:2017kh}
Johnson-Groh, M., Marois, C., De~Rosa, R.~J., {et~al.} 2017, VizieR Online Data
  Catalog, 153, 190

\bibitem[{{Jones} {et~al.}(2016){Jones}, {Jenkins}, {Brahm}, {Wittenmyer},
  {Olivares E.}, {Melo}, {Rojo}, {Jord{\'a}n}, {Drass}, {Butler}, \&
  {Wang}}]{jones:2016}
{Jones}, M.~I., {Jenkins}, J.~S., {Brahm}, R., {et~al.} 2016, \aap, 590, A38

\bibitem[{{Joy} \& {Abt}(1974)}]{1974ApJS...28....1J}
{Joy}, A.~H., \& {Abt}, H.~A. 1974, \apjs, 28, 1

\bibitem[{{Kalas} {et~al.}(2005){Kalas}, {Graham}, \& {Clampin}}]{kalas:2005}
{Kalas}, P., {Graham}, J.~R., \& {Clampin}, M. 2005, \nat, 435, 1067

\bibitem[{Kalas {et~al.}(2013)Kalas, Graham, Fitzgerald, \&
  Clampin}]{Kalas:2013hp}
Kalas, P., Graham, J.~R., Fitzgerald, M.~P., \& Clampin, M. 2013, ApJ, 775, 56

\bibitem[{{Keenan} \& {McNeil}(1989)}]{1989ApJS...71..245K}
{Keenan}, P.~C., \& {McNeil}, R.~C. 1989, \apjs, 71, 245

\bibitem[{{Kirkpatrick} {et~al.}(1991){Kirkpatrick}, {Henry}, \&
  {McCarthy}}]{1991ApJS...77..417K}
{Kirkpatrick}, J.~D., {Henry}, T.~J., \& {McCarthy}, Jr., D.~W. 1991, \apjs,
  77, 417

\bibitem[{{Konopacky} \& {Barman}(2018)}]{konopacky:2018}
{Konopacky}, Q.~M., \& {Barman}, T.~S. 2018, {HR8799: Imaging a System of
  Exoplanets}, 36

\bibitem[{Konopacky {et~al.}(2016)Konopacky, Rameau, Duch{\^e}ne, Filippazzo,
  Godfrey, Marois, Nielsen, Pueyo, Rafikov, Rice, Wang, Ammons, Bailey, Barman,
  Bulger, Bruzzone, Chilcote, Cotten, Dawson, Rosa, Doyon, Esposito,
  Fitzgerald, Follette, Goodsell, Graham, Greenbaum, Hibon, Hung, Ingraham,
  Kalas, Lafreni{\`e}re, Larkin, Macintosh, Maire, Marchis, Marley, Matthews,
  Metchev, Millar-Blanchaer, Oppenheimer, Palmer, Patience, Perrin, Poyneer,
  Rajan, Rantakyr{\"o}, Savransky, Schneider, Sivaramakrishnan, Song, Soummer,
  Thomas, Wallace, Ward-Duong, Wiktorowicz, \& Wolff}]{Konopacky:2016dk}
Konopacky, Q.~M., Rameau, J., Duch{\^e}ne, G., {et~al.} 2016, Astrophys. J.,
  829, L4

\bibitem[{{Kratter} \& {Lodato}(2016)}]{Kratter:2016}
{Kratter}, K., \& {Lodato}, G. 2016, \araa, 54, 271

\bibitem[{Kratter {et~al.}(2010)Kratter, Murray-Clay, \&
  Youdin}]{Kratter:2010gf}
Kratter, K.~M., Murray-Clay, R.~A., \& Youdin, A.~N. 2010, ApJ, 710, 1375

\bibitem[{{Kraus} {et~al.}(2008){Kraus}, {Ireland}, {Martinache}, \&
  {Lloyd}}]{kraus2008}
{Kraus}, A.~L., {Ireland}, M.~J., {Martinache}, F., \& {Lloyd}, J.~P. 2008,
  \apj, 679, 762

\bibitem[{Kroupa(2001)}]{Kroupa:2001ki}
Kroupa, P. 2001, MNRAS, 322, 231

\bibitem[{{Lafreni{\`e}re} {et~al.}(2007){Lafreni{\`e}re}, {Marois}, {Doyon},
  {Nadeau}, \& {Artigau}}]{Lafreniere:2007}
{Lafreni{\`e}re}, D., {Marois}, C., {Doyon}, R., {Nadeau}, D., \& {Artigau},
  {\'E}. 2007, \apj, 660, 770

\bibitem[{Lagrange {et~al.}(2009)Lagrange, Gratadour, Chauvin, Fusco,
  Ehrenreich, Mouillet, Rousset, Rouan, Allard, Gendron, Charton, Mugnier,
  Rabou, Montri, \& Lacombe}]{Lagrange:2009hq}
Lagrange, A.-M., Gratadour, D., Chauvin, G., {et~al.} 2009, A{\&}A, 493, L21

\bibitem[{{Lagrange} {et~al.}(2010){Lagrange}, {Bonnefoy}, {Chauvin}, {Apai},
  {Ehrenreich}, {Boccaletti}, {Gratadour}, {Rouan}, {Mouillet}, {Lacour}, \&
  {Kasper}}]{Lagrange:2010}
{Lagrange}, A.-M., {Bonnefoy}, M., {Chauvin}, G., {et~al.} 2010, Science, 329,
  57

\bibitem[{{Lagrange} {et~al.}(2012){Lagrange}, {Boccaletti}, {Milli},
  {Chauvin}, {Bonnefoy}, {Mouillet}, {Augereau}, {Girard}, {Lacour}, \&
  {Apai}}]{Lagrange:2012}
{Lagrange}, A.-M., {Boccaletti}, A., {Milli}, J., {et~al.} 2012, \aap, 542, A40

\bibitem[{{Lambrechts} {et~al.}(2014){Lambrechts}, {Johansen}, \&
  {Morbidelli}}]{Lambrechts:2014}
{Lambrechts}, M., {Johansen}, A., \& {Morbidelli}, A. 2014, \aap, 572, A35

\bibitem[{{Landsman}(1993)}]{idlastro}
{Landsman}, W.~B. 1993, in Astronomical Society of the Pacific Conference
  Series, Vol.~52, Astronomical Data Analysis Software and Systems II, ed.
  R.~J. {Hanisch}, R.~J.~V. {Brissenden}, \& J.~{Barnes}, 246

\bibitem[{{Lannier} {et~al.}(2016){Lannier}, {Delorme}, {Lagrange}, {Borgniet},
  {Rameau}, {Schlieder}, {Gagn{\'e}}, {Bonavita}, {Malo}, {Chauvin},
  {Bonnefoy}, \& {Girard}}]{lannier16}
{Lannier}, J., {Delorme}, P., {Lagrange}, A.~M., {et~al.} 2016, \aap, 596, A83

\bibitem[{{Larkin} {et~al.}(2014){Larkin}, {Chilcote}, {Aliado}, {Bauman},
  {Brims}, {Canfield}, {Cardwell}, {Dillon}, {Doyon}, {Dunn}, {Fitzgerald},
  {Graham}, {Goodsell}, {Hartung}, {Hibon}, {Ingraham}, {Johnson}, {Kress},
  {Konopacky}, {Macintosh}, {Magnone}, {Maire}, {McLean}, {Palmer}, {Perrin},
  {Quiroz}, {Rantakyr{\"o}}, {Sadakuni}, {Saddlemyer}, {Serio}, {Thibault},
  {Thomas}, {Vallee}, \& {Weiss}}]{larkin:2014}
{Larkin}, J.~E., {Chilcote}, J.~K., {Aliado}, T., {et~al.} 2014, in \procspie,
  Vol. 9147, Ground-based and Airborne Instrumentation for Astronomy V, 91471K

\bibitem[{{Laughlin} {et~al.}(2004){Laughlin}, {Bodenheimer}, \&
  {Adams}}]{Laughlin:2004}
{Laughlin}, G., {Bodenheimer}, P., \& {Adams}, F.~C. 2004, \apjl, 612, L73

\bibitem[{Lecavelier Des~Etangs \&
  Vidal-Madjar(2009)}]{LecavelierDesEtangs:2009jt}
Lecavelier Des~Etangs, A., \& Vidal-Madjar, A. 2009, A{\&}A, 497, 557

\bibitem[{{Lee} \& {Chiang}(2015)}]{Lee:2015}
{Lee}, E.~J., \& {Chiang}, E. 2015, \apj, 811, 41

\bibitem[{{Lee} \& {Chiang}(2016)}]{Lee:2016}
---. 2016, \apj, 817, 90

\bibitem[{{L{\'e}pine} {et~al.}(2013){L{\'e}pine}, {Hilton}, {Mann}, {Wilde},
  {Rojas-Ayala}, {Cruz}, \& {Gaidos}}]{2013AJ....145..102L}
{L{\'e}pine}, S., {Hilton}, E.~J., {Mann}, A.~W., {et~al.} 2013, \aj, 145, 102

\bibitem[{{Levato} {et~al.}(1987){Levato}, {Malaroda}, {Morrell}, \&
  {Solivella}}]{1987ApJS...64..487L}
{Levato}, H., {Malaroda}, S., {Morrell}, N., \& {Solivella}, G. 1987, The
  Astrophysical Journal Supplement Series, 64, 487

\bibitem[{{Levenhagen} \& {Leister}(2006)}]{2006MNRAS.371..252L}
{Levenhagen}, R.~S., \& {Leister}, N.~V. 2006, \mnras, 371, 252

\bibitem[{{Lin} {et~al.}(2018){Lin}, {Lee}, \& {Chiang}}]{Lin:2018}
{Lin}, J.~W., {Lee}, E.~J., \& {Chiang}, E. 2018, \mnras, 480, 4338

\bibitem[{{Liu} {et~al.}(2013){Liu}, {Magnier}, {Deacon}, {Allers}, {Dupuy},
  {Kotson}, {Aller}, {Burgett}, {Chambers}, {Draper}, {Hodapp}, {Jedicke},
  {Kaiser}, {Kudritzki}, {Metcalfe}, {Morgan}, {Price}, {Tonry}, \&
  {Wainscoat}}]{liu:2013}
{Liu}, M.~C., {Magnier}, E.~A., {Deacon}, N.~R., {et~al.} 2013, \apjl, 777, L20

\bibitem[{{Lloyd}(2011)}]{lloyd:2011}
{Lloyd}, J.~P. 2011, \apjl, 739, L49

\bibitem[{{Lloyd}(2013)}]{lloyd:2013}
---. 2013, \apjl, 774, L2

\bibitem[{Lovekin {et~al.}(2006)Lovekin, Deupree, \& Short}]{Lovekin:2006ij}
Lovekin, C.~C., Deupree, R.~G., \& Short, C.~I. 2006, Astrophys. J., 643, 460

\bibitem[{{Lowrance} {et~al.}(2000{\natexlab{a}}){Lowrance}, {Schneider},
  {Kirkpatrick}, {Becklin}, {Weinberger}, {Zuckerman}, {Plait}, {Malmuth},
  {Heap}, {Schultz}, {Smith}, {Terrile}, \& {Hines}}]{lowrance:2000}
{Lowrance}, P.~J., {Schneider}, G., {Kirkpatrick}, J.~D., {et~al.}
  2000{\natexlab{a}}, \apj, 541, 390

\bibitem[{{Lowrance} {et~al.}(2000{\natexlab{b}}){Lowrance}, {Schneider},
  {Kirkpatrick}, {Becklin}, {Weinberger}, {Zuckerman}, {Plait}, {Malmuth},
  {Heap}, {Schultz}, {Smith}, {Terrile}, \& {Hines}}]{2000ApJ...541..390L}
---. 2000{\natexlab{b}}, \apj, 541, 390

\bibitem[{{Lunine} {et~al.}(2008){Lunine}, {Fischer}, {Hammel}, {Henning},
  {Hillenbrand}, {Kasting}, {Laughlin}, {Macintosh}, {Marley}, {Melnick},
  {Monet}, {Noecker}, {Peale}, {Quirrenbach}, {Seager}, \&
  {Winn}}]{lunine:2008}
{Lunine}, J.~I., {Fischer}, D., {Hammel}, H., {et~al.} 2008, ArXiv e-prints,
  arXiv:0808.2754

\bibitem[{{Luyten}(1936)}]{1936ApJ....84...85L}
{Luyten}, W.~J. 1936, \apj, 84, 85

\bibitem[{Macintosh {et~al.}(2014)Macintosh, Graham, Ingraham, Konopacky,
  Marois, Perrin, Poyneer, Bauman, Barman, Burrows, Cardwell, Chilcote,
  De~Rosa, Dillon, Doyon, Dunn, Erikson, Fitzgerald, Gavel, Goodsell, Hartung,
  Hibon, Kalas, Larkin, Maire, Marchis, Marley, McBride, Millar-Blanchaer,
  Morzinski, Norton, Oppenheimer, Palmer, Patience, Pueyo, Rantakyro, Sadakuni,
  Saddlemyer, Savransky, Serio, Soummer, Sivaramakrishnan, Song, Thomas,
  Wallace, Wiktorowicz, \& Wolff}]{Macintosh:2014js}
Macintosh, B., Graham, J.~R., Ingraham, P., {et~al.} 2014, PNAS, 111, 12661

\bibitem[{Macintosh {et~al.}(2015)Macintosh, Graham, Barman, De~Rosa,
  Konopacky, Marley, Marois, Nielsen, Pueyo, Rajan, Rameau, Saumon, Wang,
  Patience, Ammons, Arriaga, Artigau, Beckwith, Brewster, Bruzzone, Bulger,
  Burningham, Burrows, Chen, Chiang, Chilcote, Dawson, Dong, Doyon, Draper,
  Duch{\^e}ne, Esposito, Fabrycky, Fitzgerald, Follette, Fortney, Gerard,
  Goodsell, Greenbaum, Hibon, Hinkley, Cotten, Hung, Ingraham, Johnson-Groh,
  Kalas, Lafreni{\`e}re, Larkin, Lee, Line, Long, Maire, Marchis, Matthews,
  Max, Metchev, Millar-Blanchaer, Mittal, Morley, Morzinski, Murray-Clay,
  Oppenheimer, Palmer, Patel, Perrin, Poyneer, Rafikov, Rantakyr{\"o}, Rice,
  Rojo, Rudy, Ruffio, Ruiz, Sadakuni, Saddlemyer, Salama, Savransky, Schneider,
  Sivaramakrishnan, Song, Soummer, Thomas, Vasisht, Wallace, Ward-Duong,
  Wiktorowicz, Wolff, \& Zuckerman}]{Macintosh:2015ew}
Macintosh, B., Graham, J.~R., Barman, T., {et~al.} 2015, Science, 350, 64

\bibitem[{{Macintosh} {et~al.}(2018){Macintosh}, {Chilcote}, {Bailey}, {De
  Rosa}, {Nielsen}, {Norton}, {Poyneer}, {Wang}, {Ruffio}, {Graham}, {Marois},
  {Savransky}, \& {Veran}}]{macintosh:2018}
{Macintosh}, B., {Chilcote}, J.~K., {Bailey}, V.~P., {et~al.} 2018, ArXiv
  e-prints, arXiv:1807.07146

\bibitem[{{Malo} {et~al.}(2013){Malo}, {Doyon}, {Lafreni{\`e}re}, {Artigau},
  {Gagn{\'e}}, {Baron}, \& {Riedel}}]{malo:2013}
{Malo}, L., {Doyon}, R., {Lafreni{\`e}re}, D., {et~al.} 2013, \apj, 762, 88

\bibitem[{{Mamajek}(2012)}]{mamajek:2012}
{Mamajek}, E.~E. 2012, \apjl, 754, L20

\bibitem[{{Mamajek} \& {Hillenbrand}(2008)}]{mamajek:2008}
{Mamajek}, E.~E., \& {Hillenbrand}, L.~A. 2008, \apj, 687, 1264

\bibitem[{{Mamajek} {et~al.}(2013){Mamajek}, {Bartlett}, {Seifahrt}, {Henry},
  {Dieterich}, {Lurie}, {Kenworthy}, {Jao}, {Riedel}, {Subasavage}, {Winters},
  {Finch}, {Ianna}, \& {Bean}}]{mamajek:2013a}
{Mamajek}, E.~E., {Bartlett}, J.~L., {Seifahrt}, A., {et~al.} 2013, \aj, 146,
  154

\bibitem[{{Manara} {et~al.}(2018){Manara}, {Morbidelli}, \&
  {Guillot}}]{Manara:2018}
{Manara}, C.~F., {Morbidelli}, A., \& {Guillot}, T. 2018, \aap, 618, L3

\bibitem[{{Marley} {et~al.}(2007){Marley}, {Fortney}, {Hubickyj},
  {Bodenheimer}, \& {Lissauer}}]{marley07}
{Marley}, M.~S., {Fortney}, J.~J., {Hubickyj}, O., {Bodenheimer}, P., \&
  {Lissauer}, J.~J. 2007, \apj, 655, 541

\bibitem[{{Marois} {et~al.}(2000){Marois}, {Doyon}, {Racine}, \&
  {Nadeau}}]{Marois2000}
{Marois}, C., {Doyon}, R., {Racine}, R., \& {Nadeau}, D. 2000, Publications of
  the Astronomical Society of the Pacific, 112, 91

\bibitem[{Marois {et~al.}(2006)Marois, Lafreni{\`e}re, Doyon, Macintosh, \&
  Nadeau}]{Marois:2006df}
Marois, C., Lafreni{\`e}re, D., Doyon, R., Macintosh, B., \& Nadeau, D. 2006,
  ApJ, 641, 556

\bibitem[{Marois {et~al.}(2008)Marois, Macintosh, Barman, Zuckerman, Song,
  Patience, Lafreni{\`e}re, \& Doyon}]{Marois:2008ei}
Marois, C., Macintosh, B., Barman, T., {et~al.} 2008, Science, 322, 1348

\bibitem[{{Marois} {et~al.}(2010){Marois}, {Macintosh}, \&
  {V{\'e}ran}}]{Marois:2010b}
{Marois}, C., {Macintosh}, B., \& {V{\'e}ran}, J.-P. 2010, in \procspie, Vol.
  7736, Adaptive Optics Systems II, 77361J

\bibitem[{Marois {et~al.}(2010)Marois, Zuckerman, Konopacky, Macintosh, \&
  Barman}]{Marois:2010gp}
Marois, C., Zuckerman, B., Konopacky, Q.~M., Macintosh, B., \& Barman, T. 2010,
  Nature, 468, 1080

\bibitem[{{Matzner} \& {Levin}(2005)}]{Matzner:2005}
{Matzner}, C.~D., \& {Levin}, Y. 2005, \apj, 628, 817

\bibitem[{{McBride} {et~al.}(2011){McBride}, {Graham}, {Macintosh}, {Beckwith},
  {Marois}, {Poyneer}, \& {Wiktorowicz}}]{McBride:2011}
{McBride}, J., {Graham}, J.~R., {Macintosh}, B., {et~al.} 2011, \pasp, 123, 692

\bibitem[{Meshkat {et~al.}(2013)Meshkat, Bailey, Rameau, Bonnefoy, Boccaletti,
  Mamajek, Kenworthy, Chauvin, Lagrange, Su, \& Currie}]{Meshkat:2013fz}
Meshkat, T., Bailey, V., Rameau, J., {et~al.} 2013, ApJL, 775, L40

\bibitem[{Meshkat {et~al.}(2015)Meshkat, Bonnefoy, Mamajek, Quanz, Chauvin,
  Kenworthy, Rameau, Meyer, Lagrange, Lannier, \& Delorme}]{Meshkat:2015hd}
Meshkat, T., Bonnefoy, M., Mamajek, E.~E., {et~al.} 2015, MNRAS, 453, 2379

\bibitem[{{Metchev} \& {Hillenbrand}(2009)}]{metchev09}
{Metchev}, S.~A., \& {Hillenbrand}, L.~A. 2009, \apjs, 181, 62

\bibitem[{Millar-Blanchaer {et~al.}(2015)Millar-Blanchaer, Graham, Pueyo,
  Kalas, Dawson, Wang, Perrin, Moon, Macintosh, Ammons, Barman, Cardwell, Chen,
  Chiang, Chilcote, Cotten, De~Rosa, Draper, Dunn, Duch{\^e}ne, Esposito,
  Fitzgerald, Follette, Goodsell, Greenbaum, Hartung, Hibon, Hinkley, Ingraham,
  Jensen-Clem, Konopacky, Larkin, Long, Maire, Marchis, Marley, Marois,
  Morzinski, Nielsen, Palmer, Oppenheimer, Poyneer, Rajan, Rantakyr{\"o},
  Ruffio, Sadakuni, Saddlemyer, Schneider, Sivaramakrishnan, Soummer, Thomas,
  Vasisht, Vega, Wallace, Ward-Duong, Wiktorowicz, \&
  Wolff}]{MillarBlanchaer:2015ha}
Millar-Blanchaer, M.~A., Graham, J.~R., Pueyo, L., {et~al.} 2015, ApJ, 811, 18

\bibitem[{Milli {et~al.}(2016)Milli, Hibon, Christiaens, Choquet, Bonnefoy,
  Kennedy, Wyatt, Absil, G{\'o}mez~Gonz{\'a}lez, del Burgo, Matr{\`a},
  Augereau, Boccaletti, Delacroix, Ertel, Dent, Forsberg, Fusco, Girard,
  Habraken, Huby, Karlsson, Lagrange, Mawet, Mouillet, Perrin, Pinte, Pueyo,
  Reyes, Soummer, Surdej, Tarricq, \& Wahhaj}]{Milli:2016fs}
Milli, J., Hibon, P., Christiaens, V., {et~al.} 2016, A{\&}A, 597, L2

\bibitem[{{Mizuno}(1980)}]{Mizuno:1980}
{Mizuno}, H. 1980, Progress of Theoretical Physics, 64, 544

\bibitem[{{Mizuno} {et~al.}(1978){Mizuno}, {Nakazawa}, \&
  {Hayashi}}]{Mizuno:1978}
{Mizuno}, H., {Nakazawa}, K., \& {Hayashi}, C. 1978, Progress of Theoretical
  Physics, 60, 699

\bibitem[{{Moe} \& {Kratter}(2018)}]{Moe:2018}
{Moe}, M., \& {Kratter}, K.~M. 2018, \apj, 854, 44

\bibitem[{Mo{\'o}r {et~al.}(2013)Mo{\'o}r, {\'A}brah{\'a}m, K{\'o}sp{\'a}l,
  Szab{\'o}, Apai, Balog, Csengeri, Grady, Henning, Juh{\'a}sz, Kiss, Pascucci,
  Szul{\'a}gyi, \& Vavrek}]{Moor:2013bg}
Mo{\'o}r, A., {\'A}brah{\'a}m, P., K{\'o}sp{\'a}l, {\'A}., {et~al.} 2013, ApJL,
  775, L51

\bibitem[{{Mordasini} {et~al.}(2017){Mordasini}, {Marleau}, \&
  {Molli{\`e}re}}]{mordasini:2017}
{Mordasini}, C., {Marleau}, G.-D., \& {Molli{\`e}re}, P. 2017, \aap, 608, A72

\bibitem[{{Morzinski} {et~al.}(2015){Morzinski}, {Males}, {Skemer}, {Close},
  {Hinz}, {Rodigas}, {Puglisi}, {Esposito}, {Riccardi}, {Pinna}, {Xompero},
  {Briguglio}, {Bailey}, {Follette}, {Kopon}, {Weinberger}, \&
  {Wu}}]{morzinski:2015}
{Morzinski}, K.~M., {Males}, J.~R., {Skemer}, A.~J., {et~al.} 2015, \apj, 815,
  108

\bibitem[{Mugrauer {et~al.}(2010)Mugrauer, Vogt, Neuh{\"a}user, \&
  Schmidt}]{Mugrauer:2010cp}
Mugrauer, M., Vogt, N., Neuh{\"a}user, R., \& Schmidt, T. O.~B. 2010, A{\&}A,
  523, L1

\bibitem[{{Mulders} {et~al.}(2015){Mulders}, {Pascucci}, \&
  {Apai}}]{Mulders:2015}
{Mulders}, G.~D., {Pascucci}, I., \& {Apai}, D. 2015, \apj, 814, 130

\bibitem[{{Murphy} {et~al.}(2013){Murphy}, {Lawson}, \&
  {Bessell}}]{2013MNRAS.435.1325M}
{Murphy}, S.~J., {Lawson}, W.~A., \& {Bessell}, M.~S. 2013, \mnras, 435, 1325

\bibitem[{{Murphy} {et~al.}(2015){Murphy}, {Corbally}, {Gray}, {Cheng}, {Neff},
  {Koen}, {Kuehn}, {Newsome}, \& {Riggs}}]{2015PASA...32...36M}
{Murphy}, S.~J., {Corbally}, C.~J., {Gray}, R.~O., {et~al.} 2015, \pasa, 32,
  e036

\bibitem[{Naylor \& Jeffries(2006)}]{Naylor:2006iz}
Naylor, T., \& Jeffries, R.~D. 2006, MNRAS, 373, 1251

\bibitem[{{Nielsen} \& {Close}(2010)}]{Nielsen:2010}
{Nielsen}, E.~L., \& {Close}, L.~M. 2010, \apj, 717, 878

\bibitem[{{Nielsen} {et~al.}(2008){Nielsen}, {Close}, {Biller}, {Masciadri}, \&
  {Lenzen}}]{Nielsen:2008}
{Nielsen}, E.~L., {Close}, L.~M., {Biller}, B.~A., {Masciadri}, E., \&
  {Lenzen}, R. 2008, \apj, 674, 466

\bibitem[{{Nielsen} {et~al.}(2013){Nielsen}, {Liu}, {Wahhaj}, {Biller},
  {Hayward}, {Close}, {Males}, {Skemer}, {Chun}, {Ftaclas}, {Alencar},
  {Artymowicz}, {Boss}, {Clarke}, {de Gouveia Dal Pino}, {Gregorio-Hetem},
  {Hartung}, {Ida}, {Kuchner}, {Lin}, {Reid}, {Shkolnik}, {Tecza}, {Thatte}, \&
  {Toomey}}]{nielsen:2013}
{Nielsen}, E.~L., {Liu}, M.~C., {Wahhaj}, Z., {et~al.} 2013, \apj, 776, 4

\bibitem[{{Nielsen} {et~al.}(2014){Nielsen}, {Liu}, {Wahhaj}, {Biller},
  {Hayward}, {Males}, {Close}, {Morzinski}, {Skemer}, {Kuchner}, {Rodigas},
  {Hinz}, {Chun}, {Ftaclas}, \& {Toomey}}]{Nielsen:2014}
---. 2014, \apj, 794, 158

\bibitem[{Nielsen {et~al.}(2016)Nielsen, Rosa, Wang, Rameau, Song, Graham,
  Macintosh, Ammons, Bailey, Barman, Bulger, Chilcote, Cotten, Doyon,
  Duch{\^e}ne, Fitzgerald, Follette, Greenbaum, Hibon, Hung, Ingraham, Kalas,
  Konopacky, Larkin, Maire, Marchis, Marley, Marois, Metchev, Millar-Blanchaer,
  Oppenheimer, Palmer, Patience, Perrin, Poyneer, Pueyo, Rajan, Rantakyr{\"o},
  Savransky, Schneider, Sivaramakrishnan, Soummer, Thomas, Wallace, Ward-Duong,
  Wiktorowicz, \& Wolff}]{Nielsen:2016ct}
Nielsen, E.~L., Rosa, R. J.~D., Wang, J., {et~al.} 2016, AJ, 152, 175

\bibitem[{{Nielsen} {et~al.}(2017){Nielsen}, {De Rosa}, {Rameau}, {Wang},
  {Esposito}, {Millar-Blanchaer}, {Marois}, {Vigan}, {Ammons}, {Artigau},
  {Bailey}, {Blunt}, {Bulger}, {Chilcote}, {Cotten}, {Doyon}, {Duch{\^e}ne},
  {Fabrycky}, {Fitzgerald}, {Follette}, {Gerard}, {Goodsell}, {Graham},
  {Greenbaum}, {Hibon}, {Hinkley}, {Hung}, {Ingraham}, {Jensen-Clem}, {Kalas},
  {Konopacky}, {Larkin}, {Macintosh}, {Maire}, {Marchis}, {Metchev},
  {Morzinski}, {Murray-Clay}, {Oppenheimer}, {Palmer}, {Patience}, {Perrin},
  {Poyneer}, {Pueyo}, {Rafikov}, {Rajan}, {Rantakyr{\"o}}, {Ruffio},
  {Savransky}, {Schneider}, {Sivaramakrishnan}, {Song}, {Soummer}, {Thomas},
  {Wallace}, {Ward-Duong}, {Wiktorowicz}, \& {Wolff}}]{nielsen:2017}
{Nielsen}, E.~L., {De Rosa}, R.~J., {Rameau}, J., {et~al.} 2017, \aj, 154, 218

\bibitem[{{North} {et~al.}(2017){North}, {Campante}, {Miglio}, {Davies},
  {Grunblatt}, {Huber}, {Kuszlewicz}, {Lund}, {Cooke}, \&
  {Chaplin}}]{north:2017}
{North}, T.~S.~H., {Campante}, T.~L., {Miglio}, A., {et~al.} 2017, \mnras, 472,
  1866

\bibitem[{{Opolski}(1957)}]{1957ArA.....2...55O}
{Opolski}, A. 1957, Arkiv for Astronomi, 2, 55

\bibitem[{{Oppenheimer} {et~al.}(2001){Oppenheimer}, {Golimowski}, {Kulkarni},
  {Matthews}, {Nakajima}, {Creech-Eakman}, \& {Durrance}}]{oppenheimer:2001}
{Oppenheimer}, B.~R., {Golimowski}, D.~A., {Kulkarni}, S.~R., {et~al.} 2001,
  \aj, 121, 2189

\bibitem[{{Ormel}(2017)}]{Ormel:2017a}
{Ormel}, C.~W. 2017, in Astrophysics and Space Science Library, Vol. 445,
  Astrophysics and Space Science Library, ed. M.~{Pessah} \& O.~{Gressel}, 197

\bibitem[{{Ormel} \& {Klahr}(2010)}]{Ormel:2010}
{Ormel}, C.~W., \& {Klahr}, H.~H. 2010, \aap, 520, A43

\bibitem[{{Owen} \& {Menou}(2016)}]{Owen:2016}
{Owen}, J.~E., \& {Menou}, K. 2016, \apjl, 819, L14

\bibitem[{Pace(2013)}]{Pace:2013cl}
Pace, G. 2013, A{\&}A, 551, L8

\bibitem[{{Pascucci} {et~al.}(2016){Pascucci}, {Testi}, {Herczeg}, {Long},
  {Manara}, {Hendler}, {Mulders}, {Krijt}, {Ciesla}, {Henning}, {Mohanty},
  {Drabek-Maunder}, {Apai}, {Sz{\H u}cs}, {Sacco}, \&
  {Olofsson}}]{Pascucci:2016}
{Pascucci}, I., {Testi}, L., {Herczeg}, G.~J., {et~al.} 2016, \apj, 831, 125

\bibitem[{{Paunzen} {et~al.}(2001){Paunzen}, {Duffee}, {Heiter}, {Kuschnig}, \&
  {Weiss}}]{2001AA...373..625P}
{Paunzen}, E., {Duffee}, B., {Heiter}, U., {Kuschnig}, R., \& {Weiss}, W.~W.
  2001, \aap, 373, 625

\bibitem[{Paxton {et~al.}(2010)Paxton, Bildsten, Dotter, Herwig, Lesaffre, \&
  Timmes}]{Paxton:2010jf}
Paxton, B., Bildsten, L., Dotter, A., {et~al.} 2010, ApJS, 192, 3

\bibitem[{{Paxton} {et~al.}(2013){Paxton}, {Cantiello}, {Arras}, {Bildsten},
  {Brown}, {Dotter}, {Mankovich}, {Montgomery}, {Stello}, {Timmes}, \&
  {Townsend}}]{Paxton:2013}
{Paxton}, B., {Cantiello}, M., {Arras}, P., {et~al.} 2013, \apjs, 208, 4

\bibitem[{{Pecaut} \& {Mamajek}(2016{\natexlab{a}})}]{Pecaut:2016}
{Pecaut}, M.~J., \& {Mamajek}, E.~E. 2016{\natexlab{a}}, \mnras, 461, 794

\bibitem[{{Pecaut} \& {Mamajek}(2016{\natexlab{b}})}]{2016MNRAS.461..794P}
---. 2016{\natexlab{b}}, \mnras, 461, 794

\bibitem[{{Perrin} {et~al.}(2014){Perrin}, {Maire}, {Ingraham}, {Savransky},
  {Millar-Blanchaer}, {Wolff}, {Ruffio}, {Wang}, {Draper}, {Sadakuni},
  {Marois}, {Rajan}, {Fitzgerald}, {Macintosh}, {Graham}, {Doyon}, {Larkin},
  {Chilcote}, {Goodsell}, {Palmer}, {Labrie}, {Beaulieu}, {De Rosa},
  {Greenbaum}, {Hartung}, {Hibon}, {Konopacky}, {Lafreniere}, {Lavigne},
  {Marchis}, {Patience}, {Pueyo}, {Rantakyr{\"o}}, {Soummer},
  {Sivaramakrishnan}, {Thomas}, {Ward-Duong}, \& {Wiktorowicz}}]{Perrin:2014}
{Perrin}, M.~D., {Maire}, J., {Ingraham}, P., {et~al.} 2014, in \procspie, Vol.
  9147, Ground-based and Airborne Instrumentation for Astronomy V, 91473J

\bibitem[{Perrin {et~al.}(2016)Perrin, Ingraham, Follette, Maire, Wang,
  Savransky, Arriaga, Bailey, Bruzzone, Chilcote, De~Rosa, Draper, Fitzgerald,
  Greenbaum, Hung, Konopacky, Macintosh, Marchis, Marois, Millar-Blanchaer,
  Nielsen, Rajan, Rameau, Rantakyr{\"o}, Ruffio, Ward-Duong, Wolff, \&
  Zalesky}]{Perrin:2016gm}
Perrin, M.~D., Ingraham, P., Follette, K.~B., {et~al.} 2016, Ground-based and
  Airborne Instrumentation for Astronomy VI, 9908, 990837

\bibitem[{{Petigura} {et~al.}(2018){Petigura}, {Marcy}, {Winn}, {Weiss},
  {Fulton}, {Howard}, {Sinukoff}, {Isaacson}, {Morton}, \&
  {Johnson}}]{Petigura:2018}
{Petigura}, E.~A., {Marcy}, G.~W., {Winn}, J.~N., {et~al.} 2018, \aj, 155, 89

\bibitem[{Pollack {et~al.}(1996)Pollack, Hubickyj, Bodenheimer, Lissauer,
  Podolak, \& Greenzweig}]{Pollack:1996jp}
Pollack, J.~B., Hubickyj, O., Bodenheimer, P., {et~al.} 1996, Icarus, 124, 62

\bibitem[{{Poyneer} {et~al.}(2014){Poyneer}, {De Rosa}, {Macintosh}, {Palmer},
  {Perrin}, {Sadakuni}, {Savransky}, {Bauman}, {Cardwell}, {Chilcote},
  {Dillon}, {Gavel}, {Goodsell}, {Hartung}, {Hibon}, {Rantakyr{\"o}}, {Thomas},
  \& {Veran}}]{poyneer:2014}
{Poyneer}, L.~A., {De Rosa}, R.~J., {Macintosh}, B., {et~al.} 2014, in
  \procspie, Vol. 9148, Adaptive Optics Systems IV, 91480K

\bibitem[{{Pueyo}(2016)}]{Pueyo2016}
{Pueyo}, L. 2016, \apj, 824, 117

\bibitem[{{Pueyo} {et~al.}(2015){Pueyo}, {Soummer}, {Hoffmann}, {Oppenheimer},
  {Graham}, {Zimmerman}, {Zhai}, {Wallace}, {Vescelus}, {Veicht}, {Vasisht},
  {Truong}, {Sivaramakrishnan}, {Shao}, {Roberts}, {Roberts}, {Rice}, {Parry},
  {Nilsson}, {Lockhart}, {Ligon}, {King}, {Hinkley}, {Hillenbrand}, {Hale},
  {Dekany}, {Crepp}, {Cady}, {Burruss}, {Brenner}, {Beichman}, \&
  {Baranec}}]{Pueyo:2015}
{Pueyo}, L., {Soummer}, R., {Hoffmann}, J., {et~al.} 2015, \apj, 803, 31

\bibitem[{{Racine} {et~al.}(1999){Racine}, {Walker}, {Nadeau}, {Doyon}, \&
  {Marois}}]{Racine1999}
{Racine}, R., {Walker}, G. A.~H., {Nadeau}, D., {Doyon}, R., \& {Marois}, C.
  1999, Publications of the Astronomical Society of the Pacific, 111, 587

\bibitem[{{Rafikov}(2005)}]{Rafikov:2005}
{Rafikov}, R.~R. 2005, \apjl, 621, L69

\bibitem[{{Rafikov}(2009)}]{Rafikov:2009}
---. 2009, \apj, 704, 281

\bibitem[{Rajan {et~al.}(2017)Rajan, Rameau, Rosa, Marley, Graham, Macintosh,
  Marois, Morley, Patience, Pueyo, Saumon, Ward-Duong, Ammons, Arriaga, Bailey,
  Barman, Bulger, Burrows, Chilcote, Cotten, Czekala, Doyon, Duch{\^e}ne,
  Esposito, Fitzgerald, Follette, Fortney, Goodsell, Greenbaum, Hibon, Hung,
  Ingraham, Johnson-Groh, Kalas, Konopacky, Lafreni{\`e}re, Larkin, Maire,
  Marchis, Metchev, Millar-Blanchaer, Morzinski, Nielsen, Oppenheimer, Palmer,
  Patel, Perrin, Poyneer, Rantakyr{\"o}, Ruffio, Savransky, Schneider,
  Sivaramakrishnan, Song, Soummer, Thomas, Vasisht, Wallace, Wang, Wiktorowicz,
  \& Wolff}]{Rajan:2017ur}
Rajan, A., Rameau, J., Rosa, R. J.~D., {et~al.} 2017, AJ, 154, 10

\bibitem[{{Rameau} {et~al.}(2013){Rameau}, {Chauvin}, {Lagrange}, {Klahr},
  {Bonnefoy}, {Mordasini}, {Bonavita}, {Desidera}, {Dumas}, \&
  {Girard}}]{Rameau:2013}
{Rameau}, J., {Chauvin}, G., {Lagrange}, A.-M., {et~al.} 2013, \aap, 553, A60

\bibitem[{Rameau {et~al.}(2013{\natexlab{a}})Rameau, Chauvin, Lagrange,
  Meshkat, Boccaletti, Quanz, Currie, Mawet, Girard, Bonnefoy, \&
  Kenworthy}]{Rameau:2013ds}
Rameau, J., Chauvin, G., Lagrange, A.-M., {et~al.} 2013{\natexlab{a}}, ApJL,
  779, L26

\bibitem[{Rameau {et~al.}(2013{\natexlab{b}})Rameau, Chauvin, Lagrange,
  Boccaletti, Quanz, Bonnefoy, Girard, Delorme, Desidera, Klahr, Mordasini,
  Dumas, \& Bonavita}]{Rameau:2013dr}
---. 2013{\natexlab{b}}, ApJL, 772, L15

\bibitem[{{Rameau} {et~al.}(2016){Rameau}, {Nielsen}, {De Rosa}, {Blunt},
  {Patience}, {Doyon}, {Graham}, {Lafreni{\`e}re}, {Macintosh}, {Marchis},
  {Bailey}, {Chilcote}, {Duchene}, {Esposito}, {Hung}, {Konopacky}, {Maire},
  {Marois}, {Metchev}, {Perrin}, {Pueyo}, {Rajan}, {Savransky}, {Wang},
  {Ward-Duong}, {Wolff}, {Ammons}, {Hibon}, {Ingraham}, {Kalas}, {Morzinski},
  {Oppenheimer}, {Rantakyear{\"o}}, \& {Thomas}}]{rameau2016}
{Rameau}, J., {Nielsen}, E.~L., {De Rosa}, R.~J., {et~al.} 2016, \apjl, 822,
  L29

\bibitem[{Riedel {et~al.}(2017)Riedel, Blunt, Lambrides, Rice, Cruz, \&
  Faherty}]{Riedel:2017dg}
Riedel, A.~R., Blunt, S.~C., Lambrides, E.~L., {et~al.} 2017, AJ, 153, 95

\bibitem[{Rizzuto {et~al.}(2011)Rizzuto, Ireland, \&
  Robertson}]{Rizzuto:2011gs}
Rizzuto, A.~C., Ireland, M.~J., \& Robertson, J.~G. 2011, MNRAS, 416, 3108

\bibitem[{{Ruffio} {et~al.}(2017{\natexlab{a}}){Ruffio}, {Macintosh}, {Wang},
  \& {Pueyo}}]{Ruffio2017SPIE}
{Ruffio}, J.-B., {Macintosh}, B., {Wang}, J.~J., \& {Pueyo}, L.
  2017{\natexlab{a}}, in Society of Photo-Optical Instrumentation Engineers
  (SPIE) Conference Series, Vol. 10400, 1040027

\bibitem[{{Ruffio} {et~al.}(2017{\natexlab{b}}){Ruffio}, {Macintosh}, {Wang},
  {Pueyo}, {Nielsen}, {De Rosa}, {Czekala}, {Marley}, {Arriaga}, {Bailey},
  {Barman}, {Bulger}, {Chilcote}, {Cotten}, {Doyon}, {Duch{\^e}ne},
  {Fitzgerald}, {Follette}, {Gerard}, {Goodsell}, {Graham}, {Greenbaum},
  {Hibon}, {Hung}, {Ingraham}, {Kalas}, {Konopacky}, {Larkin}, {Maire},
  {Marchis}, {Marois}, {Metchev}, {Millar-Blanchaer}, {Morzinski},
  {Oppenheimer}, {Palmer}, {Patience}, {Perrin}, {Poyneer}, {Rajan}, {Rameau},
  {Rantakyr{\"o}}, {Savransky}, {Schneider}, {Sivaramakrishnan}, {Song},
  {Soummer}, {Thomas}, {Wallace}, {Ward-Duong}, {Wiktorowicz}, \&
  {Wolff}}]{Ruffio2017}
{Ruffio}, J.-B., {Macintosh}, B., {Wang}, J.~J., {et~al.} 2017{\natexlab{b}},
  \apj, 842, 14

\bibitem[{{Sahade} \& {Hern{\'a}ndez}(1964)}]{1964AnAp...27...11S}
{Sahade}, J., \& {Hern{\'a}ndez}, C.~A. 1964, Annales d'Astrophysique, 27, 11

\bibitem[{{Salpeter}(1955)}]{salpeter:1955}
{Salpeter}, E.~E. 1955, \apj, 121, 161

\bibitem[{Samland {et~al.}(2017)Samland, Molli{\`e}re, Bonnefoy, Maire,
  Cantalloube, Cheetham, Mesa, Gratton, Biller, Wahhaj, Bouwman, Brandner,
  Melnick, Carson, Janson, Henning, Homeier, Mordasini, Langlois, Quanz, van
  Boekel, Zurlo, Schlieder, Avenhaus, Boccaletti, Bonavita, Chauvin, Claudi,
  Cudel, Desidera, Feldt, Galicher, Kopytova, Lagrange, Le~Coroller, Mouillet,
  Mugnier, Perrot, Sissa, \& Vigan}]{Samland:2017vi}
Samland, M., Molli{\`e}re, P., Bonnefoy, M., {et~al.} 2017, eprint
  arXiv:1704.02987, 1704.02987

\bibitem[{{Savransky} {et~al.}(2014){Savransky}, {Thomas}, {Poyneer}, {Dunn},
  {Macintosh}, {Sadakuni}, {Dillon}, {Goodsell}, {Hartung}, {Hibon},
  {Rantakyr{\"o}}, {Cardwell}, \& {Serio}}]{savransky:2014}
{Savransky}, D., {Thomas}, S.~J., {Poyneer}, L.~A., {et~al.} 2014, in
  \procspie, Vol. 9147, Ground-based and Airborne Instrumentation for Astronomy
  V, 914740

\bibitem[{{Schlaufman}(2018)}]{schlaufman:2018}
{Schlaufman}, K.~C. 2018, \apj, 853, 37

\bibitem[{{Schlieder} {et~al.}(2012){Schlieder}, {L{\'e}pine}, {Rice}, {Simon},
  {Fielding}, \& {Tomasino}}]{2012AJ....143..114S}
{Schlieder}, J.~E., {L{\'e}pine}, S., {Rice}, E., {et~al.} 2012, \aj, 143, 114

\bibitem[{{Shenavrin} {et~al.}(2011){Shenavrin}, {Taranova}, \&
  {Nadzhip}}]{2011ARep...55...31S}
{Shenavrin}, V.~I., {Taranova}, O.~G., \& {Nadzhip}, A.~E. 2011, Astronomy
  Reports, 55, 31

\bibitem[{{Siess} {et~al.}(2000){Siess}, {Dufour}, \& {Forestini}}]{siess:2000}
{Siess}, L., {Dufour}, E., \& {Forestini}, M. 2000, \aap, 358, 593

\bibitem[{{Skiff}(2014)}]{2014yCat....1.2023S}
{Skiff}, B.~A. 2014, VizieR Online Data Catalog, 1

\bibitem[{{Skuljan} {et~al.}(2004){Skuljan}, {Ramm}, \&
  {Hearnshaw}}]{2004MNRAS.352..975S}
{Skuljan}, J., {Ramm}, D.~J., \& {Hearnshaw}, J.~B. 2004, \mnras, 352, 975

\bibitem[{{Smith} \& {Terrile}(1984)}]{smith:1984}
{Smith}, B.~A., \& {Terrile}, R.~J. 1984, Science, 226, 1421

\bibitem[{{Snellen} \& {Brown}(2018)}]{Snellen:2018}
{Snellen}, I.~A.~G., \& {Brown}, A.~G.~A. 2018, Nature Astronomy,
  arXiv:1808.06257

\bibitem[{{Soderblom}(2010)}]{Soderblom:2010}
{Soderblom}, D.~R. 2010, \araa, 48, 581

\bibitem[{Soummer {et~al.}(2012)Soummer, Pueyo, \& Larkin}]{Soummer:2012ig}
Soummer, R., Pueyo, L., \& Larkin, J. 2012, ApJL, 755, L28

\bibitem[{{Soummer} {et~al.}(2007){Soummer}, {Sivaramakrishnan}, {Oppenheimer},
  {Brenner}, {Pueyo}, {Marois}, {Macintosh}, {Graham}, {Palmer}, \& {GPI
  Team}}]{soummer:2007}
{Soummer}, R., {Sivaramakrishnan}, A., {Oppenheimer}, B.~R., {et~al.} 2007, in
  Bulletin of the American Astronomical Society, Vol.~39, American Astronomical
  Society Meeting Abstracts, 968

\bibitem[{{Sparks} \& {Ford}(2002)}]{Sparks2002}
{Sparks}, W.~B., \& {Ford}, H.~C. 2002, \apj, 578, 543

\bibitem[{{Spiegel} \& {Burrows}(2012)}]{Spiegel:2012}
{Spiegel}, D.~S., \& {Burrows}, A. 2012, \apj, 745, 174

\bibitem[{{Sreedhar Rao} \& {Abhyankar}(1991)}]{1991JApA...12..133S}
{Sreedhar Rao}, S., \& {Abhyankar}, K.~D. 1991, Journal of Astrophysics and
  Astronomy, 12, 133

\bibitem[{{Stello} {et~al.}(2017){Stello}, {Huber}, {Grundahl}, {Lloyd},
  {Ireland}, {Casagrande}, {Fredslund}, {Bedding}, {Palle}, {Antoci},
  {Kjeldsen}, \& {Christensen-Dalsgaard}}]{stello:2017}
{Stello}, D., {Huber}, D., {Grundahl}, F., {et~al.} 2017, \mnras, 472, 4110

\bibitem[{{Stephenson}(1986)}]{1986AJ.....92..139S}
{Stephenson}, C.~B. 1986, \aj, 92, 139

\bibitem[{{Stock} \& {Wroblewski}(1972)}]{1972PDAUC...2...59S}
{Stock}, J., \& {Wroblewski}, H. 1972, Publications of the Department of
  Astronomy University of Chile, 2

\bibitem[{{Stone} {et~al.}(2018){Stone}, {Skemer}, {Hinz}, {Bonavita},
  {Kratter}, {Maire}, {Defrere}, {Bailey}, {Spalding}, {Leisenring},
  {Desidera}, {Bonnefoy}, {Biller}, {Woodward}, {Henning}, {Skrutskie},
  {Eisner}, {Crepp}, {Patience}, {Weigelt}, {De Rosa}, {Schlieder}, {Brandner},
  {Apai}, {Su}, {Ertel}, {Ward-Duong}, {Morzinski}, {Schertl}, {Hofmann},
  {Close}, {Brems}, {Fortney}, {Oza}, {Buenzli}, \& {Bass}}]{stone:2018}
{Stone}, J.~M., {Skemer}, A.~J., {Hinz}, P.~M., {et~al.} 2018, ArXiv e-prints,
  arXiv:1810.10560

\bibitem[{{Strassmeier} \& {Rice}(2000)}]{2000AA...360.1019S}
{Strassmeier}, K.~G., \& {Rice}, J.~B. 2000, \aap, 360, 1019

\bibitem[{Su {et~al.}(2009)Su, Rieke, Stapelfeldt, Malhotra, Bryden, Smith,
  Misselt, Moro-Martin, \& Williams}]{Su:2009ig}
Su, K. Y.~L., Rieke, G.~H., Stapelfeldt, K.~R., {et~al.} 2009, ApJ, 705, 314

\bibitem[{Su {et~al.}(2017)Su, MacGregor, Booth, Wilner, Flaherty, Hughes,
  Phillips, Malhotra, Hales, Morrison, Ertel, Matthews, Dent, \&
  Casassus}]{Su:2017kl}
Su, K. Y.~L., MacGregor, M.~A., Booth, M., {et~al.} 2017, AJ, 154, 225

\bibitem[{{Tabachnik} \& {Tremaine}(2002)}]{tabachnik:2002}
{Tabachnik}, S., \& {Tremaine}, S. 2002, \mnras, 335, 151

\bibitem[{{Toomre}(1964)}]{toomre:1964}
{Toomre}, A. 1964, \apj, 139, 1217

\bibitem[{Torres {et~al.}(2006)Torres, Quast, da~Silva, de~la Reza, Melo, \&
  Sterzik}]{Torres:2006bw}
Torres, C. A.~O., Quast, G.~R., da~Silva, L., {et~al.} 2006, A{\&}A, 460, 695

\bibitem[{{Torres} {et~al.}(2003){Torres}, {Guenther}, {Marschall},
  {Neuh{\"a}user}, {Latham}, \& {Stefanik}}]{2003AJ....125..825T}
{Torres}, G., {Guenther}, E.~W., {Marschall}, L.~A., {et~al.} 2003, \aj, 125,
  825

\bibitem[{{Triaud} {et~al.}(2017){Triaud}, {Martin}, {S{\'e}gransan},
  {Smalley}, {Maxted}, {Anderson}, {Bouchy}, {Collier Cameron}, {Faedi},
  {G{\'o}mez Maqueo Chew}, {Hebb}, {Hellier}, {Marmier}, {Pepe}, {Pollacco},
  {Queloz}, {Udry}, \& {West}}]{triaud:2017}
{Triaud}, A.~H.~M.~J., {Martin}, D.~V., {S{\'e}gransan}, D., {et~al.} 2017,
  \aap, 608, A129

\bibitem[{{Uyama} {et~al.}(2017){Uyama}, {Hashimoto}, {Kuzuhara}, {Mayama},
  {Akiyama}, {Currie}, {Livingston}, {Kudo}, {Kusakabe}, {Abe}, {Brandner},
  {Brandt}, {Carson}, {Egner}, {Feldt}, {Goto}, {Grady}, {Guyon}, {Hayano},
  {Hayashi}, {Hayashi}, {Henning}, {Hodapp}, {Ishii}, {Iye}, {Janson},
  {Kandori}, {Knapp}, {Kwon}, {Matsuo}, {Mcelwain}, {Miyama}, {Morino},
  {Moro-Martin}, {Nishimura}, {Pyo}, {Serabyn}, {Suenaga}, {Suto}, {Suzuki},
  {Takahashi}, {Takami}, {Takato}, {Terada}, {Thalmann}, {Turner}, {Watanabe},
  {Wisniewski}, {Yamada}, {Takami}, {Usuda}, \& {Tamura}}]{uyama17}
{Uyama}, T., {Hashimoto}, J., {Kuzuhara}, M., {et~al.} 2017, \aj, 153, 106

\bibitem[{van Leeuwen(2007)}]{vanLeeuwen:2007dc}
van Leeuwen, F. 2007, A{\&}A, 474, 653

\bibitem[{Vigan {et~al.}(2012)Vigan, Patience, Marois, Bonavita, De~Rosa,
  Macintosh, Song, Doyon, Zuckerman, Lafreni{\`e}re, \& Barman}]{Vigan:2012jm}
Vigan, A., Patience, J., Marois, C., {et~al.} 2012, A\&A, 544, A9

\bibitem[{{Vigan} {et~al.}(2017){Vigan}, {Bonavita}, {Biller}, {Forgan},
  {Rice}, {Chauvin}, {Desidera}, {Meunier}, {Delorme}, {Schlieder}, {Bonnefoy},
  {Carson}, {Covino}, {Hagelberg}, {Henning}, {Janson}, {Lagrange}, {Quanz},
  {Zurlo}, {Beuzit}, {Boccaletti}, {Buenzli}, {Feldt}, {Girard}, {Gratton},
  {Kasper}, {Le Coroller}, {Mesa}, {Messina}, {Meyer}, {Montagnier},
  {Mordasini}, {Mouillet}, {Moutou}, {Reggiani}, {Segransan}, \&
  {Thalmann}}]{vigan:2017}
{Vigan}, A., {Bonavita}, M., {Biller}, B., {et~al.} 2017, \aap, 603, A3

\bibitem[{{Wahhaj} {et~al.}(2013){Wahhaj}, {Liu}, {Nielsen}, {Biller},
  {Hayward}, {Close}, {Males}, {Skemer}, {Ftaclas}, {Chun}, {Thatte}, {Tecza},
  {Shkolnik}, {Kuchner}, {Reid}, {de Gouveia Dal Pino}, {Alencar},
  {Gregorio-Hetem}, {Boss}, {Lin}, \& {Toomey}}]{wahhaj13}
{Wahhaj}, Z., {Liu}, M.~C., {Nielsen}, E.~L., {et~al.} 2013, \apj, 773, 179

\bibitem[{Wang {et~al.}(2015)Wang, Ruffio, De~Rosa, Aguilar, Wolff, \&
  Pueyo}]{Wang:2015th}
Wang, J.~J., Ruffio, J.-B., De~Rosa, R.~J., {et~al.} 2015, Astrophysics Source
  Code Library, -1, 06001

\bibitem[{{Wang} {et~al.}(2016){Wang}, {Graham}, {Pueyo}, {Kalas},
  {Millar-Blanchaer}, {Ruffio}, {De Rosa}, {Ammons}, {Arriaga}, {Bailey},
  {Barman}, {Bulger}, {Burrows}, {Cardwell}, {Chen}, {Chilcote}, {Cotten},
  {Fitzgerald}, {Follette}, {Doyon}, {Duch{\^e}ne}, {Greenbaum}, {Hibon},
  {Hung}, {Ingraham}, {Konopacky}, {Larkin}, {Macintosh}, {Maire}, {Marchis},
  {Marley}, {Marois}, {Metchev}, {Nielsen}, {Oppenheimer}, {Palmer}, {Patel},
  {Patience}, {Perrin}, {Poyneer}, {Rajan}, {Rameau}, {Rantakyr{\"o}},
  {Savransky}, {Sivaramakrishnan}, {Song}, {Soummer}, {Thomas}, {Vasisht},
  {Vega}, {Wallace}, {Ward-Duong}, {Wiktorowicz}, \& {Wolff}}]{Wang:2016gl}
{Wang}, J.~J., {Graham}, J.~R., {Pueyo}, L., {et~al.} 2016, \aj, 152, 97

\bibitem[{{Wang} {et~al.}(2018{\natexlab{a}}){Wang}, {Perrin}, {Savransky},
  {Arriaga}, {Chilcote}, {De Rosa}, {Millar-Blanchaer}, {Marois}, {Rameau},
  {Wolff}, {Shapiro}, {Ruffio}, {Maire}, {Marchis}, {Graham}, {Macintosh},
  {Ammons}, {Bailey}, {Barman}, {Bruzzone}, {Bulger}, {Cotten}, {Doyon},
  {Duch{\^e}ne}, {Fitzgerald}, {Follette}, {Goodsell}, {Greenbaum}, {Hibon},
  {Hung}, {Ingraham}, {Kalas}, {Konopacky}, {Larkin}, {Marley}, {Metchev},
  {Nielsen}, {Oppenheimer}, {Palmer}, {Patience}, {Poyneer}, {Pueyo}, {Rajan},
  {Rantakyr{\"o}}, {Schneider}, {Sivaramakrishnan}, {Song}, {Soummer},
  {Thomas}, {Wallace}, {Ward-Duong}, \& {Wiktorowicz}}]{Wang:2018}
{Wang}, J.~J., {Perrin}, M.~D., {Savransky}, D., {et~al.} 2018{\natexlab{a}},
  Journal of Astronomical Telescopes, Instruments, and Systems, 4, 018002

\bibitem[{{Wang} {et~al.}(2018{\natexlab{b}}){Wang}, {Graham}, {Dawson},
  {Fabrycky}, {De Rosa}, {Pueyo}, {Konopacky}, {Macintosh}, {Marois}, {Chiang},
  {Ammons}, {Arriaga}, {Bailey}, {Barman}, {Bulger}, {Chilcote}, {Cotten},
  {Doyon}, {Duch{\^e}ne}, {Esposito}, {Fitzgerald}, {Follette}, {Gerard},
  {Goodsell}, {Greenbaum}, {Hibon}, {Hung}, {Ingraham}, {Kalas}, {Larkin},
  {Maire}, {Marchis}, {Marley}, {Metchev}, {Millar-Blanchaer}, {Nielsen},
  {Oppenheimer}, {Palmer}, {Patience}, {Perrin}, {Poyneer}, {Rajan}, {Rameau},
  {Rantakyr{\"o}}, {Ruffio}, {Savransky}, {Schneider}, {Sivaramakrishnan},
  {Song}, {Soummer}, {Thomas}, {Wallace}, {Ward-Duong}, {Wiktorowicz}, \&
  {Wolff}}]{wang:2018b}
{Wang}, J.~J., {Graham}, J.~R., {Dawson}, R., {et~al.} 2018{\natexlab{b}}, \aj,
  156, 192

\bibitem[{{Weidenschilling}(1977)}]{Weidenschilling:1977}
{Weidenschilling}, S.~J. 1977, \mnras, 180, 57

\bibitem[{{Weinberger} {et~al.}(1999){Weinberger}, {Becklin}, {Schneider},
  {Smith}, {Lowrance}, {Silverstone}, {Zuckerman}, \&
  {Terrile}}]{1999ApJ...525L..53W}
{Weinberger}, A.~J., {Becklin}, E.~E., {Schneider}, G., {et~al.} 1999, \apjl,
  525, L53

\bibitem[{Wilner {et~al.}(2018)Wilner, MacGregor, Andrews, Hughes, Matthews, \&
  Su}]{Wilner:2018jy}
Wilner, D.~J., MacGregor, M.~A., Andrews, S.~M., {et~al.} 2018, ApJ, 855, 56

\bibitem[{{Zerbi} {et~al.}(1999){Zerbi}, {Rodr{\'{\i}}guez}, {Garrido},
  {Mart{\'{\i}}n}, {Arellano Ferro}, {Sareyan}, {Krisciunas}, {Akan}, {Evren},
  {Ibano{\v g}lu}, {Keskin}, {Pekunlu}, {Tunca}, {Luedeke}, {Paparo}, {Nuspl},
  \& {Guerrero}}]{zerbi:1999}
{Zerbi}, F.~M., {Rodr{\'{\i}}guez}, E., {Garrido}, R., {et~al.} 1999, \mnras,
  303, 275

\bibitem[{Zorec \& Royer(2012)}]{Zorec:2012il}
Zorec, J., \& Royer, F. 2012, A{\&}A, 537, A120

\bibitem[{{Zuckerman} {et~al.}(2006){Zuckerman}, {Bessell}, {Song}, \&
  {Kim}}]{2006ApJ...649L.115Z}
{Zuckerman}, B., {Bessell}, M.~S., {Song}, I., \& {Kim}, S. 2006, \apjl, 649,
  L115

\bibitem[{Zuckerman {et~al.}(2011)Zuckerman, Rhee, Song, \&
  Bessell}]{Zuckerman:2011bo}
Zuckerman, B., Rhee, J.~H., Song, I., \& Bessell, M.~S. 2011, ApJ, 732, 61

\bibitem[{{Zuckerman} {et~al.}(2001){Zuckerman}, {Song}, {Bessell}, \&
  {Webb}}]{zucerkman:2001}
{Zuckerman}, B., {Song}, I., {Bessell}, M.~S., \& {Webb}, R.~A. 2001, \apjl,
  562, L87

\end{thebibliography}

\appendix

\section{Properties of Target Stars}

\begin{longrotatetable}
\begin{deluxetable}{ccccccccccccccccc}
\tablewidth{0pt}
\tabletypesize{\tiny}
\tablecaption{300-star sample properties and observing log\label{tbl:sample}}
\tablehead{ \colhead{Name} & \colhead{R.A.} & \colhead{Dec.} & \colhead{SpT} & \colhead{ref} & \colhead{$d$} & \colhead{ref}  & \colhead{$I$} & \colhead{$H$} & \colhead{Moving} & \colhead{Age} & \colhead{ref} & \colhead{Mass} & \colhead{UT Date} & \colhead{$t_{\rm exp}\times n_{\rm coadd}$} & \colhead{$t_{\rm int}$} & \colhead{$\Delta$PA} \\
& & & & & & & & & Group & & & & & & & \\
& hh:mm:ss.ss & dd:mm:ss.s & & & pc  & & mag & mag & & Myr & & M$_{\odot}$ & & sec & min & deg}
\startdata
BU Psc & 00:02:02.59 & -02:45:58.2 & A8/9V & 1 & $64.70\pm0.38$ & 2 & 6.5 & 6.3 & \nodata & 554 & 3 & 1.49 & 20160920 & $59.6\times1$ & 34.8 & 21.3\\
HD 105 & 00:05:52.54 & -41:45:11.0 & G0V & 4 & $38.845\pm0.084$ & 2 & 6.9 & 6.2 & TucHor & 45 & 5 & 1.14 & 20141110 & $59.6\times1$ & 38.8 & 31.0\\
HR 9 & 00:06:50.08 & -23:06:27.1 & F3V & 4 & $39.960\pm0.098$ & 2 & 5.8 & 5.3 & $\beta$ Pic & 26 & 6 & 1.49 & 20141109 & $59.6\times1$ & 34.8 & 44.4\\
HD 987 & 00:13:52.83 & -74:41:17.5 & G8V & 4 & $45.847\pm0.051$ & 2 & 8.0 & 7.1 & TucHor & 45 & 5 & 0.94 & 20151105 & $59.6\times1$ & 28.8 & 16.6\\
HD 984 & 00:14:10.25 & -07:11:56.8 & F7V & 1 & $45.91\pm0.12$ & 2 & 6.8 & 6.2 & Columba & 42 & 5 & 1.28 & 20150830 & $59.6\times1$ & 21.9 & 15.1\\
 & & & & & & & & & & & & & 20160919 & $59.6\times1$ & 13.9 & 9.5\\
HD 1237 A & 00:16:12.70 & -79:51:04.0 & G8V & 4 & $17.5649\pm0.0073$ & 2 & 5.9 & 5.0 & Carina-Near & 200 & 7 & 1.01 & 20150831 & $59.6\times1$ & 34.8 & 13.6\\
HD 1466 & 00:18:26.10 & -63:28:39.0 & F8V & 4 & $42.970\pm0.055$ & 2 & 6.9 & 6.2 & TucHor & 45 & 5 & 1.19 & 20150703 & $59.6\times1$ & 42.7 & 21.0\\
\enddata
\tablenotetext{}{$\dagger$ - System mass}
\tablerefs{(1) \citealp{1999MSS...C05....0H}; (2) \citealp{GaiaCollaboration:2018io}; (3) This study (SED fitting); (4) \citealp{Torres:2006bw}; (5) \citealp{bell:2015}; (6) \citealp{Nielsen:2016ct}; (7) \citealp{2006ApJ...649L.115Z}; (8) \citealp{1989ApJS...71..245K}; (9) This study (activity); (10) \citealp{2006AJ....132..161G}; (11) \citealp{vanLeeuwen:2007dc}; (12) \citealp{1987ApJS...65..581G}; (13) \citealp{1975mcts.book.....H}; (14) \citealp{1995ApJS...99..135A}; (15) No reference cited in SIMBAD; (16) \citealp{Houk:1988wv}; (17) \citealp{2004ApJ...614..386B}; (18) \citealp{1994AJ....107.1556G}; (19) \citealp{2012AJ....143..114S}; (20) \citealp{1974AJ.....79..682H}; (21) \citealp{1957ArA.....2...55O}; (22) \citealp{Houk:1982vl}; (23) \citealp{2011ARep...55...31S}; (24) \citealp{1986AJ.....92..139S}; (25) \citealp{1970TOYal..30....1H}; (26) \citealp{1989ApJS...70..623G}; (27) \citealp{1978mcts.book.....H}; (28) \citealp{2013AJ....145..102L}; (29) \citealp{2015AJ....149..106D}; (30) \citealp{2009ApJS..180..117A}; (31) \citealp{1984ApJS...55..657C}; (32) \citealp{1991ApJS...77..417K}; (33) \citealp{2016MNRAS.461..794P}; (34) \citealp{1991JApA...12..133S}; (35) \citealp{2001AA...373..625P}; (36) \citealp{2006MNRAS.371..252L}; (37) \citealp{1970AJ.....75..507H}; (38) \citealp{2013MNRAS.435.1325M}; (39) \citealp{2015PASA...32...36M}; (40) \citealp{1999ApJ...525L..53W}; (41) \citealp{2003AJ....126.2048G}; (42) \citealp{1972PDAUC...2...59S}; (43) \citealp{1974ApJS...28....1J}; (44) \citealp{2000ApJ...541..390L}; (45) \citealp{2014yCat....1.2023S}; (46) \citealp{1993yCat.3135....0C}.}
\end{deluxetable}
\end{longrotatetable}

\end{document}